\documentclass[rmp,aps,twocolumn,epsfig,showpacs,floatfix]{revtex4}
\usepackage{graphicx,epsf}
\usepackage{amsmath}
\usepackage{bm}
\usepackage{pstricks}
\usepackage{psfrag}

\usepackage[english]{babel,varioref}
\usepackage[latin1]{inputenc}
\usepackage{bbm}

\usepackage{color}
\usepackage{amssymb}

\newcommand{\BEDT}{$\alpha-$(BEDT-TTF)$_2$I$_3$}

\def\gtrsim{\mathrel{\raise.4ex\hbox{$>$}\kern-0.8em\lower.7ex\hbox{$\sim$}}}
\def\lesssim{\mathrel{\raise.4ex\hbox{$<$}\kern-0.8em\lower.7ex\hbox{$\sim$}}}

\newcommand{\Amath}{\mathcal{A}}

\newcommand{\Omath}{\mathcal{O}}
\newcommand{\Hmath}{\mathcal{H}}
\newcommand{\Smath}{\mathcal{S}}

\newcommand{\Fmath}{\mathcal{F}}
\newcommand{\Nmath}{\mathcal{N}}

\newcommand{\bone}{\mathbbm{1}}

\newcommand{\sigmab}{\mbox{\boldmath $\sigma $}}
\newcommand{\etab}{\mbox{\boldmath $\eta $}}
\newcommand{\deltab}{\mbox{\boldmath $\delta $}}

\newcommand{\Pib}{\mbox{\boldmath $\Pi $}}

\newcommand{\qbar}{\bar{q}}
\newcommand{\rhobar}{\bar{\rho}}

\newcommand{\ahat}{\hat{a}}
\newcommand{\bhat}{\hat{b}}

\newcommand{\Hhat}{\hat{H}}

\newcommand{\bp}{{\bf p}}
\newcommand{\bq}{{\bf q}}
\newcommand{\bk}{{\bf k}}
\newcommand{\br}{{\bf r}}
\newcommand{\bA}{{\bf A}}
\newcommand{\ba}{{\bf a}}

\newcommand{\be}{{\bf e}}
\newcommand{\bn}{{\bf n}}

\newcommand{\bu}{{\bf u}}
\newcommand{\bv}{{\bf v}}
\newcommand{\bw}{{\bf w}}
\newcommand{\bR}{{\bf R}}

\newcommand{\bB}{{\bf B}}
\newcommand{\bE}{{\bf E}}
\newcommand{\bK}{{\bf K}}

\newcommand{\bG}{{\bf G}}

\newcommand{\eps}{\varepsilon}

\newcommand{\atilde}{\tilde{a}}

\newcommand{\wtilde}{\tilde{w}}

\newcommand{\qtilde}{\tilde{q}}

\newcommand{\ltilde}{\tilde{\lambda}}

\newcommand{\epstilde}{\tilde{\epsilon}}
\newcommand{\gtilde}{\tilde{\gamma}}

\newcommand{\Sbar}{\bar{S}}
\newcommand{\Ibar}{\bar{I}}

\newcommand{\otilde}{\tilde{\omega}}

\newcommand{\ua}{\uparrow}
\newcommand{\da}{\downarrow}

\newcommand{\SU}{{\rm SU}}
\newcommand{\U}{{\rm U}}

\newcommand{\eei}{\end{itemize}}
\newcommand{\bei}{\begin{itemize}}
\newcommand{\beq}{\begin{equation}}
\newcommand{\beqn}{\begin{eqnarray}}
\newcommand{\eeq}{\end{equation}}
\newcommand{\eeqn}{\end{eqnarray}}
\newcommand{\nn}{\nonumber}

\begin{document}

\title{Electronic Properties of Graphene in a Strong Magnetic Field}

\author{M. O. Goerbig}

\affiliation{$^1$Laboratoire de Physique des Solides,
Univ. Paris-Sud, CNRS UMR 8502, F-91405 Orsay, France}

\date{\today}

\begin{abstract}

We review the basic aspects of electrons in graphene (two-dimensional graphite) exposed to a strong perpendicular
magnetic field. One of its most salient features is the relativistic quantum Hall effect the observation of
which has been the experimental breakthrough 
in identifying pseudo-relativistic massless charge carriers as the low-energy excitations in graphene. The effect may be understood in
terms of Landau quantisation for massless Dirac fermions, which is also the theoretical basis for the understanding of more involved 
phenomena due to electronic interactions. We present the role of electron-electron interactions both in the weak-coupling limit, where
the electron-hole excitations are determined by collective modes, and in the strong-coupling regime of partially filled relativistic Landau
levels. In the latter limit, exotic ferromagnetic phases and incompressible quantum liquids are expected to be at the origin of
recently observed (fractional) quantum Hall states. Furthermore, we discuss briefly the electron-phonon coupling in a strong magnetic
field. Although the present review has a dominating theoretical character, a close connection with available experimental observation
is intended.


\end{abstract}

\maketitle

\tableofcontents



\section{Introduction to Graphene}
\label{intro}

The experimental and theoretical study of graphene, two-dimensional (2D) graphite, 
has become a major issue of modern condensed matter research. 
A milestone was the experimental evidence of an unusual quantum Hall effect reported
in September 2005 by two different groups, the Manchester group led by Andre Geim and a
Columbia-Princeton collaboration led by Philip Kim and Horst Stormer \cite{novoselov1,zhang1}.

The reasons for this enormous scientific interest
are manyfold, but one may highlight some major motivations.
First, one may underline its possible technological potential.
One of the first publications on graphene in 2004 by the Geim group reported indeed 
an electric field effect in graphene, i.e. the possibility to
control the carrier density in the graphene sheet by simple application of a gate
voltage \cite{geimFET}. This effect is a fundamental element for the design of 
electronic devices. In a contemporary publication
Berger \textit{et al.} reported on the fabrication and the electrical contacting of monolayer graphene samples 
on epitaxially grown SiC crystals \cite{berger}.
Today's silicon-based electronics reaches its limits in 
miniaturization, which is on the order of 50 nm for an electric channel, whereas 
it has been shown that a narrow graphene strip with a width of only a few nanometers 
may be used as a transistor \cite{nanoTrans}, i.e. as the basic electronics component.

Apart from these promising technological applications, two major motivations for
fundamental research may be emphasized. Graphene is the first truely 2D crystal 
ever observed in nature and possess remarkable mechanical properties. 
Furthermore, 
electrons in graphene show relativistic behavior, and the system is therefore 
an ideal candidate for the test of quantum-field theoretical models that have 
been developed in high-energy physics. Most promenently, electrons in graphene may
be viewed as massless charged fermions living in 2D space, particles one usually does 
not encounter in our three-dimensional world. Indeed, all massless elementary 
particles happen to be electrically neutral, such as photons or 
neutrinos.\footnote{The neutrino example is only partially correct. The observed oscillation
between different neutrino flavors ($\nu_{\mu}\leftrightarrow \nu_{\tau}$) requires
indeed a tiny non-zero mass \cite{neutrino}.} Graphene is therefore
an exciting bridge between condensed-matter and high-energy physics, and the 
research on its electronic properties unites scientists with various thematic
backgrounds.

Several excellent reviews witness the enormous research achievements in graphene. 
In a first step those by
Geim and Novoselov \cite{ExpRev} and by de Heer \cite{epitaxialG} aimed at a rather global experimental 
review of exfoliated and epitaxial graphene,
respectively. Furthermore, the review by Castro Neto \cite{antonioRev} was concerned with general theoretical 
issues of electrons in graphene.
Apart from the review by Abergel \textit{et al.} \cite{AbergelRev},
more recent reviews concentrate on the subfields of graphene research,
which have themselves grown to a considerable size and that require reviews on their own. As 
an example one may cite the review by Peres \cite{peresRMP}, which is concerned with transport
properties of graphene, or that by Kotov and co-workers on interaction effects \cite{kotovRev}.
The present theoretical review deals with electronic properties of graphene in a strong magnetic field, and
its scope is delimited to mono-layer graphene. The vast amount of knowledge on bilayer graphene
certainly merits a review on its own.

\subsection{The Carbon Atom and its Hybridizations}
\label{carbon}

\begin{figure}
\centering
\includegraphics[width=6.5cm,angle=0]{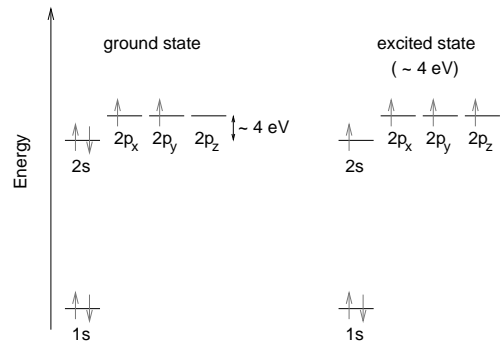}
\caption{\footnotesize{Electronic configurations for carbon in the ground state (left) and in the excited state (right).}}
\label{fig1:02}
\end{figure}

In order to understand the crystallographic structure of graphene and carbon-based materials in general, 
it is useful to review the basic chemical bonding properties of carbon atoms. The carbon atom possesses 6
electrons, which, in the atomic ground state, are in the configuration 
1s$^2$2s$^2$2p$^2$, i.e. 2 electrons fill the inner shell 1s, which is close to 
the nucleus and which is irrelevant for chemical reactions, whereas 4 electrons 
occupy the outer shell of 2s and 2p orbitals. Because the 2p orbitals (2p$_x$, 2p$_y$,
and 2p$_z$) are roughly 4 eV higher in energy than the 2s orbital, it is energetically 
favorable to put 2 electrons in the 2s orbital and only 2 of them in the 2p orbitals
(Fig \ref{fig1:02}). It turns out, however, that in the presence of other atoms, such 
as e.g. H, O, or other C atoms, it is favorable to excite one electron from the
2s to the third 2p orbital, in order to form covalent bonds with the other atoms. 

In the excited state, we therefore have four equivalent quantum-mechanical states,
$|2s\rangle$, $|2p_x\rangle$, $|2p_y\rangle$, and $|2p_z\rangle$. A quantum-mechanical
superposition of the state $|2s\rangle$ with $n$ $|2p_j\rangle$ states is called
sp$^n$ hybridization.
The sp$^1$ hybridization plays, e.g., an important role in the context of organic
chemistry (such as the formation of acetylene) and the sp$^3$ hybridization gives
rise to the formation of diamonds, a particular 3D form of carbon. Here, however, we are interested
in the planar sp$^2$ hybridization, which is the basic ingredient for the graphitic allotropes. 

\begin{figure}
\centering
\includegraphics[width=7.5cm,angle=0]{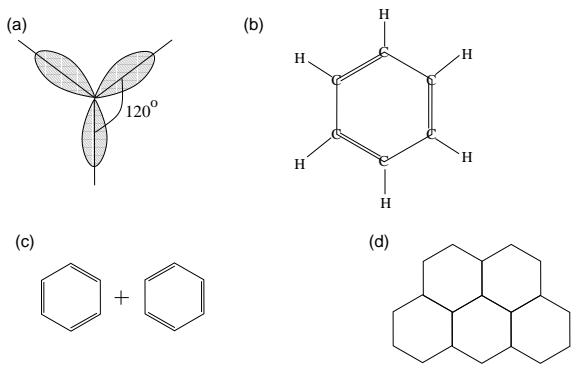}
\caption{\footnotesize{\textit{(a)} Schematic view of the sp$^2$ hybridization. 
The orbitals form angles of 120$^o$.
\textit{(b)} Benzene molecule (C$_6$H$_6$). The 6 carbon atoms are situated at the
corners of a hexagon and form covalent bonds with the H atoms. 
\textit{(c)} The quantum-mechanical ground state of the benzene ring is a superposition of the two 
configurations which differ by the position of the $\pi$ bonds. 
\textit{(d)} Graphene may be viewed as a tiling of benzene hexagons, where the H
atoms are replaced by C atoms of neighboring hexagons and where the $\pi$
electrons are delocalized over the whole structure.
}}
\label{fig1:04}
\end{figure}

As shown in Fig. \ref{fig1:04}, the three sp$^2$-hybridized orbitals are oriented in the $xy$-plane and have mutual 
120\textdegree~angles. The remaining unhybridized 2p$_z$ orbital is perpendicular to the plane.

A prominent chemical example for such hybridization is the benzene molecule the chemical structure
of which has been analyzed by August Kekul\'e in 1865 \cite{kekule1,kekule2}. 
The molecule
consists of a hexagon with carbon atoms at the corners linked by $\sigma$ bonds
[Fig. \ref{fig1:04} (b)]. Each carbon atom
has, furthermore, a covalent bond with one of the hydrogen atoms 
which stick out from the hexagon in a star-like
manner. In addition to the six $\sigma$ bonds, the remaining 2p$_z$ orbitals form three $\pi$ bonds, and the resulting
double bonds alternate with single $\sigma$ bonds around the hexagon. Because a double bond is stronger
than a single $\sigma$ bond, one may expect that the hexagon is not perfect. A double bond (C=C) yields indeed
a carbon-carbon distance of $0.135$ nm, whereas it is $0.147$ nm for a single $\sigma$ bond (C--C). However, 
the measured carbon-carbon distance in benzene is $0.142$ nm for all bonds,
which is roughly the average length of a single and a double bond. This equivalence 
of all bonds in benzene
was explained by Linus Pauling in 1931 within a quantum-mechanical treatment of the benzene ring \cite{pauling}. The 
ground state is indeed a quantum-mechanical superposition of the two possible configurations for 
the double bonds, as shown schematically in Fig. \ref{fig1:04} (c). 

These chemical considerations indicate the way towards 
carbon-based condensed matter physics -- any graphitic 
compound has indeed a sheet of graphene as its basic constituent. Such a graphene sheet may
be viewed simply as a tiling of benzene hexagons, where the hydrogen are replaced by carbon
atoms to form a neighboring carbon hexagon [Fig. \ref{fig1:04} (d)]. 
However, graphene has remained the basic constituent of graphitic systems during
a long time only on the theoretical level. From an experimental point of view, 
graphene is the youngest allotrope and accessible to electronic-transport measurements only since
2004.

For a detailed discussion of the different fabrication techniques, the most popular of which are the exfoliation
technique \cite{2Dcrystals} and thermal graphitization of epitaxially-grown SiC crystals \cite{berger}, 
we refer the reader to existing experimental reviews \cite{ExpRev,epitaxialG}. Notice that, more recently, 
large-scale graphene has been 
fabricated by chemical vapor deposition \cite{Reina09} that seems a promising technique not only for fundamental research
but also for technological applications.

\subsection{Crystal Structure of Graphene}
\label{Sec:crystal}

\begin{figure}
\centering
\includegraphics[width=8.5cm,angle=0]{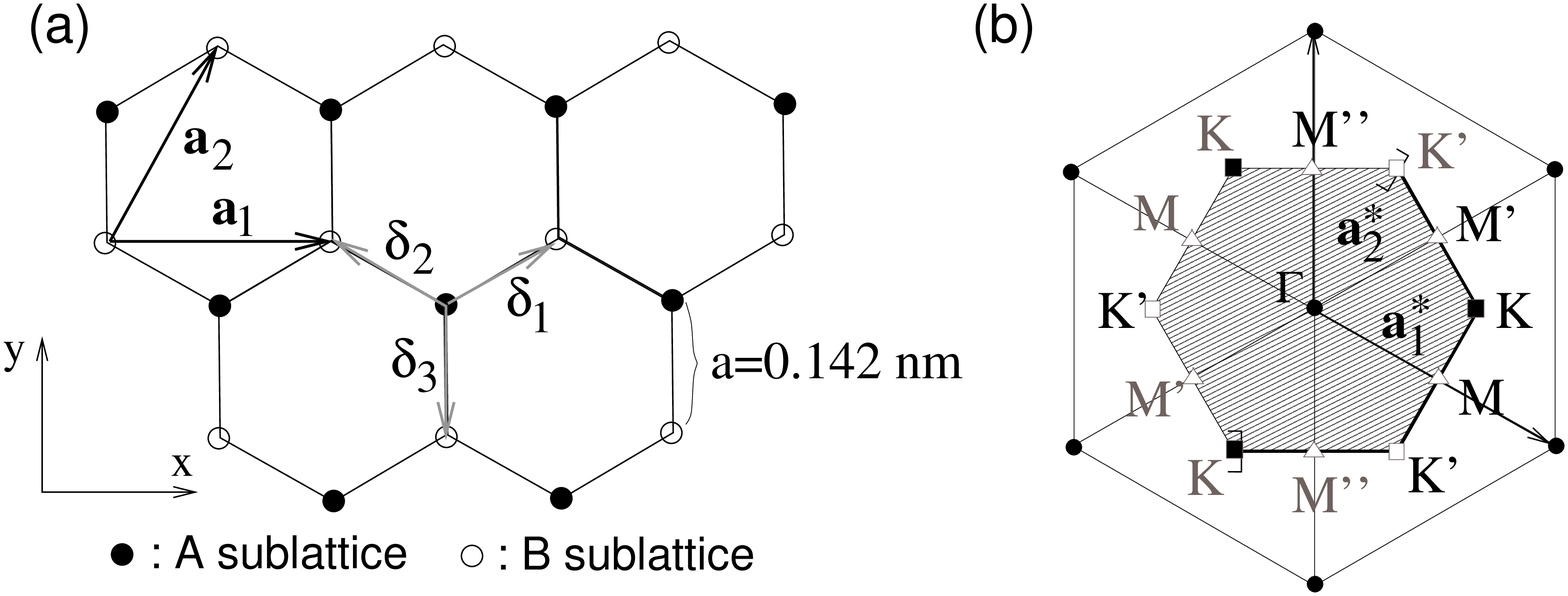}
\caption{\footnotesize{\textit{(a)} Honeycomb lattice. 
The vectors $\deltab_1$, $\deltab_2$,
and $\deltab_3$ connect \textit{nn} carbon atoms, separated by a distance $a=0.142$ nm.
The vectors $\ba_1$ and $\ba_2$ are basis vectors of the triangular Bravais
lattice. 
\textit{(b)} Reciprocal lattice of the triangular lattice. Its primitive lattice
vectors are $\ba_1^*$ and $\ba_2^*$. The shaded region represents the first Brillouin
zone (BZ), with its center $\Gamma$ and 
the two inequivalent corners $K$ (black squares)
and $K'$ (white squares). The thick part of the border of the first BZ represents
those points which are counted in its definition such that no points are
doubly counted. The first BZ, defined in a strict manner, is, thus, the shaded 
region plus the thick part of the border. For completeness, we have also
shown the three inequivalent cristallographic points $M$, $M'$, and $M''$
(white triangles).
}}
\label{fig1:07}
\end{figure}

As already mentioned in the last section, the carbon atoms in graphene condense 
in a honeycomb lattice due to their sp$^2$ hybridization. The honeycomb lattice
is not a Bravais lattice because two neighboring sites are inequivalent
from a crystallographic point of view.\footnote{This needs to be clearly distinguished from a chemical
point of view according to which they may be equivalent as in the case of graphene where both
types of sites consist of carbon atoms.} Fig.
\ref{fig1:07} (a) illustrates indeed that a site on the A sublattice has nearest
neighbors (\textit{nn}) in the directions north-east, north-west, and south, whereas
a site on the B sublattice has \textit{nn}s in the directions north, south-west, and
south-east. Both A and B sublattices, however, are triangular Bravais lattices,
and one may view the honeycomb lattice as a \textit{triangular Bravais lattice with a 
two-atom basis} (A and B).
The distance between \textit{nn} carbon atoms is $a=0.142$ nm, which is the
average of the single (C--C) and double (C=C) covalent $\sigma$ bonds, as in
the case of benzene.

The three vectors which connect a site on the A sublattice with a \textit{nn} on
the B sublattice are given by
\beq\label{eq1:02}
\deltab_1=\frac{a}{2}\left(\sqrt{3}\be_x+\be_y\right), \
\deltab_2=\frac{a}{2}\left(-\sqrt{3}\be_x+\be_y\right), \
\deltab_3=-a\be_y,
\eeq
and the triangular Bravais lattice is spanned by the basis vectors
\beq\label{eq1:03}
\ba_1=\sqrt{3}a\be_x \qquad {\rm and}\qquad \ba_2=\frac{\sqrt{3}a}{2}
\left(\be_x+\sqrt{3}\be_y\right).
\eeq
The modulus of the basis vectors yields the lattice spacing, 
$\tilde{a}=\sqrt{3}a=0.24$ nm, and
the area of the unit cell is $A_{uc}=\sqrt{3}\tilde{a}^2/2=0.051$ nm$^2$.
The density of carbon atoms is, therefore, $n_C=2/A_{uc}=39$ 
nm$^{-2}=3.9\times 10^{15}$ cm$^{-2}$. Because there is one $\pi$ electron
per carbon atom that is not involved in a covalent $\sigma$ bond, there are
as many valence electrons as carbon atoms, and their density is, thus,
$n_{\pi}=n_C=3.9\times 10^{15}$ cm$^{-2}$. As discussed in detail below,
this density is not equal to the carrier density in graphene, which
one measures in electric transport measurements.

The reciprocal lattice, which is defined with respect to the triangular Bravais
lattice, is depicted in Fig. \ref{fig1:07} (b).
It spanned by the vectors 
\beq\label{eq1:04}
\ba_1^*=\frac{2\pi}{\sqrt{3}a}\left(\be_x-\frac{\be_y}{\sqrt{3}}\right)\qquad
{\rm and}\qquad \ba_2^*=\frac{4\pi}{3a}\be_y.
\eeq
Physically, all sites of the reciprocal lattice represent equivalent wave vectors.
The first Brillouin zone [BZ, shaded region 
and thick part of the border of the hexagon in Fig. \ref{fig1:07} (b)] 
is defined as the set of inequivalent points in reciprocal space,
i.e. of points which may not be connected to one another by a reciprocal lattice
vector. The 
long wavelength excitations are situated in the vicinity of the $\Gamma$ point, in the
center of the first BZ. Furthermore, one distinguishes the six corners of the 
first BZ, which consist of the inequivalent points $K$ and $K'$ represented by 
the vectors 
\beq\label{Kpoint}
\pm \bK=\pm  \frac{4\pi}{3\sqrt{3}a}\be_x.
\eeq
The four remaining corners [shown in gray in Fig. \ref{fig1:07} (b)] may indeed
be connected to one of these points via a translation by a reciprocal lattice
vector. These cristallographic points play an essential role in the electronic
properties of graphene because their low-energy excitations are centered around
the two points $K$ and $K'$, as is discussed in detail in the following section.
We emphasise, because of some confusion in the literature on this point, that the 
inequivalence of the two BZ corners, $K$ and $K'$, has nothing to do with the
presence of two sublattices, $A$ and $B$, in the honeycomb lattice. The form of the
BZ is an intrinsic property of the \textit{Bravais} lattice, independent of the 
possible presence of more than one atom in the unit cell.
For completeness, we have also shown, in Fig. \ref{fig1:07} (b), the 
three crystallographically inequivalent $M$ points in the middle of the BZ edges.

\subsection{Electronic Band Structure of Graphene}
\label{Sec:bands}

As we have discussed in the previous section, three electrons per carbon atom in graphene
are involved in the formation of strong covalent $\sigma$ bonds, and one 
electron per atom yields the $\pi$ bonds. The $\pi$ electrons happen to be those 
responsible for the electronic properties at low energies, whereas the $\sigma$
electrons form energy bands far away from the Fermi energy \cite{SDD}. This section of the introduction is, thus,
devoted to a brief discussion of the energy bands of $\pi$ electrons within the tight-binding
approximation, which was originally calculated for the honeycomb lattice
by P. R. Wallace in 1947 \cite{wallace}.

\subsubsection{Tight-binding model for electrons on the honeycomb lattice}
\label{Sec:TB}

In the case of two atoms per unit cell, we may write down a trial wave function
\beq\label{eq2:06}
\psi_{\bk}(\br)=a_{\bk}\psi^{(A)}_{\bk}(\br)+b_{\bk}\psi^{(B)}_{\bk}(\br),
\eeq
where $a_{\bk}$ and $b_{\bk}$ are complex functions of the quasi-momentum
$\bk$. Both $\psi^{(A)}_{\bk}(\br)$ and $\psi^{(B)}_{\bk}(\br)$ 
are Bloch functions with
\beq\label{eq2:07}
\psi_{\bk}^{(j)}(\br)=\sum_{\bR_l}e^{i\bk\cdot\bR_l}\phi^{(j)}(\br+\deltab_j-\bR_l),
\eeq
where $j=A/B$ labels the atoms on the two sublattices A and B, and $\deltab_j$ is
the vector which connects the sites of the underlying Bravais lattice with the
site of the $j$ atom within the unit cell. 
The $\phi^{(j)}(\br+\deltab_j-\bR_l)$ are atomic orbital wave functions for 
electrons that are in the vicinity of the $j$ atom situated at the position
$\bR_l-\deltab_j$ at the (Bravais) lattice site $\bR_l$.
Typically one chooses the sites of one
of the sublattices, e.g. the A sublattice, to coincide with the sites of the 
Bravais lattice. Notice furthermore that there is some arbitrariness in the choice
of the phase in Eq. (\ref{eq2:07}) -- instead of choosing $\exp(i\bk\cdot\bR_l)$, one
may also have chosen $\exp[i\bk\cdot(\bR_l-\deltab_j)]$, for the
atomic wave functions. The choice, however, does not affect the physical 
properties of the system because it simply leads to a redefinition of the 
weights $a_{\bk}$ and $b_{\bk}$ which aquire a different relative phase \cite{bena}.

With the help of these wave functions, we may now search the solutions of the 
Schr\"odinger equation
$H\psi_{\bk}=\epsilon_{\bk}\psi_{\bk}$.
Multiplication of the Schr\"odinger equation by $\psi_{\bk}^*$ 
from the left yields the equation $\psi_{\bk}^*H\psi_{\bk}=\epsilon_{\bk}
\psi_{\bk}^*\psi_{\bk}$, which may be rewritten in matrix form 
with the help of Eq. (\ref{eq2:06}) 
\beq\label{eq2:08}
\left(a_{\bk}^*,b_{\bk}^*\right)\Hmath_{\bk}
\left(\begin{array}{c} a_{\bk} \\ b_{\bk}\end{array}\right)=
\epsilon_{\bk}\left(a_{\bk}^*,b_{\bk}^*\right)\Smath_{\bk}
\left(\begin{array}{c} a_{\bk} \\ b_{\bk}\end{array}\right).
\eeq
Here, the Hamiltonian matrix is defined as
\beq\label{eq2:09}
\Hmath_{\bk}\equiv \left(
\begin{array}{cc}
 \psi_{\bk}^{(A)*}H\psi_{\bk}^{(A)} & \psi_{\bk}^{(A)*}H\psi_{\bk}^{(B)} \\
\psi_{\bk}^{(B)*}H\psi_{\bk}^{(A)} & \psi_{\bk}^{(B)*}H\psi_{\bk}^{(B)} 
\end{array}\right)
=\Hmath_{\bk}^{\dagger},
\eeq
and the overlap matrix
\beq\label{eq2:10}
\Smath_{\bk}\equiv \left(
\begin{array}{cc}
 \psi_{\bk}^{(A)*}\psi_{\bk}^{(A)} & \psi_{\bk}^{(A)*}\psi_{\bk}^{(B)} \\
\psi_{\bk}^{(B)*}\psi_{\bk}^{(A)} & \psi_{\bk}^{(B)*}\psi_{\bk}^{(B)} 
\end{array}\right)
=\Smath_{\bk}^{\dagger}
\eeq
accounts for the non-orthogonality of the trial wave functions.
The eigenvalues $\epsilon_{\bk}$ of the Schr\"odinger equation yield the
electronic bands, and they may be obtained from the
secular equation
\beq\label{eq2:11}
\det\left[\Hmath_{\bk}-\epsilon_{\bk}^{\lambda}\Smath_{\bk}\right]=0,
\eeq
which needs to be satisfied for a non-zero solution of the wave functions,
i.e. for $a_{\bk}\neq 0$ and $b_{\bk}\neq 0$. The label $\lambda$ denotes 
the energy bands, and it is clear that there are as many energy bands as 
solutions of the secular equation (\ref{eq2:11}), i.e. two bands for the 
case of two atoms per unit cell.

\paragraph{Formal solution.}

Before turning to the specific case of graphene and its energy bands, we solve formally
the secular equation for an arbitrary lattice with several atoms per unit cell.
The Hamiltonian matrix (\ref{eq2:09}) may be written, with the help of Eq.
(\ref{eq2:07}), as 
\beq\label{eq2:12b}
\Hmath_{\bk}^{ij} = N\left(\epsilon^{(j)}s_{\bk}^{ij}+t_{\bk}^{ij}\right)
\eeq
where ($\deltab_{ij}\equiv \deltab_j-\deltab_i$), 
\beq\label{eq2:13b}
s_{\bk}^{ij}\equiv \sum_{\bR_l} e^{i\bk\cdot\bR_l}\int d^2r\, 
\phi^{(i)*}(\br)\phi^{(j)}(\br+\deltab_{ij}-\bR_l)
=\frac{\Smath_{\bk}^{ij}}{N}
\eeq
and we have defined the \textit{hopping matrix}
\beq\label{eq2:13}
t_{\bk}^{ij}\equiv \sum_{\bR_l} e^{i\bk\cdot\bR_l}\int d^2r\, 
\phi^{(i)*}(\br)\Delta V\phi^{(j)}(\br+\deltab_{ij}-\bR_l)\ .
\eeq
Here, $N$ is the number of unit cells, and
we have separated the Hamiltonian $H$ into an atomic orbital part
$H^{a}=-(\hbar^2/2m)\Delta+ V(\br-\bR_l+\deltab_j)$,
which satisfies the eigenvalue equation 
$H^a\phi^{(j)}(\br+\deltab_j-\bR_l)=\epsilon^{(j)}\phi^{(j)}(\br+\deltab_j-\bR_l)$
and a ``perturbative part'' $\Delta V$ which takes into account the potential term
that arises from all other atoms different from that in the atomic orbital Hamiltonian.
The last line in Eq. (\ref{eq2:12b}) has been obtained from the fact that the
atomic wave functions $\phi^{(i)}(\br)$ are eigenstates of the atomic Hamiltonian
$H^a$ with the atomic energy $\epsilon^{(i)}$ for an orbital of type $i$. This atomic
energy plays the role of an onsite energy. The secular equation now reads
$\det[t_{\bk}^{ij}-(\epsilon_{\bk}^\lambda-\epsilon^{(j)})s_{\bk}^{ij}]=0$.
Notice that, if the the atoms on the different sublattices are all of the same 
electronic configuration, one has $\epsilon^{(i)}=\epsilon_0$ for all $i$, and
one may omit this on-site energy, which yields only a constant and physically irrelevant shift of the energy bands.

\paragraph{Solution for graphene with nearest-neighbor and next-nearest-neighour
hopping.}

\begin{figure}
\centering
\includegraphics[width=4.5cm,angle=0]{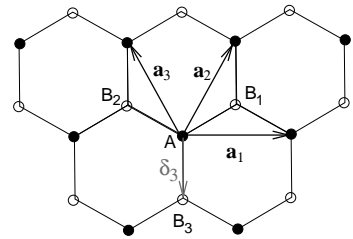}
\caption{\footnotesize{Tight-binding model for the honeycomb lattice.}}
\label{fig2:01}
\end{figure}

After these formal considerations, we now study the particular case of the 
tight-binding model on the honeycomb lattice, which yields, to great accuracy,
the $\pi$ energy bands of graphene. Because all atomic orbitals are $p_z$ 
orbitals of carbon atoms, we may omit the onsite energy $\epsilon_0$, as discussed
in the last paragraph. We choose the Bravais lattice vectors to be those of 
the A sublattice, i.e. $\deltab_A=0$, and the equivalent site on the B sublattice
is obtained by the displacement $\deltab_B=\deltab_{AB}=\deltab_3$ (see Fig. 
\ref{fig2:01}). The \textit{nn} hopping amplitude is given by the expression
\beq\label{eq2:15}
t\equiv \int d^2r\, \phi^{A*}(\br)\Delta V\phi^{B}(\br+\deltab_3),
\eeq
and we also take into account next-nearest neighbor \textit{(nnn)} hopping which connects neighboring sites 
on the \textit{same} sublattice
\beq\label{eq2:16}
t_{nnn} \equiv \int d^2r\, \phi^{A*}(\br)\Delta V\phi^{A}(\br+\ba_1)
\eeq
Notice that one may have chosen any other vector $\deltab_j$ or $\ba_2$, respectively,
in the calculation of the hopping amplitudes. Because of the normalization of
the atomic wave functions, we have $\int d^2r\phi^{(j)*}(\br)\phi^{(j)}(\br)=1$, and
we consider furthermore the overlap correction between orbitals on \textit{nn} sites,
\beq\label{eq2:17}
s\equiv \int d^2r\, \phi^{A*}(\br)\phi^{B}(\br+\deltab_3).
\eeq
We neglect overlap corrections between all other orbitals which are not \textit{nn}, as
well as hopping amplitudes for larger distances than \textit{nnn}.

If we now consider an arbitrary site $A$ on the A sublattice (Fig. \ref{fig2:01}),
we may see that the off-diagonal terms of the hopping matrix (\ref{eq2:13}) consist
of three terms corresponding to the \textit{nn} $B_1$, $B_2$, and $B_3$, all of which
have the same hopping amplitude $t$. However, only the site $B_3$ is described by the
same lattice vector (shifted by $\deltab_3$) as the site $A$ and thus yields a zero
phase to the hopping matrix. The sites $B_1$ and $B_2$ correspond to lattice vectors
shifted by $\ba_2$ and $\ba_3\equiv\ba_2-\ba_1$, respectively. Therefore, they
contribute a phase factor $\exp(i\bk\cdot\ba_2)$ and $\exp(i\bk\cdot\ba_3)$, 
respectively. The off-diagonal elements of the hopping matrix may then be
written as\footnote{The hopping matrix element $t_{\bk}^{AB}$ corresponds to a hopping 
from the $B$ to the $A$ sublattice.}
$t_{\bk}^{AB}=t\gamma_{\bk}^*=(t_{\bk}^{BA})^*$,
as well as those of the overlap matrix
$s_{\bk}^{AB}=s\gamma_{\bk}^*=(s_{\bk}^{BA})^*$,
($s_{\bk}^{AA}=s_{\bk}^{BB}=1$, due to the above-mentioned normalization of the 
atomic wave functions), where we have defined the sum of the \textit{nn} phase factors
\beq\label{eq2:18}
\gamma_{\bk}\equiv 1+e^{i\bk\cdot\ba_2}+e^{i\bk\cdot\ba_3}.
\eeq
The \textit{nnn} hopping amplitudes yield the diagonal elements of the hopping 
matrix,
\beq
t_{\bk}^{AA}=t_{\bk}^{BB} = 2t_{nnn}\sum_{i=1}^3\cos(\bk\cdot\ba_i)=t_{nnn}\left(|\gamma_{\bk}|^2-3\right),
\eeq
and one obtains, thus, the secular equation
\beq
\det\left[
\begin{array}{cc}
 t_{\bk}^{AA} - \epsilon_{\bk} & (t-s\epsilon_{\bk})\gamma_{\bk}^*\\
 (t-s\epsilon_{\bk})\gamma_{\bk} & t_{\bk}^{AA} - \epsilon_{\bk} 
\end{array}\right]=0
\eeq
with the two solutions ($\lambda=\pm$)
\beq\label{eq2:19}
\epsilon_{\bk}^{\lambda}=\frac{t_{\bk}^{AA}+ \lambda t|\gamma_{\bk}|}{
1+\lambda s|\gamma_{\bk}|}\ .
\eeq
This expression may be expanded under the reasonable assumptions $s\ll 1$ and 
$t_{nnn}\ll t$, which we further justify at the end of the paragraph,
\beqn\label{eq2:20}
\nn
\epsilon_{\bk}^{\lambda} &\simeq & t_{\bk}^{AA}+\lambda t|\gamma_{\bk}|-st|\gamma_{\bk}|^2
= t_{nnn}'|\gamma_{\bk}|^2 +\lambda t|\gamma_{\bk}|\\
\nn
&=& t_{nnn}'\left[3+2\sum_{i=1}^3\cos(\bk\cdot\ba_i)\right]\\
&&+\lambda t\sqrt{3+2\sum_{i=1}^3\cos(\bk\cdot\ba_i)},
\eeqn
where we have defined the effective \textit{nnn} hopping amplitude 
$t_{nnn}'\equiv t_{nnn}-st$,
and we have omitted the unimportant constant $-3t_{nnn}$ in the second step.
Therefore, the {\nn} overlap corrections simply yield a 
renormalization of the \textit{nnn} hopping amplitudes. The hopping amplitudes 
may be determined by fitting the energy dispersion (\ref{eq2:20}) obtained 
within the tight-binding approximation to those calculated numerically in
more sophisticated band-structure calculations \cite{partoens} or to spectroscopic
measurements \cite{mucha}. These yield a value
of $t\simeq -3$ eV for the \textit{nn} hopping amplitude and $t_{nnn}'\simeq 0.1t$, which
justifies the above-mentioned expansion for $t_{nnn}'/t\ll 1$. Notice that 
this fitting procedure does not allow for a distinction between the ``true''
\textit{nnn} hopping amplitude $t_{nnn}$ and the contribution from the overlap
correction $-st$. We, therefore, omit this distinction in the following discussion
and drop the prime at the effective \textit{nnn} hopping amplitude, but 
one should keep in mind that it is an effective parameter with a contribution
from \textit{nn} overlap corrections.

\paragraph{Energy dispersion of $\pi$ electrons in graphene.}

\begin{figure}
\includegraphics[width=5.0cm,angle=0]{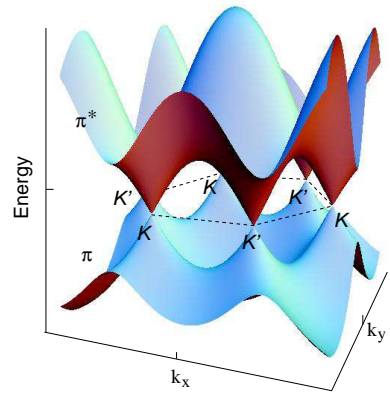}
\caption{\footnotesize{Energy dispersion as a function
of the wave-vector components $k_x$ and $k_y$, obtained within the tight-binding
approximation, for $t_{nnn}/t=0.1$. One distinguishes the valence ($\pi$) band
from the conduction ($\pi^*$) band. The Fermi level is situated at the points
where the $\pi$ band touches the $\pi^*$ band. The energy is measured in units
of $t$ and the wave vector in units of $1/a$.
}}
\label{fig2:02}
\end{figure}

The energy dispersion (\ref{eq2:20}) is plotted in Fig. \ref{fig2:02} for
$t_{nnn}/t=0.1$. It consists of two bands, labeled by the index $\lambda=\pm$, each of 
which contains the same number of states. 
Because each carbon atom contributes one
$\pi$ electron and each electron may occupy either a spin-up or a spin-down state, the
lower band with $\lambda=-$ (the $\pi$ or valence band) is completely filled and that 
with $\lambda=+$ (the $\pi^*$ or conduction band) completely empty. The Fermi
level is, therefore, situated at the points, called \textit{Dirac points}, where the $\pi$ band touches the $\pi^*$ 
band. 
Notice that only if $t_{nnn}=0$ the energy dispersion (\ref{eq2:20}) is electron-hole 
symmetric, i.e. $\epsilon_{\bk}^{\lambda}=-\epsilon_{\bk}^{-\lambda}$. This means
that 
\textit{nnn} hopping and \textit{nn} overlap corrections break the electron-hole symmetry.
The Dirac points are situated at the points $\bk^D$ where the
energy dispersion (\ref{eq2:20}) is zero, 
\beq\label{eq2:22}
\epsilon_{\bk^D}^{\lambda}=0.
\eeq
Eq. (\ref{eq2:22}) is satisfied when $\gamma_{\bk^D}=0$, i.e. when
\beqn\label{eq2:23}
{\rm Re}\gamma_{\bk^D} 
\nn
&=& 1+\cos\left[\frac{\sqrt{3}a}{2}(k_x^D+\sqrt{3}k_y^D)\right]\\
&& +\cos\left[\frac{\sqrt{3}a}{2}(-k_x^D+\sqrt{3}k_y^D)\right]=0
\eeqn
and, equally,
\beqn\label{eq2:24}
\nn
{\rm Im}\gamma_{\bk^D} &=& \sin\left[\frac{\sqrt{3}a}{2}(k_x^D+\sqrt{3}k_y^D)\right]\\
&&+\sin\left[\frac{\sqrt{3}a}{2}(-k_x^D+\sqrt{3}k_y^D)\right]=0.
\eeqn
The last equation may be satisfied by the choice $k_y^D=0$, and Eq. (\ref{eq2:23}),
thus, when 
\beq
1+2\cos\left(\frac{\sqrt{3}a}{2}k_x^D\right) =0
\qquad \Rightarrow \qquad k_x^D=\pm\frac{4\pi}{3\sqrt{3}a}.
\eeq
Comparison with Eq. (\ref{Kpoint}) shows that there are, thus, two 
inequivalent Dirac points $D$ and $D'$, which are situated at the points $K$ and
$K'$, respectively,
\beq\label{eq2:25}
\bk^D=\pm \bK=\pm \frac{4\pi}{3\sqrt{3}a}\be_x.
\eeq
Although situated at the same position in the first BZ, it is useful to make a 
clear conceptual
distinction between the Dirac points $D$ and $D'$, which are defined as the contact points between
the two bands $\pi$ and $\pi^*$, and the crystallographic
points $K$ and $K'$, which are defined as the corners of the first BZ. There are, 
indeed, situations where the Dirac points move away from the points $K$ and 
$K'$, as we will discuss in Sec. \ref{DefGraph}. 

Notice that the band Hamiltonian (\ref{eq2:09}) respects time-reversal symmetry,
$\Hmath_{\bk}=\Hmath_{-\bk}^*$, which implies  $\epsilon_{-\bk}=\epsilon_{\bk}$ 
for the dispersion relation. Therefore, if $\bk^D$ is a solution of $\epsilon_{\bk}=0$, so is 
$-\bk^D$, and Dirac points thus necessarily occur in pairs. 
In graphene, there is one pair of Dirac points, and the zero-energy states
are, therefore, doubly degenerate. One speaks of a twofold \textit{valley degeneracy}, 
which survives when we consider low-energy electronic excitations that are restricted
to the vicinity of the Dirac points, as is discussed in Sec. \ref{Sec:continuum}.

\paragraph{Effective tight-binding Hamiltonian.}

Before considering the low-energy excitations and the continuum limit, it is
useful to define an effective tight-binding Hamiltonian,
\beq\label{eq2:26} 
\Hmath_{\bk}\equiv t_{nnn}|\gamma_{\bk}|^2\bone +t \left(
\begin{array}{cc}
 0 & \gamma_{\bk}^* \\ \gamma_{\bk} & 0
\end{array}
\right)\ .
\eeq
Here, $\bone$ represents the $2\times 2$ one-matrix 
\beq
\bone=\left(
\begin{array}{cc}
 1 & 0 \\
 0 & 1
\end{array}
\right).
\eeq
This Hamiltonian effectively 
omits the problem of non-orthogonality of the wave functions by 
a simple renormalization of the \textit{nnn} hopping amplitude, as alluded to 
above. It is therefore simpler to treat than the original one (\ref{eq2:09}) the 
eigenvalue equation of which involves the overlap matrix $\Smath_{\bk}$, while it
yields the same dispersion relation (\ref{eq2:20}). 
The eigenstates of the effective Hamiltonian (\ref{eq2:26}) are the
spinors
\beq\label{eq2:27}
\Psi_{\bk}^{\lambda}=\left(
\begin{array}{c}
 a_{\bk}^{\lambda} \\ b_{\bk}^{\lambda}
\end{array}
\right)\ ,
\eeq
the components of which are the probability amplitudes of the Bloch wave function 
(\ref{eq2:06}) on the two different sublattices A and B. They may be determined
by considering the eigenvalue equation
$\Hmath_{\bk}(t_{nnn}=0)\Psi_{\bk}^{\lambda}=\lambda 
t|\gamma_{\bk}|\Psi_{\bk}^{\lambda},
$
which does not take into account the \textit{nnn} hopping correction. Indeed, these
eigenstates are also those of the Hamiltonian with $t_{nnn}\neq 0$ because
the \textit{nnn} term is proportional to the one-matrix $\bone$. The solution
of the eigenvalue equation (\ref{eq2:27}) yields 
\beq
a_{\bk}^{\lambda}=\lambda \frac{\gamma_{\bk}^*}{|\gamma_{\bk}|} b_{\bk}^{\lambda}
=\lambda e^{-i\varphi_{\bk}}b_{\bk}^{\lambda}
\eeq
and, thus, the eigenstates
\beq\label{eq2:28}
\Psi_{\bk}^{\lambda}=\frac{1}{\sqrt{2}}\left(
\begin{array}{c}
1 \\ \lambda e^{i\varphi_{\bk}}
\end{array}
\right),
\eeq
where 
$\varphi_{\bk}=\arctan(
{\rm Im}\gamma_{\bk}/{\rm Re} \gamma_{\bk})$.

As one may have expected, the spinor represents an equal probability to find an
electron in the state $\Psi_{\bk}^{\lambda}$ on the A as on the B sublattice because
both sublattices are built from carbon atoms with the same onsite energy 
$\epsilon^{(i)}$.

\subsubsection{Continuum limit}
\label{Sec:continuum}

In order to describe the low-energy excitations, i.e. electronic excitations with an
energy that is much smaller than the band width $\sim |t|$, one may
restrict the excitations to quantum states in the vicinity of the Dirac points
and expand the energy dispersion around $\pm \bK$. The wave vector 
is, thus, decomposed as $\bk=\pm \bK+\bq$, where $|\bq|\ll |\bK|\sim 1/a$. The 
small parameter, which governs the expansion of the energy dispersion, is therefore
$|\bq|a\ll 1$. 

It is evident from the form of the energy dispersion (\ref{eq2:20}) and
the effective Hamiltonian that the basic entity to be expanded is the sum of
the phase factors $\gamma_{\bk}$. As we have already mentioned, there is some
arbitrariness in the definition of $\gamma_{\bk}$, as a consequence of the arbitrary choice
of the relative phase between the two sublattice components -- indeed, a change 
$\gamma_{\bk}\rightarrow \gamma_{\bk}\exp(if_{\bk})$ in Eq. (\ref{eq2:18}) for a real and non-singular function
$f_{\bk}$ does not affect the dispersion relation (\ref{eq2:20}), which only depends on the modulus
of the phase-factor sum. For the series expansion, it turns out to be more convenient not to use the expression (\ref{eq2:18}),
but one with $f_{\bk}=\bk\cdot\deltab_3$, which renders the expression more symmetric \cite{bena},
\beq\label{eq:phasesB1}
e^{i\bk\cdot\deltab_3}\gamma_{\bk}=e^{i\bk\cdot\deltab_1}+e^{i\bk\cdot\deltab_2}+e^{i\bk\cdot\deltab_3}
\eeq

In the series expansion, we need to distinguish furthermore the sum at the $K$ point
from that at the $K'$ point,
\beqn\label{eq2:30}
\nn
\gamma_{\bq}^{\pm}&\equiv& e^{i\bk\cdot\deltab_3}\gamma_{\bk=\pm \bK+\bq} = \sum_{j=1}^3e^{\pm i\bK\cdot\deltab_j}e^{i\bq\cdot\deltab_j} 
\\
\nn
&\simeq& e^{\pm i 2\pi/3}\left[1 + i\bq\cdot\deltab_1 - 
\frac{1}{2}(\bq\cdot\deltab_1)^2\right] \\
\nn
&& + e^{\mp i 2\pi/3}\left[1 + i\bq\cdot\deltab_2 - 
\frac{1}{2}(\bq\cdot\deltab_2)^2\right]\\
\nn
&& + \left[1 + i\bq\cdot\deltab_3 - 
\frac{1}{2}(\bq\cdot\deltab_3)^2\right] \\
&=& \gamma_{\bq}^{\pm (0)}+\gamma_{\bq}^{\pm (1)}+\gamma_{\bq}^{\pm (2)}
\eeqn
By definition of the Dirac points and their
position at the BZ corners $K$ and $K'$, we have 
$\gamma_{\bq}^{\pm (0)}=\gamma_{\pm\bK}=0$. We limit the expansion to second order
in $|\bq|a$.

\paragraph{First order in $|\bq|a$.}

The first-order term is given by 
\beqn\label{eq2:31}
\nn
\gamma_{\bq}^{\pm (1)} &=& i\frac{a}{2}\left[(\sqrt{3}q_x+q_y)e^{\pm i2\pi/3}
- (\sqrt{3}q_x-q_y)e^{\mp i2\pi/3} \right]\\
&& -iq_ya=\mp \frac{3a}{2}(q_x \pm i q_y),
\eeqn
which is obtained with the help of $\sin(\pm 2\pi/3)=\pm \sqrt{3}/2$ and
$\cos(\pm 2\pi/3)=-1/2$. This yields the effective low-energy Hamiltonian
\beq\label{eq2:32}
\Hmath_{\bq}^{{\rm eff},\xi}=\xi\hbar v_F(q_x\sigma^x + \xi q_y\sigma^y),
\eeq
where we have defined the Fermi velocity\footnote{The minus sign in the definition is added to 
render the Fermi velocity positive because the hopping parameter $t\simeq -3$ eV happens to be
negative, as mentioned in the last section.}
\beq\label{eq2:32b}
v_F \equiv -\frac{3ta}{2\hbar}=\frac{3|t|a}{2\hbar}
\eeq
and used the Pauli matrices 
\beq
\sigma^x=\left(\begin{array}{cc} 0 & 1 \\ 1 & 0 \end{array}\right)
\qquad {\rm and} \qquad
\sigma^y=\left(\begin{array}{cc} 0 & -i \\ i & 0 \end{array}\right).
\eeq
Furthermore, we have introduced the \textit{valley pseudospin} $\xi=\pm$, where
$\xi=+$ denotes the $K$ point at $+\bK$ and $\xi=-$ the $K'$ point at $-\bK$ modulo a reciprocal lattice vector.
The low-energy Hamiltonian (\ref{eq2:32}) does not take into account 
\textit{nnn}-hopping corrections, which are proportional to $|\gamma_{\bk}|^2$
and, thus, occur only in the second-order expansion of the energy dispersion
[at order $\Omath(|\bq|a)^2$]. The energy dispersion (\ref{eq2:20}) therefore reads
\beq\label{eq2:32t}
\epsilon_{\bq,\xi=\pm}^{\lambda} = \lambda\hbar v_F |\bq|,
\eeq
independent of the valley pseudospin $\xi$. We have already alluded to this
twofold valley degeneracy in Sec. \ref{Sec:TB}, in the framework of the discussion
of the zero-energy states at the BZ corners. From Eq. (\ref{eq2:32t}) it is apparent that the continuum limit
$|\bq|a\ll 1$ coincides with the limit $|\epsilon|\ll |t|$, as described above, 
because $|\epsilon_{\bq}|= 3ta|\bq|/2 \ll |t|$ then.

It is convenient to swap the spinor components at the $K'$ point (for $\xi=-$),
\beq
\Psi_{\bk,\xi=+} = \left(\begin{array}{c}\psi_{\bk,+}^A \\ \psi_{\bk,+}^B \end{array}\right), \qquad 
\Psi_{\bk,\xi=-} = \left(\begin{array}{c}\psi_{\bk,-}^B \\ \psi_{\bk,-}^A \end{array}\right)\ ,
\eeq
i.e. to invert the role of the two sublattices. In this case,
the effective low-energy Hamiltonian may be represented as 
\beq\label{eq2:33}
\Hmath_{\bq}^{{\rm eff},\xi}=\xi\hbar v_F(q_x\sigma^x + q_y\sigma^y) =
\hbar v_F \tau^z \otimes \bq\cdot \sigmab,
\eeq
i.e. as two copies of the \textit{2D Dirac Hamiltonian} $H_D=v_F \bp\cdot\sigmab$ (with the momentum $\bp=\hbar\bq$),
where we have introduced the four-spinor representation 
\beq
\Psi_{\bq} = \left(\begin{array}{c}\psi_{\bq,+}^A \\ \psi_{\bq,+}^B \\
\psi_{\bq,-}^B \\ \psi_{\bq,-}^A
\end{array}\right)
\eeq
in the last line via the $4\times4$ matrices
\beq
\tau^z\otimes\sigmab = \left(\begin{array}{cc} \sigmab & 0 \\ 0 & -\sigmab
\end{array}\right),
\eeq
and $\sigmab\equiv(\sigma^x,\sigma^y)$.
In this four-spinor representation, the first two components represent the 
lattice components at the $K$ point and the last two components those at
the $K'$ point. We emphasise that one must clearly distinguish both types of pseudospin: (a) the \textit{sublattice pseudospin} is represented by the Pauli matrices $\sigma^j$, 
where ``spin up'' corresponds to the component on one sublattice and ``spin down'' to
that on the other one. A rotation within the SU(2) sublattice-pseudospin space yields
the band indices $\lambda=\pm$, and the band index is, thus, 
intimitely related to the sublattice pseudospin. (b) The \textit{valley pseudospin}, which
is described by a second set of Pauli matrices $\tau^j$, the $z$-component 
of which appears in the Hamiltonian (\ref{eq2:33}), is due to the twofold
valley degeneracy and is only indirectly related to the presence of two sublattices.

The eigenstates of the Hamiltonian (\ref{eq2:33}) are the four-spinors
\beq\label{eq2:34}
\Psi_{\bq,\lambda}^{\xi=+} = \frac{1}{\sqrt{2}}\left(\begin{array}{c}
1 \\ \lambda  e^{i\varphi_{\bq}}\\ 0 \\ 0
\end{array}\right), ~  ~
\Psi_{\bq,\lambda}^{\xi=-} = \frac{1}{\sqrt{2}}\left(\begin{array}{c}
 0 \\ 0 \\ 1 \\ -\lambda e^{i\varphi_{\bq}}
\end{array}\right),
\eeq
where we have, now,
\beq\label{eq2:35}
\varphi_{\bq}=\arctan\left(\frac{q_y}{q_x}\right)\ .
\eeq

\paragraph{Chirality.}

In high-energy physics, one defines the helicity of a particle as the
projection of its spin onto the direction of propagation \cite{weinberg}, 
\beq\label{eq3:25}
\eta_{\bq}=\frac{\bq\cdot\sigmab}{|\bq|},
\eeq
which is a Hermitian and unitary operator with the eigenvalues $\eta=\pm $,
$\eta_{\bq}|\eta=\pm\rangle = \pm |\eta=\pm\rangle$.
Notice that $\sigmab$ describes, in this case, the true physical spin of the particle. 
In the absence of a mass term, the helicity operator commutes with the Dirac Hamiltonian, and 
the helicity is, therefore, a good quantum number, e.g. in the description of neutrinos, which
have approximately zero mass. One finds indeed, in nature, that all neutrinos are ``left-handed''
($\eta=-$), i.e. their spin is antiparallel to their momentum, whereas all anti-neutrinos 
are ``right-handed'' ($\eta=+$).

For massive Dirac particles, the helicity operator (\ref{eq3:25}) no longer commutes with the 
Hamiltonian. One may, however, decompose a quantum state $|\Psi\rangle$ describing a massive Dirac particle into 
its \textit{chiral} components, with the help of the projectors
\beq
|\Psi_L\rangle=\frac{1-\eta_{\bq}}{2}|\Psi\rangle \qquad {\rm and}\qquad |\Psi_R\rangle=\frac{1+\eta_{\bq}}{2}|\Psi\rangle .
\eeq
In the case of massless Dirac particles, with a well-defined helicity $|\Psi\rangle =|\eta =\pm\rangle$, one simply
finds 
\beq
|\Psi_L^+\rangle=\frac{1-\eta_{\bq}}{2}|+\rangle=0, \qquad 
|\Psi_R^+\rangle=\frac{1+\eta_{\bq}}{2}|+\rangle=|+\rangle
\eeq
and 
\beq
|\Psi_L^-\rangle=\frac{1-\eta_{\bq}}{2}|-\rangle=|-\rangle, \qquad 
|\Psi_R^-\rangle=\frac{1+\eta_{\bq}}{2}|-\rangle=0,
\eeq
such that one may then identify helicity and chirality. Because we are concerned with
massless particles in the context of graphene, we make this identification in the remainder of this
review and use the term \textit{chirality}.

\begin{figure}
\includegraphics[width=6.5cm,angle=0]{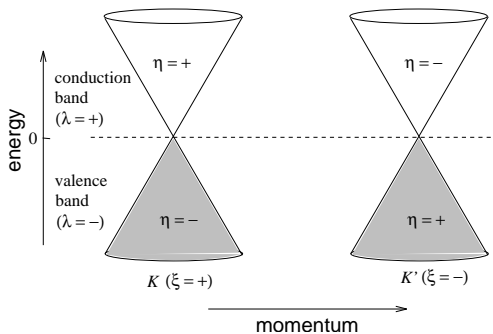}
\caption{\footnotesize{Relation between band index $\lambda$, valley pseudospin $\xi$, and chirality 
$\eta$ in graphene.
}}
\label{fig3:02}
\end{figure}

For the case of graphene,  one may
use the same definition (\ref{eq3:25}), but the
Pauli matrices define now the sublattice pseudospin instead of the true spin.
The operator $\eta_{\bq}$
clearly commutes with the massless 2D Dirac Hamiltonian (\ref{eq2:33}), and one may even express 
the latter as
\beq
\Hmath_{\bq}^{{ \rm eff},\xi}=\xi\hbar v_F |\bq| \eta_{\bq}\ ,
\eeq
which takes into account the two-fold valley degeneracy, in terms of the
valley pseudospin $\xi=\pm$. The band index $\lambda$, which describes the valence and the 
conduction band, is therefore entirely determined by the chirality and the valley pseudospin,
and one finds
\beq\label{eq3:27}
\lambda=\xi\eta\ ,
\eeq
which is depicted in Fig. \ref{fig3:02}.

We notice finally that the chirality is a preserved quantum number in elastic scattering processes induced by 
impurity potentials $V_{imp}=V(\br)\bone$ that vary smoothly on the lattice scale. In this case, inter-valley 
scattering is suppressed, and the chirality thus conserved, as a consequence of Eq. (\ref{eq3:27}). This effect
gives rise to the absence of backscattering in graphene \cite{shon98} and is at the origin of Klein tunneling
according to which a massless Dirac particle is fully transmitted, under normal incidence, through a 
high electrostatic barrier without being reflected \cite{kats06}. This rather counter-intuitive result
was first considered as a paradox and led to the formulation of a \textit{charged vacuum} in the potential 
barrier \cite{klein}, which may be indentified in the framework of band theory with a Fermi level in the 
valence band.

\paragraph{Higher orders in $|\bq|a$.}

Although most of the fundamental properties of graphene are captured
within the effective model obtained at first order in the 
expansion of the energy dispersion, it is useful to take into account
second-order terms. These corrections include \textit{nnn} hopping corrections
and off-diagonal second-order contributions from the expansion of
$\gamma_{\bk}$. The latter yield the so-called \textit{trigonal warping},
which consists of an anisotropy in the energy dispersion around the
Dirac points.

The diagonal second-order term, which stems from the \textit{nnn}
hopping, is readily obtained from Eq. (\ref{eq2:31}),
\beq\label{eq2:36}
\Hmath_{nnn}^{\xi}=t_{nnn}|\gamma_{\bq}^{\xi}|^2\bone
\simeq t_{nnn}|\gamma_{\bq}^{\xi(1)}|^2\bone =\frac{9a^2}{4}t_{nnn}|\bq|^2\bone,
\eeq 
independent of the valley index $\xi$.

The off-diagonal second-order terms are 
$
t\gamma_{\bq}^{\xi(2)}= -\hbar v_F a(q_x - i\xi q_y)^2/4$.
Notice that there is a natural energy hierarchy between the diagonal and off-diagonal second-order 
terms when compared to the leading linear term; whereas the off-diagonal terms are on the order $\Omath(|\bq|a)$
as compared to the energy scale $\hbar v_F|\bq|$, the diagonal term is on the order 
$\Omath((t_{nnn}/t)|\bq|a)$ and thus roughly an order of magnitude smaller. We therefore take into account also
the off-diagonal third order term
$
t\gamma_{\bq}^{\xi(3)}= -\xi\hbar v_F a^2 (q_x + i\xi q_y)|\bq|^2/8$,
which also needs to be considered when calculating the high-energy corrections of the
energy levels in a magnetic field (see Sec. \ref{Sec:DispCorr}). Up to third order, the off-diagonal terms
therefore read
\beqn\label{eq:Disp3}
\nn
t\gamma_{\bq}^{\xi} &=& \xi\hbar v_F\left[(q_x+i\xi q_y) - \xi \frac{a}{4}(q_x-i\xi q_y)^2 \right.\\
&& \left.- \frac{a^2}{8}|\bq|^2(q_x+i\xi q_y)\right],
\eeqn
where one may omit the valley-dependent sign before the $y$-components of the wave vector by sweeping the sublattice components
in the spinors when changing the valley.

In order to appreciate the influence of the second-order off-diagonal terms on the energy bands, 
we need to calculate the modulus of $\gamma_{\bq}^{\xi}$,
\beq
|\gamma_{\bq}^{\xi}|
\simeq  \frac{3a}{2}|\bq|\left[1-\xi\frac{|\bq|a}{4}
\cos(3\varphi_{\bq})\right],
\eeq
where we have used the parametrization
$q_x=|\bq|\cos\varphi_{\bq}$ and $q_y=|\bq|\sin\varphi_{\bq}$,
and where we have restricted the expansion to second order.
Finally, the energy dispersion (\ref{eq2:20}) expanded to second order in
$|\bq|a$ reads
\beq\label{eq2:37}
\epsilon_{\bq,\xi}^{\lambda}=\frac{9a^2}{4}t_{nnn}|\bq|^2 +
\lambda\hbar v_F|\bq|\left[1-\xi\frac{|\bq|a}{4}\cos(3\varphi_{\bq})\right]\ .
\eeq

As mentioned in Sec. \ref{Sec:TB}, it is apparent from Eq. (\ref{eq2:37})
that the \textit{nnn} correction 
breaks the electron-hole symmetry $\epsilon_{\bq,\xi}^{-\lambda}=
-\epsilon_{\bq,\xi}^{\lambda}$. This is, however, a rather small correction,
of order $|\bq|a t_{nnn}/t$, to the first-order effective Hamiltonian (\ref{eq2:33}).
The second-order expansion of the phase factor sum $\gamma_{\bq}$ yields a more 
relevant correction -- the third term in Eq. (\ref{eq2:37}), that is of order $|\bq|a\gg 
|\bq|a t_{nnn}/t$ -- to the linear theory. It depends explicitly on the valley
pseudospin $\xi$ and renders the energy dispersion anisotropic in $\bq$ around the $K$ and
$K'$ point. The tripling of the period, due to the term 
$\cos(3\varphi_{\bq})$, is a consequence of the symmetry of the underlying lattice
and is precisely the origin of trigonal warping.

\begin{figure}
\includegraphics[width=8.5cm,angle=0]{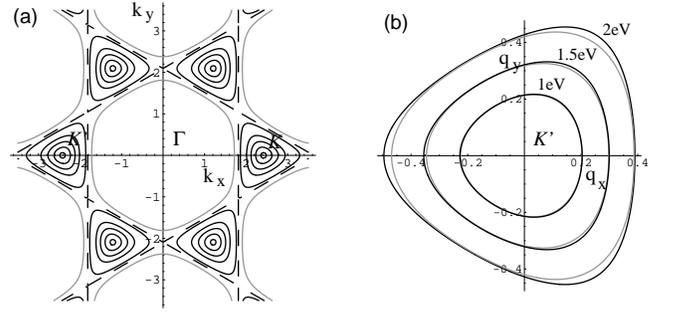}
\caption{\footnotesize{Contours of constant (positive) energy in reciprocal
space. \textit{(a)} Contours obtained from the full dispersion relation (\ref{eq2:20}).
The dashed line corresponds to the energy $t+t_{nnn}$, which separates closed orbits
around the $K$ and $K'$ points (black lines, with energy $\epsilon 
<t+t_{nnn}$) from those 
around the $\Gamma$ point (gray line, with energy $\epsilon >t+t_{nnn}$).
\textit{(b)} Comparison of the contours at energy $\epsilon=1$ eV, 1.5 eV, and 2 eV
around the $K'$ point. The black lines correspond to the energies calculated from
the full dispersion relation (\ref{eq2:20}) and the gray ones to those calculated
to second order within the continuum limit (\ref{eq2:37}).
}}
\label{fig2:04}
\end{figure}

The trigonal warping of the dispersion relation is visualized in Fig. \ref{fig2:04},
where we have plotted the contours of constant (positive) energy in Fourier space. The 
closed energy contours around the $K$ and $K'$ points at low energy are 
separated by the high-energy contours around the $\Gamma$ point by the
dashed lines in Fig. \ref{fig2:04} (a) at energy $|t+t_{nnn}|$ the crossing points 
of which correspond to the $M$ points. As mentioned above, the dispersion relation
has saddle points at these points at the border of the first BZ, which yield 
van Hove singularities in the density of states.
In Fig. \ref{fig2:04} (b), we compare constant-energy contours of the full 
dispersion relation to those obtained from Eq. (\ref{eq2:37}) calculated within
a second-order expansion. The contours are indistinguishable for an energy 
of $\epsilon=|t|/3\simeq 1$ eV, and the continuum limit yields rather accurate 
results up to energies as large as 2 eV. Notice that, in today's exfoliated graphene
samples on SiO$_2$ substrates, one may probe, by field-effect doping of the
graphene sheet, energies which are on the order of $100$ meV. Above these energies
the capacitor breaks down, and Fig. \ref{fig2:04} (a) indicates that the 
continuum limit (\ref{eq2:37}) yields extremely accurate results at these energies.

We finally mention that, when higher-order terms in $|\bq|a$ are taken into account, the chirality
operator (\ref{eq3:25}) no longer commutes with the Hamiltonian. Chirality is therefore only
a good quantum number in the vicinity of the Dirac points.

\subsection{Deformed Graphene}
\label{DefGraph}

\begin{figure}
\epsfysize+4.0cm
\epsffile{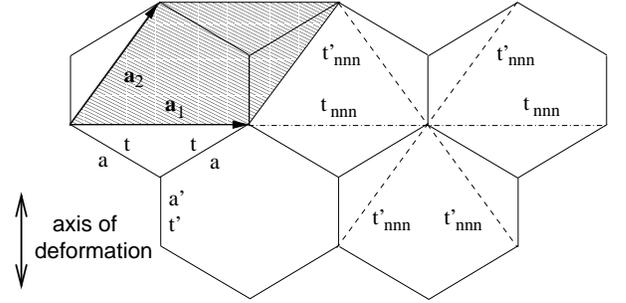}
\caption{\footnotesize Quinoid-type deformation of the honeycomb lattice -- the bonds parallel
to the deformation axis (double arrow) are modified. The shaded region indicates
the unit cell of the oblique lattice, spanned by the lattice vectors $\ba_1$ and
$\ba_2$. Dashed and dashed-dotted lines indicate next-nearest neighbors, with
characteristic hopping integrals $t_{nnn}$ and $t_{nnn}'$, respectively, which
are different due to the lattice deformation.}
\label{fig1:08}
\end{figure}

In the previous section, we have considered a perfect honeycomb lattice which is invariant under
a $2\pi/3$ rotation. As a consequence, all hopping parameters along the \textit{nn} bonds $\deltab_j$ 
were equal. An interesting situation arises when the graphene sheet is deformed, such that rotation
symmetry is broken. In order to illustrate the consequences, we may apply a uniaxial strain in the 
$y$-direction,\footnote{In our simplified model, we only consider one bond length changed by the strain.
The more general case has been considered by Peirera \textit{et al.} \cite{peirera}.
However, the main effects are fully visible in the simplified model.} 
$a\rightarrow a'=a+\delta a$, 
in which case one obtains a quinoid-type deformation (Fig. \ref{fig1:08}).
The hopping $t'$ along $\deltab_3$ is then different from that $t$
along $\deltab_1$ and $\deltab_2$  \cite{hasegawa,zhu,dietl,wunsch2,GFMP,farjam},
\beq\label{eq:ModHopp}
t\rightarrow t'= t + \frac{\partial t}{\partial a} \delta a.
\eeq
Furthermore, also four of six \textit{nnn} hopping integrals are affected by the strain (see Fig. \ref{fig1:08}), 
\beq\label{eq:ModHoppP}
t_{nnn}\rightarrow t_{nnn}'=t_{nnn}+\frac{\partial t_{nnn}}{\partial a}\delta a.
\eeq


If one considers a moderate deformation $\epsilon \equiv \delta a/a\ll 1$, the effect on the hopping amplitudes may
be estimated with the help of Harrison's law \cite{harrison}, according to which $t=C \hbar^2/m a^2$, where $C$ is 
a numerical prefactor of order 1. One therefore finds a value
\beq\label{eq:def_t}
\frac{\partial t}{\partial a}=-\frac{2t}{a}\sim -4.3\, {\rm eV/\AA}\qquad {\rm and} \qquad t'=t(1-2\epsilon)
\eeq
which coincides well with the value $\partial t/\partial a\simeq 5$ eV/\AA, which may be found in the literature \cite{SDD,dillon}.
The estimation of the modified \textit{nnn} hopping integral $t_{nnn}'$ is slightly more involved. One may use a law 
$t_{nnn}(b,a) \approx t(a) \exp[-(b-a)/d(a)]$
familiar in the context of the extended H\"uckel model \cite{huckel}, where $b$ is the \textit{nnn} distance, 
and $d\approx a/3.5\approx 0.4$~\AA \, is a caracteristic
distance related to the overlap of atomic orbitals. In
undeformed graphene one has $b=a\sqrt{3}$, whereas in quinoid-type
graphene $b'=b(1+\varepsilon/2)$, which
gives 
\beq\label{TnnnP}
t_{nnn}'=t_{nnn}(1-2\varepsilon +b\varepsilon/2d).
\eeq

\begin{figure}
\epsfysize+5.0cm
\epsffile{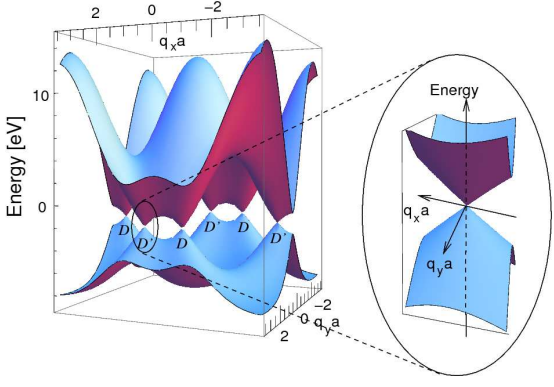}
\caption{\footnotesize Band dispersion of the quinoid-type deformed the honeycomb lattice,
for a lattice distortion of $\delta a/a=-0.4$, with $t=3$ eV, $t_{nnn}/t=0.1$,
$\partial t/\partial a=-5$ eV/\AA, and
$\partial t_{nnn}/\partial a=-0.7$ eV/\AA.
The inset shows a zoom on one of the Dirac points, $D'$. 
}
\label{fig1:09}
\end{figure}

The electronic properties of quinoid-type graphene may then be described in terms of an effective Hamiltonian of the type
(\ref{eq2:26}),
\beq\label{eq:HamEffQ}
\Hmath_{\bk} = t_{nnn}h_\bk \bone + t\left(
\begin{array}{cc}
 0 & \gtilde_{\bk}^* \\ \gtilde_{\bk} & 0
\end{array}
\right)\ ,
\eeq
with \cite{GFMP}
\beqn\label{HTBD}
\nn
h_\bk&=& 2\cos\sqrt{3}k_xa + 2\frac{t_{nnn}'}{t_{nnn}}
\left\{\cos\left[\frac{\sqrt{3}k_xa}{2}+
k_ya\left(\frac{3}{2}+\epsilon\right)\right]\right.\\
&&\left.+\cos\left[-\frac{\sqrt{3}k_xa}{2}+
k_ya\left(\frac{3}{2}+\epsilon\right)\right]\right\},
\eeqn
and the off-diagonal elements
\beq\label{eq:ODQuin}
\gtilde_{\bk}= 2 e^{ik_ya(3/2+\epsilon)}\cos\left(\frac{\sqrt{3}}{2}k_xa\right)+(1-2\epsilon).
\eeq
The resulting energy dispersion 
\beq\label{eq:enDispQ}
\epsilon_{\bk}^{\lambda}=t_{nnn}h_{\bk} + \lambda t |\gtilde_{\bk}|
\eeq
is plotted in Fig. \ref{fig1:09} for an unphysically large deformation, $\epsilon=0.4$, for illustration reasons.
Notice that the reversible deformations are limited by a value of $\epsilon\sim 0.1 ... 0.2$ beyond which the
graphene sheet cracks \cite{strain}. One notices, in Fig. \ref{fig1:09}, two effects of the deformation: i) the Dirac
points no longer coincide with the corners of the first BZ, the form of which is naturally also modified by the deformation; 
and ii) the cones in the vicinity of the Dirac points are tilted, i.e. the \textit{nnn} hopping term
(\ref{HTBD}) breaks the electron-hole symmetry already at \textit{linear} order in $|\bq|a$. These two points are discussed in more detail
in the following two subsections.

\subsubsection{Dirac point motion}
\label{Sec:DPMot}

In order to evaluate quantitatively the position of the Dirac points, which are defined as the contact points between the valence 
($\lambda=-$) and the conduction ($\lambda=+$) bands, one needs to solve the equation $\gtilde_{\bk^D}=0$, in analogy with the
case of undeformed graphene discussed in Sec. \ref{Sec:TB}. One then finds 
\beq\label{DPquin}
k_y^D=0 \qquad {\rm and} \qquad k_x^Da=\xi \frac{2}{\sqrt{3}}\arccos\left(
-\frac{t'}{2t}\right),
\eeq
where the valley index $\xi=\pm$ denotes again the two inequivalent Dirac points $D$ and $D'$,
respectively. As already mentioned, the Dirac points $D$ and $D'$ coincide, for undistorted graphene, with the
crystallographic points $K$ and $K'$, respectively, at the corners of the first BZ.
The distortion makes both pairs of points move in the same direction due to the
negative value of $\partial t/\partial a$. However, unless the parameters are
fine-tuned, this motion is different, and the two pairs of points no longer
coincide.

\begin{figure}
\epsfysize+6.0cm
\epsffile{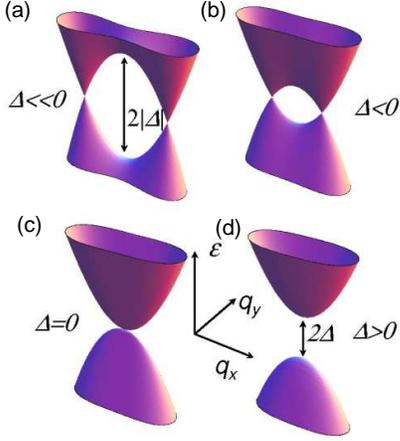}
\caption{\footnotesize
Topological semi-metal insulator transition in the model (\ref{eq:UnHamMerg}) driven by the gap parameter $\Delta$. 
\textit{(a)} Two well-separated Dirac cones for $\Delta\ll 0$, as for graphene.
\textit{(b)} When lowering the modulus of the (negative) gap parameter, the Dirac points move towards a single point. 
\textit{(c)} The two Dirac points merge into a single point at the transition ($\Delta=0$). The band dispersion remains linear 
in the $q_y$-direction while it becomes parabolic in the $q_x$-direction.
\textit{(d)} Beyond the transition ($\Delta> 0$), the (parabolic) bands are separated by a band gap $\Delta$ (insulating phase).
From \onlinecite{MPFG1}.
}
\label{fig1:10}
\end{figure}

One further notices that Eq. (\ref{DPquin}) has (two) solutions only for $t'\leq 2t$. Indeed, the two Dirac points merge at the
characteristic point $M''$ at the border of the first BZ (see Fig. \ref{fig1:07}). The point $t'=2t$ is special insofar as it 
characterizes a topological phase transition between a semi-metallic phase (for $t'<2t$) with a pair of Dirac cones and a band insulator
(for $t'>2t$) \cite{peirera,dietl,wunsch2,MPFG1,MPFG2,esaki09}. 
In the vicinity of the transition, one may expand the Hamiltonian (\ref{eq:HamEffQ}) around the merging point $M''$ 
\cite{MPFG1,MPFG2}, and one
finds\footnote{We do not consider the diagonal part of the Hamiltonian, here, i.e. we choose $t_{nnn}=0$, because it does not
affect the position of the Dirac points.}
\beq\label{eq:UnHamMerg}
\Hmath_{\bq}^{M}=\left(
  \begin{array}{cc}
    0 & \Delta+ \frac{\hbar^2q_x^2}{2 m^*} -  i \hbar c q_y  \\
 \Delta + \frac{\hbar^2 q_x^2}{2 m^*} +  i\hbar c q_y & 0 \\
  \end{array}
\right) ,
\eeq
in terms of the mass $m^*=2\hbar^2/3ta^2$ and the velocity $c=3ta/\hbar$ \cite{MPFG2}.
The gap parameter $\Delta= t'-2t$ changes its sign at the transition -- it is negative in the semi-metallic and positive in the 
insulating phase, where it describes a true gap (Fig. \ref{fig1:10}).

The Hamiltonian (\ref{eq:UnHamMerg}) has quite a particular form in the vicinity of the merging points: it is linear in the 
$q_y$-direction, as one would expect for Dirac points, but it is quadratic in the $q_x$-direction \cite{dietl}. This is a 
general feature of merging points, which may only occur at the $\Gamma$ point or else at half a reciprocal lattice vector
$\bG/2$, i.e. in the center of a BZ border line (such as the $M$ points) \cite{MPFG1}. Indeed, one may show that in the case
of a time-reversal symmetric Hamiltonian, the Fermi velocity in the $x$-direction then vanishes such that one must take into
account the quadratic order in $q_x$ in the energy band. Notice that such hybrid semi-Dirac points, with a linear-parabolic dispersion 
relation, are unaccessible in graphene because unphysically large strains would be required \cite{peirera,strain}. However, 
such points may exist in other physical systems such as cold atoms in optical lattices \cite{zhu,wunsch2,zhao,hou,KLlee}, the 
quasi-2D organic material \BEDT~\cite{katayama06,kobayashi07} or VO$_2$/TiO$_2$ heterostructures \cite{banerjee09}.

\subsubsection{Tilted Dirac cones}

Another aspect of quinoid-type deformed graphene and a consequence of the fact that the Dirac points
no longer coincide with the BZ corners $K$ and $K'$ of high crystallographic symmetry is the tilt of the Dirac cones. This
may be appreciated when expanding the Hamiltonian (\ref{eq:HamEffQ}) to linear order around the Dirac points $\xi\bk_D$, 
instead of an expansion around the point $M''$ as in the last subsection. In contrast to the undeformed case (\ref{eq2:36}),
the diagonal components $h_\bk$ now yield a linear contribution \cite{GFMP},
$
t_{nnn}h_{\xi\bk^D + \bq}\bone \simeq \xi\hbar\bw_0\cdot\bq\,\bone$,
in terms of the \textit{tilt velocity}
\beq\label{eq:tiltV}
w_{0x}=\frac{2\sqrt{3}}{\hbar}(t_{nnn}a\sin2\theta + t_{nnn}'a\sin\theta)~~ {\rm and} ~~ w_{0y}=0,
\eeq
where we have defined $\theta\equiv \arccos(-t'/2t)$. The linear model is therefore described by the 
Hamiltonian,\footnote{This model may be viewed as the minimal form of the generalized Weyl Hamiltonian (with $\sigma^0\equiv \bone$)
$$
H_W=\sum_{\mu=0,...,3}\hbar\bv_{\mu}\cdot\bq \,
\sigma^{\mu},
$$
which is the most general $2\times 2$ matrix Hamiltonian that yields a linear dispersion relation.}
\beq\label{eq:Weyl}
\Hmath_{\bq}^{\xi}=\xi\hbar(\bw_0\cdot\bq \bone + w_xq_x\sigma^x + w_yq_y\sigma^y),
\eeq
with the renormalized anisotropic velocities 
$$
w_x = \frac{\sqrt{3}ta}{\hbar}\sin\theta \qquad {\rm and} \qquad w_y=\frac{3}{2}\frac{t'a}{\hbar}\left(1+\frac{2}{3}\epsilon\right).
$$

Diagonalizing the Hamiltonian (\ref{eq:Weyl}) yields the dispersion relation
\beq\label{eq:EndispWeyl}
\epsilon_{\lambda}^{\xi}(\bq)=\hbar\bw_0\cdot\bq+\lambda\hbar\sqrt{w_x^2q_x^2+w_y^2q_y^2},
\eeq
and one notices that the first term ($\hbar\bw_0\cdot\bq$) breaks indeed the symmetry $\epsilon_{\lambda}^{\xi}(\bq)=\epsilon_{\lambda}^{\xi}(-\bq)$ in each valley, i.e.
it tilts the Dirac cones in the direction opposite to $\bw_0$, as well as the electron-hole symmetry 
$\epsilon_{\lambda}(\bq)=-\epsilon_{-\lambda}(\bq)$ at the same wave vector.\footnote{In the absence of the tilt term 
$\hbar\bw_0\cdot\bq\bone$, this is a consequence of the symmetry $\sigma^z\Hmath\sigma^z= -\Hmath$, which is satisfied both by the 
effective Hamiltonian (\ref{eq2:26}) for $t_{nnn}=0$ and the linearised version (\ref{eq2:33})
in each valley for undeformed graphene.}
Indeed, the linearity in $\bq$ of the generalized Weyl Hamiltonian (\ref{eq:Weyl}) satisfies only the symmetry 
$\Hmath_{\bq}^{\xi}=-\Hmath_{-\bq}^{\xi}$ inside each valley. 

Furthermore, one notices that the chiral symmetry is preserved even in the presence of the tilt term if one redefines the 
chirality operator (\ref{eq3:25}) as 
$
\eta_{\bq}=(w_xq_x\sigma^x+w_yq_y\sigma^y)/\sqrt{w_x^2q_x^2+w_y^2q_y^2}$,
which naturally commutes with the Hamiltonian (\ref{eq:Weyl}). The eigenstates of the chirality operator are still given by
\beq
\psi_{\eta}=\frac{1}{\sqrt{2}}\left(
\begin{array}{c}
1\\
\eta e^{-i\varphi_{\bq}}
\end{array}\right)\ ,
\eeq
with $\tan\varphi_{\bk}\equiv w_yq_y/w_xq_x$, and one notices that these states are also the natural eigenstates of the Hamiltonian
(\ref{eq:Weyl}).

One finally notices that not all values of the tilt parameter $\bw_0$ are indeed physical. In
order to be able to associate $\lambda=+$ to a positive and $\lambda=-$ to a
negative energy state, one must fulfill the condition
\beq\label{TiltCond}
\wtilde_0<1,
\eeq
in terms of the \textit{tilt parameter}
\beq\label{TiltParam}
\wtilde_0 \equiv \sqrt{\left(\frac{w_{0x}}{w_x}\right)^2+\left(\frac{w_{0y}}{w_y}\right)^2}.
\eeq
In the particular case of the deformation in the $y$-axis, which is discussed here and in which case $w_{0y}=0$ [see Eq. 
(\ref{eq:tiltV})], the general form of the tilt parameter reduces to $\wtilde_0=w_{0x}/w_x$. 
Unless this condition is fulfilled, the iso-energetic lines are no longer ellipses
but hyperbolas. In quinoid-type deformed graphene, the tilt parameter may be evaluated as \cite{GFMP}
\beq\label{TiltQG}
\wtilde_0 = 2 \left(\frac{t_{nnn}}{t}\frac{\sin 2\theta}{\sin\theta}+\frac{t_{nnn}'}{t}
\right)
\simeq \frac{2}{t^2} (t t_{nnn}'-t' t_{nnn}) \simeq 0.6\epsilon,
\eeq
where we have used Eqs. (\ref{eq:def_t}) and (\ref{TnnnP}). Even at moderate deformations ($\epsilon<0.1$), the tilt of the Dirac cones 
is on the order of 5\%, and one may therefore hope to observe the effect, e.g. in angle-resolved photoemission spectroscopy
(ARPES) measurements \cite{arpesRev}
that have been successfully applied to epitaxial graphene \cite{arpes01} and graphitic samples \cite{arpes02}. Notice that 
the Dirac cones are naturally tilted in \BEDT~\cite{katayama06,kobayashi07}, 
where the Dirac points occur at positions of low crystallographic symmetry 
within the first BZ.



\section{Dirac Equation in a Magnetic Field and the Relativistic Quantum Hall Effect}

As already mentioned in the introduction, a key experiment in graphene research was the discovery of a particular quantum Hall effect 
\cite{novoselov1,zhang1}, which unveiled the relativistic nature of low-energy electrons in graphene. For a deeper understanding
of this effect and as a basis for the following parts, we discuss here relativistic massless 2D fermions in a strong quantizing
magnetic field (Sec. \ref{Sec:DiracB}). The limits of the Dirac equation in the treatment of the high-field properties of graphene
are discussed in Sec. \ref{Sec:DispCorr}, and we terminate this section with a discussion of the relativistic Landau level spectrum
in the presence of an in-plane electric field (Sec. \ref{Sec:DiracE}) and that of deformed graphene (Sec. \ref{Sec:DiracDef}).

\subsection{Massless 2D Fermions in a Strong Magnetic Field}
\label{Sec:DiracB}

In order to describe free electrons in a magnetic field, 
one needs to replace the canonical momentum $\bp$ by the gauge-invariant kinetic momentum \cite{jackson}
\beq\label{mom}
\bp \rightarrow \Pib = \bp + e\bA(\br), 
\eeq
where $\bA(\br)$ is the vector potential that generates the magnetic field $\bB=\nabla\times \bA(\br)$. The 
kinetic momentum is proportional to the electron velocity ${\bf v}$, which must naturally be 
gauge-invariant because it is a physical quantity.

In the case of electrons on a lattice, the substitution (\ref{mom}), which 
is then called \textit{Peierls substitution}, remains correct as long as the lattice spacing 
$\atilde$ is much smaller than the \textit{magnetic length}
\beq\label{lB}
l_B = \sqrt{\frac{\hbar}{eB}}\ ,
\eeq
which is the fundamental length scale in the presence of a magnetic field. Because $\atilde=0.24$ nm 
and $l_B\simeq 26\, {\rm nm}/\sqrt{B{\rm [T]}}$, this condition is fulfilled in graphene for the magnetic fields, which may
be achieved in today's high-field laboratories ($\sim 45$ T in the continuous regime and $\sim 80$ T in the pulsed 
regime).

With the help of the (Peierls) substitution (\ref{mom}), one may thus immediately write down the Hamiltonian for charged particles in 
a magnetic field if one knows the Hamiltonian in the absence of the field,
\beq
\Hmath(\bp) \rightarrow H(\Pib)  = \Hmath(\bp + e\bA)= H^B(\bp,\br). 
\eeq
Notice that because of the spatial dependence of the vector potential, the resulting Hamiltonian is no longer translation invariant, 
and the (canonical) momentum $\bp=\hbar\bq$ is no longer a conserved quantity. 
For the Dirac Hamiltonian (\ref{eq2:33}), which we have derived in the preceding section to lowest
order in $|\bq| a$, the Peierls substitution yields
\beq\label{DiracB}
\Hmath_{B}^{\xi}=\xi\hbar v_F(q_x\sigma^x + q_y\sigma^y) \rightarrow
\Hmath_{B}^{{\rm eff},\xi}=\xi v_F(\Pi_x\sigma^x + \Pi_y\sigma^y).
\eeq

We further notice that, because electrons do not only possess a charge but also a spin, each energy level resulting from the
diagonalization of the Hamiltonian (\ref{DiracB}) is split into two spin branches separated by the Zeeman effect 
$\Delta_Z=g\mu_B B$, where $g$ is the $g$-factor of the host material [$g\sim 2$ for graphene \cite{zhang}] and 
$\mu_B=e\hbar/2m_0$ is the Bohr magneton, in terms of the bare electron mass $m_0$. In the remainder of this section,
we concentrate on the orbital degrees of freedom which yield the characteristic level structure of electrons in a magnetic field
and therefore neglect the spin degree of freedom, i.e. we consider \textit{spinless} fermions. Effects related to the 
internal degrees of freedom are discussed in a separate section (Sec. \ref{Chap:FQHE}) in the framework of the quantum-Hall 
ferromagnet.

\subsubsection{Quantum-mechanical treatment}
\label{Sec:QMtreat}

One may easily treat the Hamiltonian (\ref{DiracB}) quantum-mechanically with the help of the standard \textit{canonical quantization} 
\cite{CT}, according to which the components of the position $\br=(x,y)$ and the associated canonical momentum $\bp=(p_x,p_y)$
satisfy the commutation relations
$[x,p_x]= [y,p_y]=i\hbar$ and $[x,y]=[p_x,p_y]=[x,p_y]=[y,p_x]=0$.
As a consequence of these relations, the components of the kinetic momentum no longer commute,
and, with the help of the commutator relation \cite{CT}
\beq\label{Haus}
[\Omath_1,f(\Omath_2)] = \frac{d f}{d \Omath_2} [\Omath_1,\Omath_2]
\eeq
between two arbitrary operators, the commutator of which is an operator that commutes itself with both 
$\Omath_1$ and $\Omath_2$,  one finds
\beq\label{ComMom}
\left[\Pi_x,\Pi_y\right] = -ie\hbar \left(\frac{\partial A_y}{\partial x} - \frac{\partial A_x}{\partial y}\right) =
-i \frac{\hbar^2}{l_B^2}\ ,
\eeq
in terms of the magnetic length (\ref{lB}).

For the quantum-mechanical solution of the Hamiltonian (\ref{DiracB}), it is convenient to use the pair of conjugate operators $\Pi_x$ 
and $\Pi_y$ to introduce \textit{ladder operators} in the same manner as in the quantum-mechanical treatment of the one-dimensional
harmonic oscillator. These ladder operators play the role of a \textit{complex} gauge-invariant 
momentum (or velocity), and they read 
\beq\label{ladder}
\ahat = \frac{l_B}{\sqrt{2}\hbar}\left(\Pi_x - i\Pi_y\right) ~~ {\rm and} ~~
\ahat^{\dagger} = \frac{l_B}{\sqrt{2}\hbar}\left(\Pi_x + i\Pi_y\right),
\eeq
where we have chosen the appropriate normalization such as to obtain the usual commutation relation
\beq\label{ComLad}
[\ahat,\ahat^{\dagger}]=1.
\eeq
It turns out to be helpful for practical calculations to invert the expression for the ladder operators (\ref{ladder}),
\beq\label{ladder1}
\Pi_x = \frac{\hbar}{\sqrt{2}l_B}\left(\ahat^{\dagger}+\ahat\right) ~~ {\rm and} ~~
\Pi_y = \frac{\hbar}{i\sqrt{2}l_B}\left(\ahat^{\dagger}-\ahat\right).
\eeq

\subsubsection{Relativistic Landau levels}
\label{Sec:RelLL}

In terms of the ladder operators (\ref{ladder}), the Hamiltonian (\ref{DiracB}) becomes
\beq\label{HamLadD}
H_B^{\xi}= 
\xi\sqrt{2}\frac{\hbar v_F}{l_B}\left(\begin{array}{cc}
0 & \ahat \\ \ahat^{\dagger} & 0
        \end{array}\right) .
\eeq
One remarks the occurence of a characteristic frequency $\omega'=\sqrt{2}v_F/l_B$, which plays the role of the cyclotron frequency
in the relativistic case. Notice, however, that this frequency may not be written in the form $eB/m_b$ because the band mass 
is strictly zero in graphene, such that the frequency would diverge.\footnote{Sometimes, a 
density-dependent \textit{cyclotron mass} $m_C$ 
is formally introduced via the equality $\omega'\equiv eB/m_C$.}

\begin{figure}
\centering
\includegraphics[width=8.0cm,angle=0]{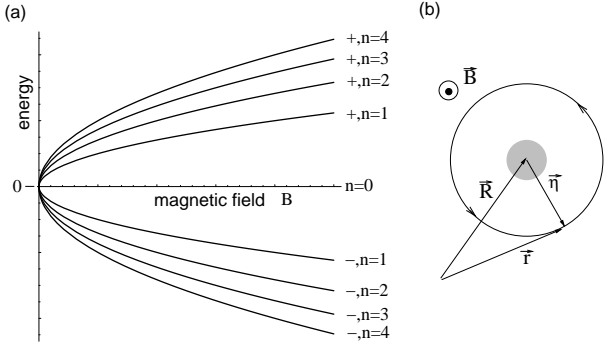}
\caption{\footnotesize \textit{(a)} Relativistic Landau levels as a function of the magnetic field.
\textit{(b)} Semi-classical picture of cyclotron motion described 
by the cyclotron coordinate $\etab$, where the charged particle turns around the guiding center $\bR$.
The gray region depicts the uncertainty on the guiding center, as indicated by Eq. (\ref{minsurf}).}
\label{fig09}
\end{figure}

The eigenvalues and the eigenstates of the Hamiltonian (\ref{HamLadD}) are readily obtained by solving the eigenvalue equation
$H_B^{\xi}\psi_n = \epsilon_n \psi_n$, in terms of the 2-spinors, 
\beq
\psi_n=\left(\begin{array}{c} u_n \\ v_n \end{array} \right).
\eeq
We thus need to solve the system of equations
\beq\label{eigen}
\xi\hbar\omega'\ahat\, v_n = \epsilon_n \, u_n ~~ {\rm and} ~~ \xi\hbar\omega' \ahat^{\dagger}\, u_n= \epsilon_n\, v_n ,
\eeq
which yields the equation 
\beq\label{eigen2}
\ahat^{\dagger}\ahat\, v_n = \left(\frac{\epsilon_n}{\hbar\omega'}\right)^2 v_n
\eeq
for the second spinor component. 
One may therefore identify, up to a 
numerical factor, the second spinor component $v_n$ with the eigenstate 
$|n\rangle$ of the usual number operator
$\ahat^{\dagger}\ahat$, with $\ahat^{\dagger}\ahat |n\rangle = n|n\rangle$ in terms of the integer $n\geq 0$.
Furthermore, one observes that the square of the energy is proportional to this quantum number,
$\epsilon_n^2 = (\hbar\omega')^2 n$. This equation has two solutions, a positive and a negative one, and one needs to
introduce another quantum number $\lambda=\pm$, which labels the states of positive and negative energy, respectively.
This quantum number plays the same role as the band index ($\lambda=+$ for the conduction and $\lambda=-$ for the
valence band) in the zero-$B$-field case discussed in the preceding section. One thus obtains the spectrum \cite{mcclure}
\beq\label{RelLLs}
\epsilon_{\lambda,n} = \lambda \frac{\hbar v_F}{l_B}\sqrt{2n}
\eeq
of \textit{relativistic Landau levels} (LLs) 
that disperse as $\lambda\sqrt{Bn}$ as a function of the magnetic field [see Fig. \ref{fig09}(a)]. 
Notice that, as in the $B=0$ case, the 
level spectrum is two-fold valley-degenerate.

Once we know the second spinor component, the first component is obtained from Eq. (\ref{eigen}), which
reads $u_n\propto \ahat\,v_n\sim \ahat|n\rangle \sim |n-1\rangle$ because of the usual equations 
\beq\label{nlad}
\ahat^{\dagger}|n\rangle = \sqrt{n+1}|n+1\rangle ~~ {\rm and} ~~
\ahat|n\rangle = \sqrt{n} |n-1\rangle
\eeq
for the ladder operators, where the last equation is valid for $n>0$.
One then needs to distinguish the zero-energy LL
($n=0$) from all other levels. Indeed, for $n=0$, the first component is zero because
\beq\label{n0lad}
\ahat|n=0\rangle = 0
\eeq
In this
case one obtains the spinor 
\beq\label{spinN0}
\psi_{n=0} = \left(\begin{array}{c} 0 \\ |n=0\rangle  \end{array}\right).
\eeq
In all other cases ($n\neq 0$), one has positive and negative energy solutions, which differ among each other by a relative sign
in one of the components. A convenient representation of the associated spinors is given by
\beq\label{spinN}
\psi_{\lambda,n\neq 0}^{\xi} = \frac{1}{\sqrt{2}}\left(\begin{array}{c} |n-1\rangle \\ \xi\lambda |n\rangle  \end{array}\right).
\eeq

The particular form of the $n=0$ spinor (\ref{spinN0}) associated with zero-energy states merits a more detailed comment. One notices
that only the second spinor component is non-zero. Remember that this component corresponds to the B sublattice in the $K$-valley 
($\xi=+$) and to the A sublattice in the $K'$-valley ($\xi=-$) -- the valley pseudospin therefore coincides with the sublattice pseudospin, 
and the two sublattices are decoupled at zero energy. Notice that this is also the case in the absence of a magnetic field,
where the relation (\ref{eq3:27}) between the chirality, the band index and the valley pseudospin is only valid at non-zero values of the 
wave vector, i.e. not exactly at zero energy. Indeed, the chirality can no longer be defined as the projection of the sublattice
pseudospin on the direction of propagation $\bq/|\bq|$, which is singular at $\bq=0$. 
At zero energy, it is therefore useful to identify the chirality 
with the valley pseudospin. Notice, however, that this particularity concerns, in the absence of a magnetic field, only a non-extensive
number of states (only two) because of the vanishing density of states at zero energy, whereas the zero-energy LL $n=0$ is
macroscopically degenerate, as discussed in the following paragraphs.

\paragraph{LL degeneracy.}

A particular feature of LLs, both relativistic and non-relativistic ones, consists of their large degeneracy, which equals the number 
of flux quanta $N_B= \Amath\times B/(h/e)$ threading the 2D surface $\Amath$ occupied by the electron gas. From the classical
point of view, this degeneracy is related to the existence of a constant of motion, namely the position of the \textit{guiding center},
i.e. the center of the classical cyclotron motion. Indeed, due to translation invariance in a uniform magnetic field, the energy
of an electron does not depend on the position of this guiding center. Translated to quantum mechanics, this means that the
operator corresponding to this guiding center $\bR=(X,Y)$ commutes with the Hamiltonian $\Hmath(\bp+e\bA)$.

In order to understand how the LL degeneracy is related to the guiding-center operator, we formally decompose the position 
operator 
\beq\label{eq:PosOp}
\br=\bR + \etab
\eeq
into its guiding center $\bR$ and the cyclotron variable $\etab=(\eta_x,\eta_y)$, as depicted 
in Fig. \ref{fig09}(b). Whereas the guiding center is a constant of
motion, as mentioned above, the cyclotron variable describes the dynamics of the electron in a magnetic field and is, classically,
the time-dependent component of the position. Indeed, the cyclotron variable is perpendicular to the electron's velocity and thus
related to the kinetic momentum $\Pib$ by 
\beq\label{CyclVar}
\eta_x= \frac{\Pi_y}{eB} ~~~ {\rm and} ~~~ \eta_y = - \frac{\Pi_x}{eB}\ ,
\eeq
which, as a consequence of the commutation relations (\ref{ComMom}), satisfy
\beq\label{ComCV}
[\eta_x,\eta_y] = \frac{[\Pi_x,\Pi_y]}{(eB)^2} = -il_B^2,
\eeq
whereas they commute naturally with the guiding-center components $X$ and $Y$. Equation (\ref{ComCV}) thus induces the
commutation relation
\beq\label{ComGC}
[X,Y]=-[\eta_x,\eta_y]=il_B^2,
\eeq
in order to satisfy $[x,y]=0$.

These commutation relations indicate that the components of the guiding-center operator form a pair of conjugate variables, and 
one may introduce, in the same manner as for the kinetic momentum operator $\Pib$, the ladder operators 
\beq\label{laddGC}
\bhat=\frac{1}{\sqrt{2}l_B}(X+iY)~~ {\rm and} ~~ \bhat^{\dagger}=\frac{1}{\sqrt{2}l_B}(X-iY),
\eeq
which again satisfy the usual commutation relations $[\bhat,\bhat^{\dagger}]=1$ and which naturally commute with
the Hamiltonian. One may then introduce a number operator $\bhat^{\dagger}\bhat$ associated with these ladder operators, the eigenstates of 
which satisfy the eigenvalue equation
\beq
\bhat^{\dagger}\bhat|m\rangle = m|m\rangle.
\eeq
One thus obtains a second quantum number, an integer $m\geq 0$, 
which is necessary to describe the full quantum states in addition to the LL quantum number $n$, and the 
completed quantum states (\ref{spinN0}) and (\ref{spinN}) then read
\beq\label{QstateN0}
\psi_{n=0,m}^{\xi} = \psi_{n=0}^{\xi}\otimes |m\rangle = \left(\begin{array}{c} 0 \\ |n=0,m\rangle  \end{array}\right)
\eeq
and
\beq\label{QstateR}
\psi_{\lambda n,m}^{\xi} = \psi_{\lambda n}^{\xi}\otimes |m\rangle = \frac{1}{\sqrt{2}}\left(\begin{array}{c} |n-1,m\rangle \\ 
\xi \lambda |n,m\rangle  \end{array}\right),
\eeq
respectively.

One may furthermore use the commutation relation (\ref{ComGC}) for counting the number of states, i.e. the degeneracy, in each
LL. Indeed, this relation indicates that one may not measure both components of the guiding center simultaneously, which is therefore
smeared out over a surface 
\beq\label{minsurf}
\Delta X\Delta Y=2\pi l_B^2,
\eeq 
as it is depicted in Fig. \ref{fig09}(b). The result (\ref{minsurf}) for the surface occupied by a quantum state may be calculated
rather simply if one chooses a particular gauge, such as the Landau or the symmetric gauge for the vector potential, but its 
general derivation is rather involved \cite{Imry}.
This minimal surface plays the same role as the surface (action) $h$ in phase space and therefore allows us to count the number of 
possible quantum states of a given (macroscopic) surface $\Amath$,
\beq
N_B=\frac{\Amath}{\Delta X\Delta Y} =  \frac{\Amath}{2\pi l_B^2}= n_B \times \Amath,
\eeq
where we have introduced the flux density
\beq\label{fluxdens}
n_B= \frac{1}{2\pi l_B^2} = \frac{B}{h/e},
\eeq 
which is nothing other than the magnetic field measured in units of the flux quantum $h/e$, as already mentioned above.
The ratio between the electronic density $n_{el}$ and this flux density then defines the \textit{filling factor}
\beq\label{filling}
\nu=\frac{n_{el}}{n_B}=\frac{h n_{el}}{eB},
\eeq
which characterizes the filling of the different LLs.

\paragraph{The relativistic quantum Hall effect.}

The integer quantum Hall effect (IQHE) in 2D electron systems \cite{KDP}
is a manifestation of the LL quantization and the macroscopic
degeneracy (\ref{fluxdens}) of each level, as well as of semi-classical electron localization
due to the sample impurities.\footnote{Strictly speaking, the IQHE requires only the breaking of 
translation invariance, which in a diffusive sample is due to impurities. In a ballistic sample, translation
invariance is broken via the sample edges \cite{butt}. } 
In a nutshell, this energy quantization yields a quantization of the Hall resistance 
\beq\label{HallRes}
R_H=\frac{h}{e^2N},
\eeq
where $N=[\nu]$ is the integer part of the filling factor (\ref{filling}), while the longitudinal resistance 
vanishes.\footnote{A simultaneous measurement of the Hall and the longitudinal resistance requires a particular geometry with at
least four electric contacts [for a recent review on the quantum Hall effect, see Ref. \cite{goerbigQHE}].}
The resistance quantization reflects the presence of an incompressible quantum liquid with gapped single-particle and density excitations.
In the case of the IQHE, at integer filling factors, the gap is simply given by the energy difference between adjacent LLs which must
be overcome by an electron that one adds to the system. Notice that if one takes into account the electron spin and a vanishing Zeeman
effect, the condition for the occurence of the IQHE is satisfied when both spin branches of the last LL $n$ 
are completely filled, and one thus 
obtains the Hall-resistance quantization at the filling factors
\beq\label{fillingIQHE}
\nu^{IQHE}=2n,
\eeq
i.e. for even integers. Odd integers may principally be observed at higher magnetic fields when the Zeeman effect becomes prominent, and
the energy gap is then no longer given by the inter-LL spacing but by the Zeeman gap. This picture is naturally simplistic and needs to
be modified if one takes into account electronic interactions -- their consequences, such as the fractional quantum Hall effect or 
ferromagnetic states are discussed, in the context of graphene, in Sec. \ref{Chap:FQHE}.

The phenomenology of the relativistic quantum Hall effect (RQHE) in graphene is quite similar to that of the IQHE. Notice, however, that 
one is confronted not only with the two-fold spin degeneracy of electrons in graphene (in the absence of a strong Zeeman effect), but also with the two-fold valley degeneracy due to the presence of the $K$ and $K'$ points in the first BZ, which govern the low-energy
electronic properties. The filling factor therefore changes by steps of 4 between adjacent plateaus in the Hall resistance. Furthermore,
the filling factor (\ref{filling}) is defined in terms of the carrier density which vanishes at the Dirac point. This particle-hole 
symmetric situation naturally corresponds to a half-filled zero-energy LL $n=0$, whereas all levels with $\lambda=-$ are completely
filled and all $\lambda=+$ levels are unoccupied. In the absence of a Zeeman effect and electronic interactions, there is thus
no quantum Hall effect at $\nu=0$, and the condition of a completely filled (or empty) $n=0$ LL is found for $\nu=2$ ($\nu=-2$). As
a consequence, the signature of the RQHE is a Hall-resistance quantization at the filling factors \cite{gusynin05,gusynin06,peres06}
\beq\label{fillingRQHE}
\nu^{RQHE}=2(2n+1),
\eeq
which needs to be contrasted to the series (\ref{fillingIQHE}) of the IQHE in non-relativistic 2D electron systems. The series 
(\ref{fillingRQHE}) has indeed been observed in 2005 within the quantum Hall measurements \cite{novoselov1,zhang1}, which 
thus revealed the relativistic character of electrons in exfoliated graphene. More recently, the RQHE has been observed also
in epitaxial graphene with moderate mobilities \cite{shen09,wu09,jobst09}.

\paragraph{Experimental observation of relativistic Landau levels.}

The $\sqrt{Bn}$ dispersion of
relativistic LLs has been observed experimentally in transmission spectroscopy, where one shines 
monochromatic light on the sample and measures
the intensity of the transmitted light. Such experiments have been performed both on epitaxial 
\cite{sadowski} and  exfoliated graphene \cite{jiang}.

When the monochromatic light is in resonance with a dipole-allowed transition from the (partially) filled $(\lambda,n)$ 
to the (partially) unoccupied LL $(\lambda',n\pm 1)$,
it is absorbed due to an electronic excitation between the two levels.
Notice that, in a non-relativistic 2D electron gas,
the only allowed dipolar transition is that from the last occupied LL $n$ to the first unoccupied one $n+1$. The transition
energy is $\hbar\omega_C$, independently of $n$, and one therefore observes a single absorption line (cyclotron resonance) that is
robust to electron-electron interactions, as a consequence of Kohn's theorem \cite{kohn}.

\begin{figure}
\centering
\includegraphics[width=6.5cm,angle=0]{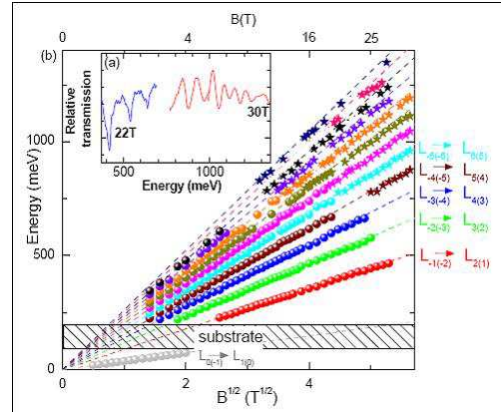}
\caption{\footnotesize 
Transmission spectroscopy on epitaxial multilayer graphene \cite{plochocka}. The inset shows a representative
transmission spectrum. The main figure represents the positions of the absorption lines as a function of the square-root of the
magnetic field. The dashed lines correspond to transitions calculated at linear order, in agreement with the Dirac equation, 
whereas one notices downward deviations in the high-energy limit.
}
\label{fig:LLcorr}
\end{figure}

In graphene, however, there are many more allowed transitions due to the presence 
of two electronic bands, the conduction and the valence band, and the transitions have the energies
\beq\label{eq:dipoleT}
\Delta_{n,\lambda}=\frac{\hbar v_F}{l_B}\left[\sqrt{2(n+1)}-\xi \sqrt{2n}\right],
\eeq
where $\lambda=+$ denotes an intraband and $\lambda=-$ an interband transition \cite{sadowski,iyengar,abergel}. 
One therefore obtains families of resonances
the energy of which disperses as $\Delta_{n,\lambda}\propto \sqrt{B}$, as it has been observed in the experiments  
[see Fig. \ref{fig:LLcorr}, where we show the results from Ref. \cite{plochocka}]. 
Notice that the dashed lines in Fig. \ref{fig:LLcorr} are fits with a single fitting parameter 
(the Fermi velocity $v_F$), which matches well all experimental points for different values of $n$ in the low-energy regime.

Moreover, the relativistic LLs have later been directly observed in scanning-tunneling spectroscopy in graphene on a graphite 
substrate\footnote{With the help of the same technique, relativistic LLs had before been identified even in graphite \cite{li07}.}
\cite{liSTM09} as well as on epitaxial graphene \cite{song10}.

\subsection{Limits of the Dirac Equation in the Description of Graphene Landau Levels}
\label{Sec:DispCorr}

Transmission spectroscopy is an ideal tool for the study of the high-energy part of the LL
spectrum when considering the transitions $(\lambda=-,n)\rightarrow(\lambda=+,n\pm 1)$, for $n\gg 1$.
As discussed in Sec. \ref{Sec:continuum}, one expects deviations [of order $\Omath(|\bq|^2a^2)$] from the
linear dispersion in this limit. These deviations renormalize the energy of the LLs and thus the transition 
energies. 

In order to quantify the effect \cite{plochocka}, we may use the Peierls substitution (\ref{mom}) and the expressions
(\ref{ladder1}) in the terms (\ref{eq2:36})
and (\ref{eq:Disp3}) corresponding to the higher-order diagonal and off-diagonal band terms, respectively. This yields the 
Hamiltonian 
\beq\label{eq:HamTWB}
H_B^{\xi}=\left(\begin{array}{cc}
         h' & h_{\xi}^*\\ h_{\xi} & h'
        \end{array}\right), 
\eeq
where the diagonal elements read
\beq\label{eq:TWBdiag}
h'= \hbar\omega'\frac{3t_{nnn}a}{\sqrt{2}tl_B}\ahat^{\dagger}\ahat,
\eeq
and the off-diagonal ones are 
\beq\label{eq:TWBodiag}
h_{\xi}=\xi \hbar\omega'\left(\ahat^{\dagger} - \xi\frac{aw_1}{2\sqrt{2}l_B}\ahat^2 -\frac{a^2w_2^2}{4l_B^2} 
\ahat^{\dagger 2}\ahat\right).
\eeq
Naturally, to lowest order in $a/l_B$, one obtains again the Hamiltonian (\ref{HamLadD}). The dimensionless parameters $w_1$ and $w_2$ are
artificially added to the expressions and play the role of fitting parameters in the comparison with experimental measurements, as will
be discussed below. They measure the deviation from the tight-binding-model expectation, $w_1=w_2=1$.  Notice furthermore that, since we 
are interested in the $n\gg 1$ limit, we do not care about corrections related to the ordering of the ladder operators, such that 
we identify $a^{\dagger 2}a^2\simeq a^2a^{\dagger 2}\simeq (a^{\dagger}a)^2$ in the following parts.

In the calculation of the LL spectrum, one may proceed in the same manner as in Sec. \ref{Sec:RelLL} -- the eigenvalue equation
(\ref{eigen2}) for the second spinor component now becomes
\beq\label{eigenTW}
h_{\xi}h_{\xi}^{\dagger}\, v_n\simeq (\epsilon_n - h')^2\, v_n,
\eeq
which is asymptotically correct in the large-$n$ limit, where we may neglect the commutator $[h_{\xi},h']$ on the right-hand side 
of the equation.\footnote{The commutator would yield relative corrections that are on the order of $1/n$ and $a/l_B$ as compared to
the energy scale $(t_{nnn}/t)(a/l_B)n$ that dominates $h'$.}
The combination $\Hhat_{\xi}\equiv h_{\xi}h_{\xi}^{\dagger}$ is now interpreted as some \textit{fake} Hamiltonian which needs to 
be diagonalized in order to obtain the modified LLs. Notice that $n$ remains a good quantum number if one considers $h'$ on the 
right-hand side of the eigenvalue equation. The left-hand side consists of a term 
\beq
\Hhat_0 \simeq (\hbar\omega')^2\left[\ahat^{\dagger}\ahat - \frac{4w_2^2 -w_1^2}{8}\left(\frac{a}{l_B}\right)^2
\left(\ahat^{\dagger}\ahat\right)^2\right]
\eeq
that contains powers of 
$\ahat^{\dagger}\ahat$ and thus respects the quantum number $n$, but in addition it contains the trigonal-warping term
\beq\label{eq:fakeTW}
\Hhat_{t.w.}=-\xi\frac{w_1(\hbar\omega')^2a}{2\sqrt{2}l_B}(\ahat^{\dagger 3} + \ahat^3),
\eeq
which does not commute with $\ahat^{\dagger}\ahat$ and which
needs to be treated apart. If we neglect this trigonal-warping term for a moment, the LL energies are obtained from the 
quadratic equation
\beq\label{eq:QE}
(\hbar\omega')^2\left[n - \frac{4w_2^2 -w_1^2}{8}\frac{a^2}{l_B^2}
n^2\right] \simeq \left(\epsilon_n - \hbar\omega'\frac{3t_{nnn}a}{\sqrt{2} tl_B} n \right)^2.
\eeq

In order to account for the trigonal-warping term in the eigenvalue equation (\ref{eigenTW}), 
we may use a perturbative treatment, which is justified because of 
the small parameter $a/l_B$. There
is no contribution at first order since $\langle n|\ahat^{(\dagger)3}|n\rangle=0$
due to the orthogonality of the eigenstates $\langle n|n'\rangle=\delta_{n,n'}$.
At second order, one obtains
\beq
\delta_n=-\frac{(\hbar\omega')^2}{8}\left(\frac{a}{l_B}\right)^2\times 3n\left[1+\mathcal{O}(1/n)\right],
\eeq
which needs to be added to the right-hand side in Eq. (\ref{eq:QE}).
Interestingly, trigonal warping thus yields the same correction to the energies of
the relativistic LLs as the third-order term in the expansion of the band dispersion, although trigonal warping occurs at second order in the absence of a magnetic field, as we have discussed in Sec. \ref{Sec:continuum}. This effect is due to the anisotropy 
of the band correction; in the presence of a magnetic field, the $\cos(3\varphi_{\bq})$ term in Eq. (\ref{eq2:37}) is averaged over the angle $\varphi_{\bq}$, and therefore 
only contributes at second order in the perturbation theory described above. This
eventually yields a correction of order $(a/l_B)^2$ to the LL energy, as does
the third-order term in the correction of the band dispersion.

One finally obtains, in the large-$n$ limit, where these corrections become relevant,
the energies of the relativistic LLs \cite{plochocka}
\beqn\label{eq:LLTW}
\epsilon_{\lambda n} &=& \hbar\frac{v_F}{l_B} \frac{3t_{nnn}}{t}\frac{a}{l_B}n\\
\nn
&&+\lambda\hbar\frac{v_F}{l_B}\sqrt{2n}\left\{1-\frac{3w^2}{8}\left(\frac{a}{l_B}\right)^2n
\left[1+\mathcal{O}(1/n)\right]\right\},
\eeqn
independent of the valley index $\xi$, 
where $\mathcal{O}(1/n)$ stands for corrections of order $1/n$. Notice that the fitting parameters $w_1$ and $w_2$
cannot be determined independently from a fit to the spectroscopic measurement, but only the combination
$w^2\equiv (w_1^2+2w_2^2)/3$. 
Equation (\ref{eq:LLTW}) generalizes a calculation for the relativistic LLs when only \textit{nnn} hopping is
taken into account \cite{peres06}.

In Fig.
(\ref{fig:LLcorr}), we show experimental results obtained from high-field transmission spectroscopy on multi-layer epitaxial graphene 
\cite{plochocka}. Qualitatively, one notices a downward renormalization of the transition energies 
\beq
\Delta_{n}=\epsilon_{\lambda=+, n} - \epsilon_{\lambda=-,n}
\eeq
in the interband regime for large values of $n$, in agreement with Eq. \ref{eq:LLTW}.
Notice that because transmission spectroscopy is sensitive to energy-level differences,
the \textit{nnn} correction in Eq. (\ref{eq:LLTW}) yields only a correction on the order of $(t_{nnn}/t)(a/l_B)/n\lesssim 1\%/n$ 
at $B\sim 25$ T as 
compared to the energy scale $t(a/l_B)n$ of the transition, whereas the other term yields a correction on the order of
$(a/l_B)^2 \times n\sim 0.5\%\times n$. The latter corrections thus become more relevant in the large-$n$ limit than the \textit{nnn} 
correction. Indeed, the experiment \cite{plochocka} was not capable of probing the electron-hole symmetry breaking associated with
the \textit{nnn} term, whereas a quantitative study of the high-energy transitions revealed a good semi-quantitative agreement with
the calculated LL spectrum (\ref{eq:LLTW}). However, it has been shown that the simple-minded tight-binding approach 
(with $w=1$) underestimates
the higher-order band corrections and that the best fit to Eq. (\ref{eq:LLTW}) is obtained for a value of $w=2.8$. The origin
of this discrepancy is yet unexplained, and it may be interesting to perform high-field transmission spectroscopy measurement also 
on single-layer exfoliated graphene in order to understand whether the stronger downward renormalization of the LLs is due to interlayer
couplings in the epitaxial multi-layer sample.

\subsection{Landau Level Spectrum in the Presence of an Inplane Electric Field}
\label{Sec:DiracE}

A remarkable consequence of the relativistic character of electrons in graphene and the Lorentz invariance of the Dirac equation
is their behavior in crossed magnetic and electric fields, where the magnetic field remains perpendicular to the graphene
sheet and the electric field is applied in the plane. Remember that in a non-relativistic 2D electron systems, the electric field
$\bE=E\be_y$ (in the $y$-direction) simply lifts the LL degeneracy and adds a term $\hbar(E/B)k$ to the LL energies, where $k$ is the 
wave vector in the $x$-direction. At a fixed wave vector $k$, the LL spacing is unaffected by the in-plane field. 

The situation is different for relativistic electrons in graphene, as a consequence of the Lorentz invariance of the
Dirac equation. One may choose a reference frame in which the electric field vanishes as long as the \textit{drift} velocity 
$v_D=E/B$ is smaller than the Fermi velocity, which plays the role of an upper bound for the physically significant 
velocities in the same manner as the speed of light in relativity 
(magnetic regime).\footnote{In the opposite case, $v_D>v_F$, one may choose a reference system in which the magnetic field vanishes 
(electric regime) \cite{jackson}.} In addition to the lifted LL degeneracy, the LL spacing is 
reduced \cite{lukose,peres07}, as may be seen from a Lorentz boost into the reference frame which moves at the drift velocity
and in which the electric field vanishes. In this reference frame, the magnetic field is reduced by the factor
\beq\label{eq:LBoost}
\sqrt{1-(E/v_FB)^2},
\eeq
such that the LLs (\ref{RelLLs}), which scale as $\sqrt{B'}=\sqrt{B}[1-(E/v_FB)^2]^{1/4}$ with the magnetic field, read
\beq
\epsilon_{\lambda,n}^{\prime} = \lambda\frac{\hbar v_F}{l_B}[1-(E/v_FB)^2]^{1/4}\sqrt{2n},
\eeq
where the primes indicate the physical quantities in the moving frame of reference. When measuring the energy in the original
(lab) frame of reference, the above energy spectrum also needs to be transformed into this frame of reference, which amounts
to being multiplied by another factor (\ref{eq:LBoost}), such that the spectrum of relativistic LLs in the presence of an
in-plane electric field becomes \cite{lukose}
\beq\label{eq:LLEfield}
\epsilon_{\lambda,n;k} = \lambda\frac{\hbar v_F}{l_B}[1-(E/v_FB)^2]^{3/4}\sqrt{2n} + \hbar \frac{E}{B} k .
\eeq
The quantum-mechanical derivation of this result will be discussed in more detail in Sec. \ref{Sec:GenWeylB} in the
context of the generalized Weyl Hamiltonian in a magnetic field.

\subsection{Landau Levels in Deformed Graphene}
\label{Sec:DiracDef}

As we have discussed in Sec. \ref{DefGraph}, a uniaxial strain deforms the graphene sheet and modifies the electronic 
structure. The induced anisotropy of the Fermi velocity $w_x\neq w_y$ is essentially washed out by the magnetic field, which yields an
effective averaging over the Fermi surface, $v_F\rightarrow v_F'=\sqrt{w_xw_y}$. More spectacular are the two further consequences 
of the deformation; (a) the tilt of the Dirac cones accounted for in the generalized Weyl Hamiltonian (\ref{eq:Weyl}) and
(b) the topological phase transition due to the Dirac point motion. The implication for the LL spectrum are briefly reviewed in the
following sections.

\subsubsection{The generalized Weyl Hamiltonian in a magnetic field}
\label{Sec:GenWeylB}

With the help of the Peierls substitution (\ref{mom}) and the expression of the kinetic momentum in terms of ladder 
operators (\ref{ladder1}), the generalized Weyl Hamiltonian (\ref{eq:Weyl}) may be cast into the form
\beq\label{WeylB}
H_B^{\xi}=\xi\frac{\hbar \sqrt{ 2 w_xw_y}}{l_B}\left(\begin{array}{cc}
 \frac{\wtilde_0}{2}(\ahat e^{i \varphi}+{\rm H.c.}) & \ahat \\
 \ahat^{\dagger} & \frac{\wtilde_0}{2}(\ahat e^{i \varphi}+{\rm H.c.} )
\end{array}
\right).
\eeq
where
\beq
\wtilde_0 e^{i\varphi}\equiv \frac{w_{0x}}{w_x}+i\frac{w_{0y}}{w_y} ,
\eeq
in terms of the effective tilt parameter (\ref{TiltParam}) and the angle $\varphi$ between the $x$-axis and
the direction of the effective tilt $(w_{0x}/w_x,w_{0y}/w_y)$, renormalized by the Fermi velocities $w_x$ and $w_y$ 
in the $x$- and $y$-direction, respectively.

The Hamiltonian (\ref{WeylB}) may be solved quantum-mechanically in a straight-forward, but lengthy manner \cite{peres07,morinari}.
Instead, one may also obtain the result in a simpler semi-classical treatment \cite{GFMP}, with the help of the Onsager 
relation \cite{onsager,onsager2} according to which the surface $S(\epsilon)$ enclosed by a trajectory of constant
energy $\epsilon$ in reciprocal space is quantized as
\beq
S(\epsilon)l_B^2=(2\pi)^2\int_0^{\epsilon}d\epsilon'\, \rho(\epsilon')=2\pi(n+\gamma),
\eeq
where $n$ is an integer denoting the energy level which coincides
with the Landau level in the full quantum treatment. The additional
contribution $\gamma$ is related to a Berry phase acquired by an
electron during its cyclotron orbit. Usually, one has $\gamma=1/2$
except if there is an extra Berry phase of $\pi$, which in our case
yields $\gamma=0$, as for graphene with no
tilt \cite{mikitik}. If one considers a density of states which
scales as $\rho(\epsilon)\propto \epsilon^{\alpha}$, the energy
levels thus scale as 
\beq\label{scaling} 
\epsilon_n \sim [B(n+\gamma)]^{1/(1+\alpha)}, 
\eeq 
in the large-$n$ limit.

Because the density of states vanishes linearly at the Dirac point, as in the case of no tilt, i.e. $\alpha=1$,
the scaling argument (\ref{scaling}) yields the energy levels,
\beq\label{LLraw}
\epsilon_{\lambda,n}\simeq\lambda\sqrt{2}\frac{\hbar v_F^*}{l_B}\sqrt{n},
\eeq
as for unconstrained graphene, apart from a renormalization of the
Fermi velocity. The latter is readily obtained from the calculation of
the total number of  states
below the energy $\epsilon$ within the positive energy cone, 
\beq
N_+(\epsilon)=\frac{1}{(2\pi)^2\hbar^2w_xw_y}\int_{\epsilon_+(\qtilde)\leq\epsilon}
d\qtilde_x d\qtilde_y=\frac{1}{2\pi\hbar^2 v_F^{*2}}\frac{\epsilon^2}{2}, 
\eeq 
where we have defined $\qtilde_{x/y}\equiv w_{x/y}q_{x/y}$, and the
renormalized Fermi velocity is 
\beq\label{FermiVel}
v_F^{*2}=
\left[w_x w_y(1-\wtilde_0^2)^{3/2}\right], \eeq 
in terms of the effective tilt parameter (\ref{TiltParam}).
This yields the result
\beq\label{eq02}
\epsilon_{\lambda, n}=\lambda \frac{\hbar\sqrt{w_xw_y}}{l_B}(1-\wtilde_0^2)^{3/4}\sqrt{2n}\, 
\eeq
for the LL spectrum in the presence of a tilt, which coincides with the one obtained from the full
quantum treatment \cite{peres07,morinari}.
One finally notices that the LL spacing becomes zero for $\wtilde_0=1$, which corresponds to the condition (\ref{TiltCond})
of maximal tilt for the Dirac cones, as discussed in Sec. \ref{DefGraph} -- indeed for values of $\wtilde_0$ larger than 1, 
the isoenergetic lines are no longer closed elliptic orbits but open hyperbolas, for which the energy is not quantised.

\subsubsection{Tilted Dirac cones in a crossed magnetic and electric field}

One notices that the form (\ref{eq02}) of LLs for tilted Dirac cones is the same as that of the LL spectrum (\ref{eq:LLEfield})
if one interprets the drift velocity $v_D=E/v_FB$ as an effective electric-field induced tilt. The magnetic regime 
$E/B < v_F$ corresponds then to the regime of closed orbits ($\wtilde_0<1$) and the open hyperbolic orbits may be identified
with the electric regime $E/B > v_F$. Mathematically, the generalized Weyl Hamiltonian with an in-plane electric field
may still be cast into the form (\ref{WeylB})
\beq
H_B^{\xi} \rightarrow H_{E/B}^{\xi} = H_B^{\xi\prime} + \hbar \frac{E}{B} k \bone,
\eeq
where $H_B^{\xi\prime}$ is the same as that of Eq. (\ref{WeylB}) if one replaces the tilt parameter 
$\wtilde_0\exp(i\varphi)$ by \cite{GFMP2}
\beq\label{eq03}
\wtilde_{\xi}(E) e^{i\varphi_{\xi}(E)}\equiv \frac{w_{\xi x}}{w_x}+i\frac{w_{\xi y}}{w_y} .
\eeq
Here, the renormalized tilt velocity is given by
\beq\label{eq04}
\bw_{\xi}=(w_{\xi x},w_{\xi y})\equiv \bw_0 - \xi \frac{\bE\times\bB}{B^2},
\eeq
and the angle $\varphi_{\xi}$ is the angle between this velocity and the $x$-axis.

The resulting energy spectrum is given by 
\beq\label{eq05}
\epsilon_{\lambda, n;k}^{\xi}(E)=\lambda \frac{\hbar \sqrt{w_xw_y}}{l_B}\left[1-\wtilde_{\xi}(E)^2\right]^{3/4}\sqrt{2n} 
+ \hbar \frac{E}{B} k\, .
\eeq
Naturally one obtains the result (\ref{eq:LLEfield}) for undeformed graphene in an in-plane electric field, for $w_x=w_y=v_F$ and
$\bw_0=0$, as well as the LL spectrum (\ref{eq02}) for the generalized Weyl Hamiltonian with tilted Dirac cones for zero
in-plane field ($E=0$). However, the most interesting situation arises when both the tilt and an in-plane field are present, in which
case one observes a lifting of the valley degeneracy that is maximal when the electric field is applied perpendicular to the
tilt velocity, $\bE\perp \bw_0$ \cite{GFMP2}. 

Notice that, in order to obtain an effect on the order of $\sim 1\%$, extremely large electric fields would be required 
(on the order of $10^6$ V/m) for a 10\% deformation of the lattice \cite{GFMP2}. 
It seems therefore difficult to observe the effect in graphene,
e.g. in high-field transmission spectroscopy or transport measurments, whereas the effect may be more visible in \BEDT, where
the Dirac cones are naturally tilted \cite{katayama06,kobayashi07} and where lower electric fields would be required for a 
comparable effect due to a roughly ten times smaller effective Fermi velocity.



\section{Electronic Interactions in Graphene -- Integer Quantum Hall Regime}
\label{Chap:Ints}

In the preceding sections, we have discussed the electronic properties of graphene within a one-particle model, i.e. we have 
neglected the Coulomb interaction between the electrons. In many materials, the one-particle picture yields the correct 
qualitative description of the electronic properties and is modified only quantitatively if one includes the electron-electron
interactions within perturbation theory \cite{mahan,GV}.
Notice, however, that there exists a class of materials -- strongly correlated 
electron systems -- the electronic properties of which may not be described correctly, not even on the qualitative level, within
a one-particle picture. 

In order to quantify the role of the electronic interactions, i.e. the correlations, in graphene, one needs to compare the 
characteristic Coulomb energy $E_{int}= e^2/\eps \ell$ at the average inter-electronic distance
($\eps$ is the dielectric
constant describing the environment the 2D electron gas is embedded in) to the kinetic one $E_{kin}(k_F)$ at the same length 
scale, given in terms of the Fermi wave vector $k_F$,  $\ell\sim k_F^{-1}$,
\beq\label{rs}
r_s=\frac{E_{int}}{E_{kin}}.
\eeq
If this dimensionless interaction parameter becomes very large, $r_s\gg 1$, the electrons are strongly correlated. In 
non-relativistic 2D metals with a parabolic band dispersion, $E_{kin}\sim \hbar^2 k_F^2/m_b$,
the dimensionless parameter reads
\beq\label{rs:nonrel}
r_s=\frac{m_b e^2}{\hbar^2\eps} \ell \sim \frac{1}{a_0^*k_F},
\eeq
in terms of the effective Bohr radius $a_0^*=a_0 \eps m_0/m_b$, where $a_0=0.5$ \AA~is the Bohr radius in vacuum and 
$m_b/m_0$ the ratio between the band and the bare electron mass. The relevance of electronic correlations therefore increases
in the dilute limit when $\ell \gg a_0^*$. Notice that the parameter $r_s$, which is also called Wigner-Seitz radius, 
plays the role of a length measured in units of the effective Bohr radius $a_0^*$.

The same argument applied to graphene yields a completely different result. Whereas the scaling of the Coulomb energy remains the
same, that of the kinetic energy is changed due to the linearity of the band dispersion. As a consequence the dimensionless 
interaction parameter in graphene reads
\beq\label{rs:graph}
\alpha_G= \frac{E_{int}}{E_{kin}}= \frac{e^2}{\hbar\eps v_F}\simeq \frac{2.2}{\eps},
\eeq
independent of the carrier density.\footnote{In contrast to an electron system with a parabolic band dispersion, this parameter
can no longer be interpreted as a dimensionless radius, and we therefore use the notation $\alpha_G$ rather than $r_s$.} 
The correlations are therefore in an intermediate regime but may be decreased if the graphene 
sheet is embedded in an environment with a large dielectric constant. Notice that the expression (\ref{rs:graph}) is the same
as that of the fine structure constant $\alpha=e^2/\hbar \eps c=1/137$ in quantum electrodynamics \cite{weinberg} if one replaces the
Fermi velocity by the velocity of light, which is roughly 300 times larger. One therefore calls $\alpha_G$ alternatively the
\textit{graphene fine structure constant}.

\paragraph*{Long-range versus short-range interactions.}

Another important aspect of interacting electrons is the range of the interaction potential. Whereas the underlying Coulomb
potential $e^2/\eps r$ is long-range, short-range interaction models such as the Hubbard model are often -- and successfully -- 
used in the description of correlated metals. The use of such a short-range interaction may be justified by the screening
properties of interacting electrons, which are correctly captured in a Thomas-Fermi approach \cite{mahan,GV} according to
which the Coulomb interaction potential is screened above a characteristic screening length 
$\lambda_{TF} \sim 1/k_{TF}$.\footnote{Notice that the Thomas-Fermi approach is restricted to static screening effects, 
whereas dynamic screening require a more complex treatment, e.g. in the framework of the random-phase approximation.} In
2D, the Thomas-Fermi wave vector 
\beq\label{TFwv}
k_{TF}\simeq r_s k_F
\eeq
is given in terms of the dimensionless interaction parameter (\ref{rs}) and the Fermi wave vector
$k_F$.\footnote{In three space dimensions, the relation reads $k_{TF}^2\simeq r_s k_F^2$.}

One notices that, for metals with a parabolic dispersion relation, the Thomas-Fermi wave vector is simply given in terms of
the inverse effective Bohr radius, $k_{TF}\sim 1/a_0^*$, independent of the electronic density. Unless the band mass is very small
as compared to the bare electron mass or the dielectric constant of the host material very large, the Coulomb interaction is
therefore screened on the atomic scale. A description of such systems in the framework of short-range
interaction models, such as the Hubbard model, then becomes better justified than in systems with a small band mass or a prominent
dielectric constant (such as in 2D electron systems in GaAs heterostructures).
Typical examples, where short-range interaction model yields valuable physical insight, 
are heavy-fermion compounds [for a review see Ref. \cite{coleman}].

The situation is again drastically different in graphene where the Thomas-Fermi wave vector (\ref{TFwv}) becomes
\beq\label{TFwvG}
k_{TF}^G\simeq \alpha_G k_F \simeq \frac{2.2}{\eps} k_F \sim \sqrt{n_{el}},
\eeq
i.e. it vanishes at the Dirac points where the carrier density goes to zero, and the screening length then diverges.\footnote{Due
to this divergence of the screening length, one principally needs to describe screening beyond the level of linear-response theory
\cite{katsnelson06}.} 
Notice that even for doped graphene, where one may typically induce carrier densities on the order of $10^{12}$ cm$^{-2}$, the screening
length is $\lambda_{TF}\gtrsim 10$ nm, i.e. much larger than the lattice scale. 

One thus comes to the conclusion that the relevant electronic interactions in graphene are long-range Coulomb interactions that may
not be captured, in contrast to other materials with a parabolic band dispersion, within models such as the Hubbard model
\cite{herbut06,herbut07}. We 
therefore investigate, in this section, the fate of the long-range Coulomb interaction in a strong magnetic field. 
In Sec. \ref{Sec:CoulB},
we decompose the Coulomb interaction Hamiltonian in the two-spinor basis of the low-energy electronic wave functions in graphene and 
comment on its symmetry with respect to the valley pseudospin. The role of these interactions in the particle-hole excitation spectrum
is studied in Sec. \ref{Sec:PHES}, where we discuss the resulting collective excitations in the IQHE regime, which allows for a
perturbative treatment. The strong-correlation regime of partially filled LLs (regime of the fractional quantum Hall effect) 
is presented separately in Sec. \ref{Chap:FQHE}.

\subsection{Decomposition of the Coulomb interaction in the Two-Spinor Basis}
\label{Sec:CoulB}

Generally, the Coulomb interaction for 2D electrons may be accounted for by the Hamiltonian
\beq\label{eq:IntHam0}
H_{int}=\frac{1}{2}\sum_{\bq} v(\bq)\rho(-\bq)\rho(\bq),
\eeq
in terms of the Fourier components $\rho(\bq)=\int d^2r\exp(-i\bq\cdot\br)\psi^{\dagger}(\br)\psi(\br)$ 
of the electronic density $\psi^{\dagger}(\br)\psi(\br)$ and the 2D Fourier transformed
$1/r$ Coulomb potential, $v(\bq)=2\pi e^2/\eps |\bq|$. If one takes into account the electronic spin $\sigma=\ua,\da$,
the Coulomb interaction respects the associated SU(2) symmetry, and the Fourier components are then simply the sum 
of the densities $\rho_{\sigma}(\bq)$ in both spin orientations, $\rho(\bq)=\rho_{\ua}(\bq) + \rho_{\da}(\bq)$. For notational
convenience, we neglect the spin index in the following discussion keeping in mind that the spin SU(2) symmetry is respected.
The density operators may be decomposed in the 
basis of the spinor wave functions (\ref{QstateN0}) and (\ref{QstateR}) for relativistic electrons in 
graphene, 
\beq\label{eq:DensDec}
\rho(\bq)=\sum_{\begin{subarray}{c} \lambda n, m;\xi \\ \lambda' n',m';\xi' \end{subarray}}
\psi_{\lambda n,m;\xi}^{\dagger} e^{-i\bq\cdot\br} \psi_{\lambda' n',m';\xi'}\, 
c_{\lambda n,m;\xi}^{\dagger}c_{\lambda' n',m';\xi'},
\eeq
where $c_{\lambda n,m;\xi}^{(\dagger)}$ are fermion operators in second quantization that annihilate (create) an electron
in the quantum state 
\beqn\label{eq:QstateB}
\nn
\psi_{\lambda n,m;\xi=+} &=& \left(
\begin{array}{c} 1_n^* |n-1,m\rangle \\ \lambda 2_n^* |n,m\rangle \end{array} \right) e^{i\bK\cdot\br} 
\\
{\rm and} ~~
\psi_{\lambda n,m;\xi=-} &=& \left(
\begin{array}{c}  -\lambda 2_n^* |n,m\rangle \\ 1_n^* |n-1,m\rangle \end{array} \right) e^{-i\bK\cdot\br} .
\eeqn
In order to avoid confusion in the case of inter-valley coupling, we use now a representation
in which the first spinor component represents the amplitude on the A sublattice and the second on the B sublattice for both
valleys.
Contrary to the expressions (\ref{QstateN0}) and (\ref{QstateR}), the state (\ref{eq:QstateB}) is valid for both $n=0$ and 
$n\neq 0$ by using the short-hand notation $1_n^*\equiv \sqrt{(1-\delta_{n,0})/2}$ and $2_n^*\equiv \sqrt{(1+\delta_{n,0})/2}$.
Furthermore, we have explicitly taken into account the rapidly oscillating part $\exp(i\xi\bK\cdot\br)$ due to the
two different valleys, whereas the expressions (\ref{QstateN0}) and (\ref{QstateR}) are only concerned with the slowly varying
envelope function. Explicitly, the Fourier components of the density operator (\ref{eq:DensDec}) then read
\beq\label{eq:DensB}
\rho(\bq)=\sum_{\begin{subarray}{c} \lambda n,\lambda' n' \\ \xi,\xi' \end{subarray}}\Fmath_{\lambda n,\lambda' n'}^{\xi,\xi'}(\bq)
\rhobar_{\lambda n,\lambda' n'}^{\xi,\xi'}(\bq),
\eeq
in terms of the \textit{reduced} density operators
\begin{widetext}
\beq\label{eq:RedDens}
\rhobar_{\lambda n,\lambda' n'}^{\xi,\xi'}(\bq) = \sum_{m,m'}\left\langle m\left| e^{-i[\bq + (\xi-\xi')\bK]\cdot\bR} \right| 
m'\right\rangle c_{\lambda n,m;\xi}^{\dagger}c_{\lambda' n',m';\xi'},
\eeq
which may also be interpreted as \textit{magneto-exciton} operators associated with a particular inter-LL 
transition (see Sec. \ref{sec:ElPhon}), and the \textit{form factors}
\beq\label{eq:form2}
\Fmath_{\lambda n,\lambda' n'}(\bq) \equiv \Fmath_{\lambda n,\lambda' n'}^{\xi,\xi}(\bq) = 1_n^*1_{n'}^*\left\langle n-1\left|
e^{-i\bq\cdot\etab} \right| n'-1\right\rangle +\lambda\lambda' 2_n^*2_{n'}^*\left\langle n\left|e^{-i\bq\cdot\etab} \right| n'\right\rangle
\eeq
for intra-valley and
\beq\label{eq:form}
\Fmath_{\lambda n,\lambda' n'}^{+,-}(\bq) = \lambda 1_{n'}^*2_{n}^*\left\langle n\left|
e^{-i(\bq + 2\bK)\cdot\etab} \right| n'-1\right\rangle
-\lambda' 1_n^*2_{n'}^*\left\langle n-1\left|
e^{-i(\bq + 2\bK)\cdot\etab} \right| n'\right\rangle = \left[\Fmath_{\lambda' n', \lambda n}^{-,+}(-\bq)\right]^*
\eeq
for inter-valley processes. 
Here, we have used the decomposition $\br=\bR+\etab$ of the position operator into its guiding center and cyclotron coordinate
(see Sec. \ref{Sec:QMtreat}) and the fact that $f_1(\etab)f_2(\bR)|n,m\rangle = f_1(\etab)|n\rangle \otimes f_2(\bR)|m\rangle$,
for two arbitrary functions $f_1$ and $f_2$. The full expressions for the matrix elements in Eqs. (\ref{eq:RedDens}), 
(\ref{eq:form2}), and (\ref{eq:form}) may be found in Appendix \ref{app:ME}.

In terms of the reduced density operators (\ref{eq:RedDens}), the interaction Hamiltonian (\ref{eq:IntHam0}) reads
\beq\label{eq:IntHam1}
H_{int} = \frac{1}{2} \sum_{\bq}\sum_{\begin{subarray}{c}{\lambda_1 n_1 \hdots \lambda_4 n_4}\\ {\xi_1 \hdots \xi_4}\end{subarray}}
v_{\lambda_1 n_1 \hdots \lambda_4 n_4}^{\xi_1\hdots\xi_4}(\bq)\rhobar_{\lambda_1 n_1,\lambda_3 n_3}^{\xi_1,\xi_3}(-\bq)
\rhobar_{\lambda_2 n_2,\lambda_4 n_4}^{\xi_2,\xi_4}(\bq),
\eeq
\end{widetext}
where the interaction vertex is defined as 
\beq\label{eq:IntVert}
v_{\lambda_1 n_1 \hdots \lambda_4 n_4}^{\xi_1\hdots\xi_4}(\bq)= \frac{2\pi e^2}{\eps |\bq|} 
\Fmath_{\lambda_1 n_1,\lambda_3 n_3}^{\xi_1,\xi_3}(-\bq)\Fmath_{\lambda_2 n_2,\lambda_4 n_4}^{\xi_2,\xi_4}(\bq).
\eeq

\subsubsection{SU(2) valley symmetry}

\begin{figure}
\centering
\includegraphics[width=7.5cm,angle=0]{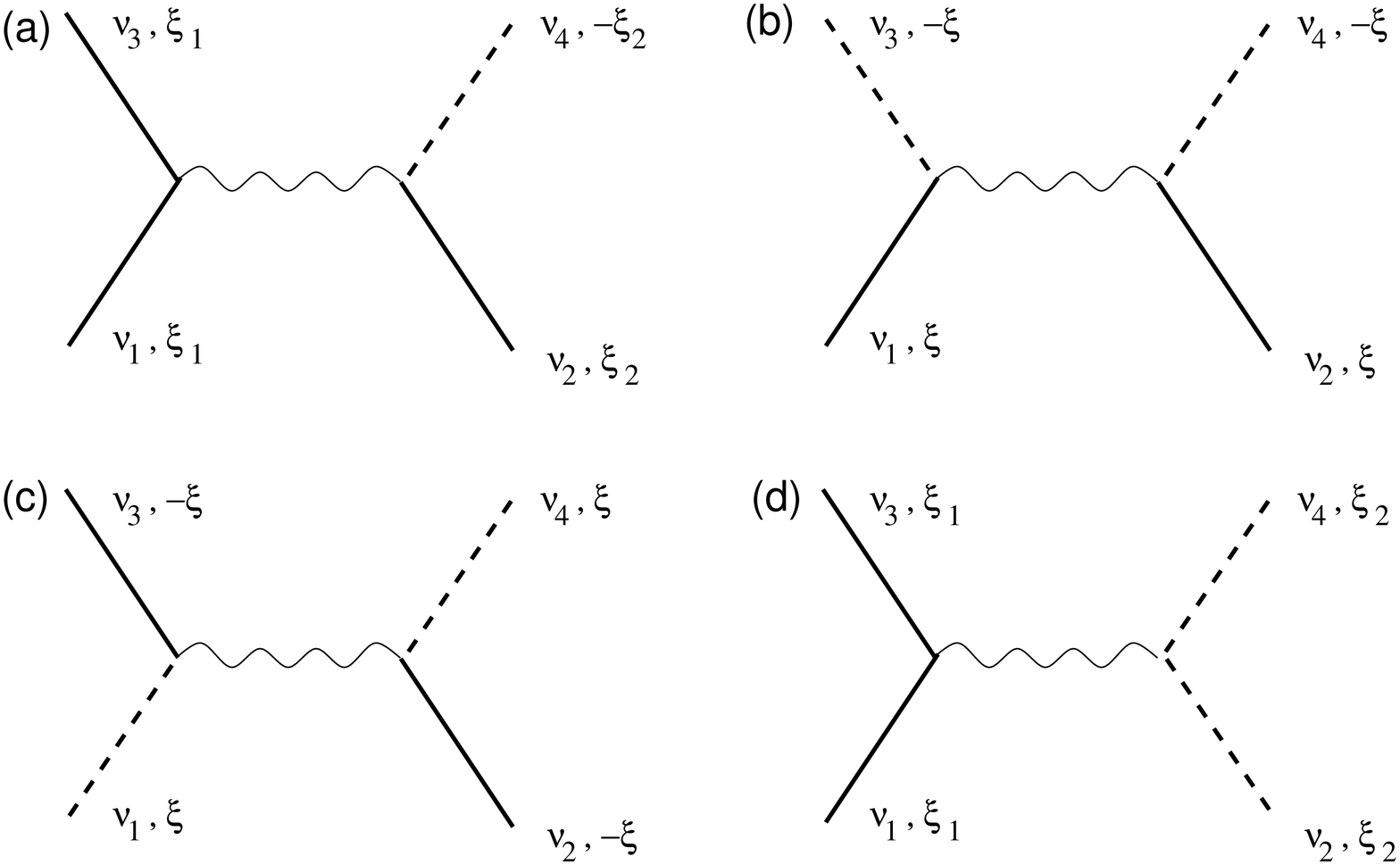}
\caption{\footnotesize{Diagrammatic representation of the interaction vertex (we use the short-hand notation $\nu_i=(\lambda_i n_i,m_i)$
for the quantum numbers);
\textit{(a)} vertex associated with terms of the form $v_{\lambda_1 n_1\hdots \lambda_4 n_4}^{\xi,\xi,\xi',-\xi'}(\bq)$ 
or $v_{\lambda_1 n_1\hdots \lambda_4 n_4}^{\xi,-\xi,\xi',\xi'}(\bq)$, \textit{(b)}
vertex of \textit{Umklapp} type, $v_{\lambda_1 n_1\hdots \lambda_4 n_4}^{\xi,\xi,-\xi,-\xi}(\bq)$,
\textit{(c)} vertex of \textit{backscattering} type, $v_{\lambda_1 n_1\hdots \lambda_4 n_4}^{\xi,-\xi,-\xi,\xi}(\bq)$
and \textit{(d)}
vertex respecting the SU($2$) valley-pseudospin symmetry $v_{\lambda_1 n_1\hdots \lambda_4 n_4}^{\xi,-\xi,\xi',-\xi'}(\bq)$.
}}
\label{fig:vertex}
\end{figure}

One notices that, in contrast to the SU(2) symmetry associated with the physical spin, the Hamiltonian (\ref{eq:IntHam1}) does not 
respect a similar valley-pseudospin symmetry due to possible inter-valley couplings. An SU(2) valley-pseudospin symmetry would 
be respected for the case $\xi_1=\xi_3$ and $\xi_2=\xi_4$, i.e. if the interaction vertex (\ref{eq:IntVert}) 
\beq\label{eq:IntVertbis}
v_{\lambda_1 n_1 \hdots \lambda_4 n_4}^{\xi_1\hdots\xi_4}(\bq) \propto \delta_{\xi_1,\xi_3}\delta_{\xi_2,\xi_4}.
\eeq
One may show, however, that the SU(2) valley-pseudospin symmetry is \textit{approximately} respected when considering the different
classes of interaction vertices depicted in Fig. \ref{fig:vertex}. 

\begin{itemize}

\item Consider the diagram in Fig. \ref{fig:vertex}(a), which represents a vertex of the type 
$v_{\lambda_1 n_1\hdots \lambda_4 n_4}^{\xi,\xi,\xi',-\xi'}(\bq)$ or $v_{\lambda_1 n_1\hdots \lambda_4 n_4}^{\xi,-\xi,\xi',\xi'}(\bq)$. 
In this case, the particle on the left remains in the same valley whereas that on the right changes the valley. Such a process
would require a momentum transfer of $\pm \bK$, i.e. of the wave vector connecting the two valleys, and therefore does not
respect momentum conservation, in the absence of a magnetic field. Naturally, momentum is not a good quantum number here due to
to the magnetic field, but momentum conservation manifests itself by an exponential suppression of such processes. In order
to appreciate this point, we need to consider the Gaussian in the form factors (\ref{eq:form2}) and (\ref{eq:form}), 
\beq
\Fmath_{\lambda n,\lambda'n'}^{\xi,\xi'}(\bq) \propto e^{-|\bq + (\xi-\xi')\bK|^2l_B^2/4},
\eeq
as discussed in Appendix \ref{app:ME} [see Eq. (\ref{eqnA02})]. One therefore sees that the interaction vertex contains a 
Gaussian term 
\beqn
v_{\lambda_1 n_1\hdots \lambda_4 n_4}^{\xi,\xi,\xi',-\xi'}(\bq) &\propto& e^{-(q^2 + |\bq \pm \bK|^2)l_B^2/4}\\
\nn
&&\sim e^{-(|\bq'|^2+|\bK|^2/4)l_B^2/2}\\
\nn
&&\sim e^{-|\bK|^2l_B^2/8}\sim  e^{-\# l_B^2/a^2},
\eeqn
where $\#$ represents an unimportant numerical factor and where we have shifted the momentum $\bq'=\bq\pm \bK/2$
in the second step. The processes associated with the diagram in Fig. \ref{fig:vertex}(a)
are thus exponentially suppressed in $l_B^2/a^2\simeq 10^4/B{\rm[T]}$ and may safely be neglected in the range of
physically accessible magnetic fields.

\item The same fate is reserved for the diagram in Fig. \ref{fig:vertex}(b), which represents a process of \textit{Umklapp} type.
In this case, the vertex reads
\beqn
v_{\lambda_1 n_1\hdots \lambda_4 n_4}^{\xi,\xi,-\xi,-\xi}(\bq) &\propto& e^{-(|\bq+\bK|^2+|\bq- \bK|^2)l_B^2/4}\\
\nn
&&\sim e^{-|\bK|^2l_B^2/2}\sim  e^{-\# l_B^2/a^2},
\eeqn
which is again exponentially small in $l_B^2/a^2$.

\item The situation is different for \textit{backscattering}-type diagrams [Fig. \ref{fig:vertex}(c)], in which case the 
interaction vertex is
\beq
v_{\lambda_1 n_1\hdots \lambda_4 n_4}^{\xi,-\xi,-\xi,\xi}(\bq) \propto e^{-(|\bq\pm \bK|^2+|\bq\pm\bK|^2)l_B^2/4}.
\eeq
One may then redefine the wave vector $\bq'=\bq\pm\bK$, which is eventually an integration variable in the interaction Hamiltonian
(\ref{eq:IntHam1}), and the interaction vertex becomes 
\beqn
v_{\lambda_1 n_1\hdots \lambda_4 n_4}^{\xi,-\xi,-\xi,\xi}(\bq') &\propto& \frac{2\pi e^2}{\eps |\bq'\mp\bK|}e^{-{q'}^2l_B^2/2}\\
\nn
&&\sim \frac{2\pi e^2}{\eps |\bK|}e^{-{q'}^2l_B^2/2}.
\eeqn
As an order of magnitude, with $|\bK|\sim 1/a$, one then notices that the backscattering interaction vertex is
suppressed by a factor of $a/l_B\sim 0.005\times \sqrt{B{\rm [T]}}$ as compared to the leading energy scale $e^2/\eps l_B$.

\item The leading interaction vertex is therefore the SU(2) valley-pseudospin symmetric one depicted in Fig. \ref{fig:vertex}(d),
for which the rapidly oscillating contribution at $\bK$ vanishes, as may be seen directly from the form factors (\ref{eq:form}).

\end{itemize}

The above argument, which generalizes symmetry considerations for the interactions in a single relativistic LL
\cite{GMD,alicea,herbut,doretto},
shows that although the valley SU(2) symmetry is not an exact symmetry, such as the SU(2) symmetry associated with 
the physical spin, it is \textit{approximately} respected by the long-range Coulomb interaction. Valley-symmetry breaking terms 
are due to lattice effects beyond the continuum limit and therefore suppressed by the small factor $a/l_B$, which quantifies precisely
corrections due to effects on the lattice scale. If one takes into account the additional spin degree of freedom, the resulting
four-fold spin-valley degeneracy may then be described within the larger SU(4) symmetry group, which turns out to be relevant
in the description of strong-correlation effects in partially filled LLs (Sec. \ref{Chap:FQHE}).

\begin{widetext}

\subsubsection{SU(4) spin-valley symmetric Hamiltonian}

The SU(4)-symmetric part of the interaction Hamiltonian (\ref{eq:IntHam1}) finally reads
\beq\label{eq:IntHamSU4}
H_{int}^{sym}=\frac{1}{2}\sum_{\bq}\sum_{\lambda_1 n_1 \hdots \lambda_4 n_4}v_{\lambda_1 n_1 \hdots \lambda_4 n_4}^{sym}(\bq)
\rhobar_{\lambda_1 n_1,\lambda_3 n_3}(-\bq)\rhobar_{\lambda_2 n_1,\lambda_4 n_4}(\bq),
\eeq
where the symmetric interaction vertex is
\beq\label{eq:IntVertSU4}
v_{\lambda_1 n_1 \hdots \lambda_4 n_4}^{sym}(\bq)=\frac{2\pi e^2}{\eps |\bq|} \Fmath_{\lambda_1 n_1,\lambda_3 n_3}(-\bq)
\Fmath_{\lambda_2 n_2,\lambda_4 n_4}(\bq),
\eeq
in terms of the reduced density operators
\beq\label{eq:RedDens2}
\rhobar_{\lambda n,\lambda' n'}(\bq) \equiv \sum_{\xi=\pm}\rho_{\lambda n,\lambda' n'}^{\xi,\xi}(\bq)
=\sum_{\xi=\pm}\sum_{\sigma=\ua,\da}\sum_{m,m'}\left\langle m\left| e^{-i\bq\cdot\bR} \right| 
m'\right\rangle c_{\lambda n,m;\xi,\sigma}^{\dagger}c_{\lambda' n',m';\xi,\sigma},
\eeq
\end{widetext}
where we have explicitly taken into account the spin index $\sigma=\ua,\da$ in the last line.

We finally notice that the graphene form factors (\ref{eq:form2}) may also be rewritten in terms
of the LL form factors
\beq\label{eq:form2DEG}
F_{n,n'}(\bq) = \left\langle n\left|e^{-i\bq\cdot\etab} \right| n'\right\rangle,
\eeq
which arise in a similar decomposition of the Coulomb interaction in Landau states in the non-relativistic
2D electron gas, as 
\beq
\Fmath_{\lambda n,\lambda' n'}(\bq)=1_n^*1_{n'}^*F_{n-1,n'-1}(\bq) + \lambda\lambda' 2_n^*2_{n'}^*
F_{n,n'}(\bq).
\eeq
To summarize the differences and the similarities between the interaction Hamiltonians in graphene and the 
non-relativistic 2D electron system, one first realizes that its structure is the same if one replaces the 
LL form factor (\ref{eq:form2DEG}) by the graphene form factors (\ref{eq:form2}) and if one takes into account
the larger (approximate) internal symmetry SU(4), due to the spin-valley degeneracy, instead of the spin
SU(2) symmetry.

In the remainder of this section, we neglect the symmetry-breaking part of the Hamiltonian and consider the Coulomb
interaction to respect the SU(2) valley symmetry.

\subsection{Particle-Hole Excitation Spectrum}
\label{Sec:PHES}

The considerations of the previous subsection allow us to discuss the role of the Coulomb interaction within a perturbative 
approach in the IQHE regime for $\nu=\pm 2(2n+1)$, where the (non-interacting) ground state is non-degenerate and separated by
the cyclotron gap $\sqrt{2}(\hbar v_F/l_B)(\sqrt{n+1}-\sqrt{n})$ from its excited states. Quite generally, 
the inter-LL transitions evolve into coherent collective excitations, as a consequence of these Coulomb interactions. Prominent
examples in the non-relativistic 2D electron gas are the upper-hybrid mode (sometimes also called \textit{magneto-plasmon}), which 
is the magnetic-field counterpart of the usual 2D plasmon \cite{GV}, and \textit{magneto-excitons} \cite{KH}. In the present
subsection, we discuss how these modes manifest themselves in graphene in comparison with the non-relativistic 2D electron gas.

\subsubsection{Graphene particle-hole excitation spectrum at $B=0$}
\label{sec:PHES_B0}

\begin{figure}
\centering
\includegraphics[width=8.5cm,angle=0]{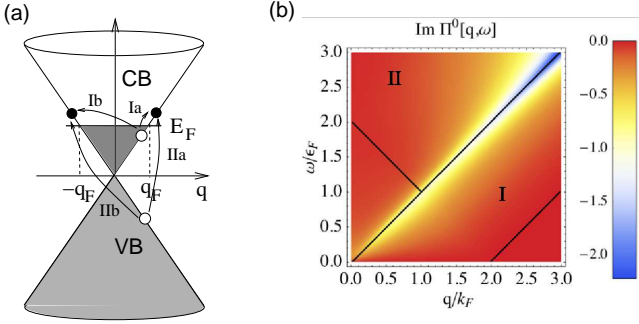}
\caption{\footnotesize{Zero-field particle-hole exctiation spectrum for doped graphene.
\textit{(a)} Possible intraband (I) and interband (II) single-pair excitations in doped graphene. The exctiations close to the 
Fermi energy may have a wave-vector transfer comprised between $q=0$ (Ia) and $q=2q_F$ (Ib), in terms of the Fermi wave vector $q_F$.
\textit{(b)} Spectral function ${\rm Im}\,\Pi_0(\bq,\omega)$ in the wave-vector/energy plane. The regions corresponding to intra- and
interband excitations are denoted by (I) and (II), respectively.
}}
\label{fig:PHESB0}
\end{figure}

Before discussing the particle-hole excitation spectrum (PHES) for graphene in the IQHE regime, we briefly review the one for $B=0$
as well as its associated collective modes \cite{shung,ando06,wunsch,hwang07}. Quite generally, the PHES is determined by the 
spectral function
\beq\label{eq:SpecFun}
S(\bq,\omega)=-\frac{1}{\pi}{\rm Im}\,\Pi(\bq,\omega),
\eeq
which may be viewed as the spectral weight of the allowed particle-hole excitations, in terms of the polarizability $\Pi(\bq,\omega)$,
which plays the role of a density-density response function \cite{mahan,GV}.

The particle-hole excitations for non-interacting electrons
in doped graphene are depicted in Fig. \ref{fig:PHESB0}.\footnote{We consider here only the case of 
a Fermi energy $\epsilon_F$ in the conduction band, for simplicity.} In contrast to the PHES of electrons in a 
single parabolic band (the non-relativistic 2D electron gas), there are two different types of excitations: intraband excitations
[labeled by I in Fig. \ref{fig:PHESB0}(a)],
where both the electron and the hole reside in the conduction band (CB), and interband excitations
[labeled by II in Fig. \ref{fig:PHESB0}(a)], where an electron is promoted from the valence band (VB) to the CB.
In undoped graphene, there exist naturally only interband excitations (II). If the electron and the hole have an energy close
to the Fermi energy, the allowed excitations imply a wave-vector transfer that lies in between $q=0$ (Ia) and $q=2q_F$ (Ib). 
At non-zero values $\epsilon$ of the transfered energy, one needs to search for available quantum states at larger wave vectors, and
the particle-hole pair wave vector is then restricted to $\epsilon/\hbar v_F < q < 2q_F + \epsilon/\hbar v_F$, as a consequence
of the linear dispersion relation in graphene. This gives rise to the region I, which describes the \textit{intraband particle-hole
continuum}, and its linear boundaries in the PHES described by the spectral function in Fig. \ref{fig:PHESB0}(b).

In addition to intraband excitations, one notices that interband excitations become possible above a threshold energy of $\epsilon_F$,
where an electron at the top of the VB (at $\bq=0$) may be promoted to an empty state slightly above the Fermi energy. The associated
wave-vector transfer is naturally $q=q_F$. The point $(q_F,\epsilon_F)$ marks the bottom of the region II in Fig. \ref{fig:PHESB0}(b),
which determines the region of allowed interband excitations \textit{(interband particle-hole continuum)}. 
Direct interband excitations with zero wave-vector transfer are possible above an energy of $2\epsilon_F$. 

Another aspect of the PHES in Fig. \ref{fig:PHESB0} is the strong concentration of spectral weight around the central diagonal 
$\omega=\hbar v_F|\bq|$. This concentration is a particularity of graphene due to the electrons' chirality \cite{polini1}.
Indeed, if one considers
a $2q_F$ backscattering
process in the vicinity of the Fermi energy in the CB, Eq. (\ref{eq3:27}) indicates that the chirality, i.e. the
projection of the sublattice-pseudospin on the direction of the wave vector, is preserved. The inversion of the direction of propagation in
the $2q_F$ process would therefore require an inversion of the A and B sublattices that is not supported by most of the 
scattering or interaction processes. This effect is reflected by a strong suppression of the spectral weight when approaching
the right boundary of the region I in the PHES associated with processes of the type Ib in Fig. \ref{fig:PHESB0}(a). Similarly,
the conservation of the electrons' chirality (\ref{eq3:27})
favors $2q_F$ processes in the interband region (II) and the suppression 
of direct $\bq=0$ interband excitations of the type IIa in Fig. \ref{fig:PHESB0}(a). Notice that, although the direction of the
wave vector is inverted in a $2q_F$ process, this indicates still the absence of backscattering because the group velocity 
$\bv = \nabla_{\bq} \epsilon_{\bq}^{\lambda}/\hbar = \lambda v_F \bq/|\bq|$ remains unchanged -- the change in the sign due to 
the inversion of the wave vector is indeed canceled by the one associated with the change of the band index.

\paragraph{Formal calculation of the spectral function.}

\begin{figure}
\centering
\includegraphics[width=8.5cm,angle=0]{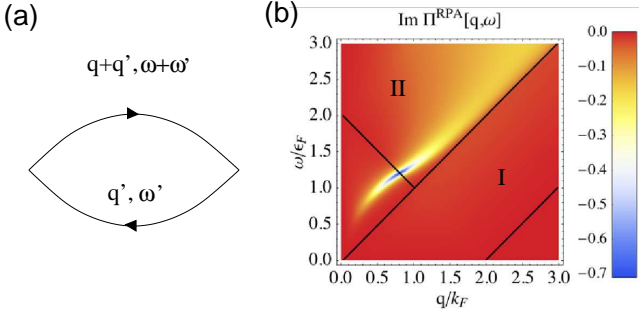}
\caption{\footnotesize{\textit{(a)} Particle-hole bubble diagram (polarizability), in terms of Green's functions 
$G(\bq,\omega)$ (lines).
\textit{(b)} Spectral function ${\rm Im}\,\Pi_{RPA}(\bq,\omega)$ for doped graphene in the wave-vector/energy plane.
The electron-electron interactions are taken into account within the RPA. We have chosen $\alpha_G=1$ here.
}}
\label{fig:PHESB0RPA}
\end{figure}

In order to obtain the spectral function, it is apparent from Eq. (\ref{eq:SpecFun}) that one needs to calculate the 
polarizability $\Pi(\bq,\omega)$ of the 2D system, which may be found with the help of the Green's functions $G(\bq,\omega)$,
\beq\label{eq:polar}
\Pi(\bq,\omega) = -i {\rm Tr}\int \frac{d\omega'}{2\pi}\sum_{\bq'}G(\bq',\omega')G(\bq+\bq',\omega+\omega'),
\eeq
where Tr means the trace since the Greens functions are $2\times 2$ matrices as a consequence of the matrix
character of the kinetic Hamiltonian. Diagrammatically, the polarizability may be represented by the so-called bubble
diagram shown in Fig. \ref{fig:PHESB0RPA}(a), and one finds for non-interacting electrons in 
graphene \cite{shung,ando06,wunsch,hwang07,polini1}
\begin{equation}\label{eq:Lind}
\Pi^0(\bq,\omega)=\frac{g}{{\cal A}}\sum_{\bq',\lambda,\lambda'}\frac{n\left(\epstilde_{\bq'}^{\lambda}\right)-n\left(\epstilde_{\bq'+\bq}^{\lambda'}\right)}{
\epstilde_{\bq'}^{\lambda}-\epstilde_{\bq'+\bq}^{\lambda'}
+\hbar\omega+i\delta}C_{\lambda\lambda'}(\bq',\bq'+\bq),
\end{equation}
where $\epstilde_{\bq}^{\lambda}=\lambda\hbar v_F|\bq| - \epsilon_F$ is the energy of the quantum state $\psi_{\lambda}(\bq)$
measured from the Fermi energy $\epsilon_F$, $g=4$ takes into account the four-fold spin-valley degeneracy, and
$n(\tilde{\epsilon}_{\bq}^{\lambda})$ is the Fermi-Dirac distribution function that reduces to a Heavyside step function
$n(\tilde{\epsilon}_{\bq}^{\lambda})=\Theta(-\tilde{\epsilon}_{\bq}^{\lambda})$ at zero temperature. 
Equation (\ref{eq:Lind}) is nothing other than the Lindhard function \cite{mahan,GV}, apart from the factor
\beq
C_{\lambda\lambda'}(\bq',\bq'+\bq)\equiv \frac{1+\lambda\lambda'\cos\theta_{\bq',\bq'+\bq}}{2},
\eeq
in terms of the angle $\theta_{\bq',\bq'+\bq}$ between $\bq'$ and $\bq'+\bq$, which takes into account the particular chirality
properties of graphene -- as already mentioned above, this chirality factor vanishes for backscattering processes, i.e. for
intraband ($\lambda=\lambda'$)
$2q_F$ processes with $\theta_{\bq',\bq'+\bq}=\pi$ as well as for interband ($\lambda=-\lambda'$) $q=0$
processes with $\theta_{\bq',\bq'+\bq}=0$ or $2\pi$.

Notice finally that the quantity $\delta$ in Eq. (\ref{eq:Lind}) is an infinitesimal energy in the case of pure graphene and
may be used (for finite values) as a phenomenological measure of the impurity broadening $\delta\simeq \hbar/\tau$, in terms
of a life time $\tau$ of the excitations.

\paragraph{Polarizability in the random-phase approximation.}

The diagrammatic approach is particularly adapted for taking into account the electronic interactions on the level
of the random-phase approximation (RPA), which amounts to calculating a geometric series of bubble diagrams
and which has shown to yield reliable results for doped graphene \cite{wunsch,hwang07,sabio08}. The RPA has also
been applied to undoped graphene \cite{gonzalez94,gonzalez99}, but its validity has been 
questioned \cite{gangad08,uchoa09} because of the vanishing density
of states, which would require to take into account diagrams beyond the RPA \cite{katsnelson06}.
The RPA polarizability then becomes
\beq\label{eq:PolRPA}
\Pi^{RPA}(\bq,\omega)=\frac{\Pi^0(\bq,\omega)}{\eps^{RPA}(\bq,\omega)},
\eeq
in terms of the polarizability (\ref{eq:Lind}) for non-interacting electrons and the dielectric
function
\beq\label{eq:dielFon}
\eps^{RPA}(\bq,\omega)= 1 -\frac{2\pi e^2}{\eps|\bq|}\Pi^0(\bq,\omega).
\eeq

The spectral function associated with the RPA polarizability (\ref{eq:PolRPA}), which is shown in Fig. \ref{fig:PHESB0RPA}(b),
reveals the characteristic coherent 2D \textit{plasmon} mode, which corresponds to the solution of the 
implicit equation $\eps^{RPA}(\bq,\omega_{pl})=0$ and the dispersion relation of which reads 
\begin{equation}\label{EqPlasmonB02DEG}
\omega_{pl}(q)\simeq\sqrt{\frac{2e^2\epsilon_F}{\hbar^2\eps}q}
\end{equation}
in the small-$q$ limit \cite{shung,wunsch}.
Interestingly, this equation
is valid also for non-relativistic electrons in conventional 2D electron systems \cite{stern67} 
if one takes into account the difference in the density dependence of the Fermi energy ($\epsilon_F=\pi n_{el}/m_b$
for non-relativistic 2D electrons and $\epsilon_F=\hbar v_F\sqrt{\pi n_{el}}$ in graphene) as well as that in the Fermi velocity 
($v_F=\sqrt{2\epsilon_F/m_b}$ for non-relativistic electrons, as compared to a constant $v_F$ in graphene).
Notice that the dispersion relation is restricted to small values of $q$ (as compared to the Fermi wave vector $k_F$), whereas 
the numerical solution presented in Fig. \ref{fig:PHESB0RPA} indicates that 
the asymptotic dependence of the plasmon mode is indeed given by the central diagonal
$\omega_{uh}(q) \gtrsim v_F q$ \cite{shung,wunsch}. 
Therefore, contrary to the plasmon in 2D metals with a parabolic dispersion relation,
the plasmon in graphene does not enter region I, but only the interband particle-hole continuum (region II). 
In this region, the Landau damping is less
efficient, and the coherence of the mode thus survives to a certain extent without decaying into incoherent particle-hole
excitations.

\subsubsection{Polarizability for $B\neq 0$}
\label{Sec:PolBnon0}

In the case of a strong magnetic field applied perpendicular to the graphene sheet, one needs to take into 
account the quantization of the kinetic energy into relativistic LLs described in Sec. \ref{Sec:QMtreat}, as well as 
the spinorial eigenfunctions $\psi_{n\lambda,m}^{\xi}$ in the calculation of the polarizability. One finds a similar
expression for the zero-temperature polarizability of non-interacting electrons as in Eq. (\ref{eq:Lind}),
\begin{widetext}
\beq\label{eq:LindB}
\Pi_B^0(\bq,\omega) = g\sum_{\lambda n,\lambda'n'}\frac{\Theta(\epsilon_F-\lambda\hbar\omega'\sqrt{n})-
\Theta(\epsilon_F-\lambda^{\prime}\hbar\omega'\sqrt{n'})}{\lambda\hbar \omega'\sqrt{n}-
\lambda^{\prime}\hbar \omega'\sqrt{n'}+\hbar\omega+i\delta} |\Fmath_{\lambda n,\lambda'n'}(\bq)|^2,
\eeq
\end{widetext}
in terms of the graphene form factors (\ref{eq:form2}) and the characteristic frequency $\omega'=\sqrt{2}v_F/l_B$ introduced 
in Sec. \ref{Sec:QMtreat}. One notices that the first part is nothing other than a Lindhard 
function \cite{mahan,GV} for relativistic LLs filled up to the Fermi energy 
$\epsilon_F=\hbar (v_F/l_B)\sqrt{2N_F}$, which is chosen to be situated between
a completely filled ($N_F$) and a completely empty ($N_F+1$) LL in the CB (IQHE regime). The second factor is the 
modulus square of the graphene form factors which plays the role of the chirality factor
$C_{\lambda,\lambda'}(\bq',\bq'+\bq)$ in the absence of a magnetic field \cite{RFG,shizuya,RGF}.\footnote{A similar expression
for the polarizability has also been obtained in Refs. \cite{berman,tahir} though with approximate form factors.}

As for the zero-field case, one may distinguish two contributions to the polarizability, one that may be viewed as a 
vacuum polarizability $\Pi^{vac}(\bq,\omega)$ and that stems from interband excitations when the Fermi level is at the Dirac 
point, and a second one that comes from intraband excitations in the case of doped graphene. Because undoped graphene with
zero carrier density does not correspond to an IQHE situation -- as we have already discussed in Sec. \ref{Sec:DiracB}, the zero-energy 
LL $n=0$ is only half-filled then --, we define, here, the vacuum polarizability with respect to the completely filled zero-energy
level.

In order to describe more explicitly the different contributions to the polarizability, we define the auxiliary quantities \cite{RFG}
\beqn
\nn
\Pi_{\lambda n,\lambda'n'}(\bq,\omega) &=& \frac{\left|\Fmath_{\lambda n,\lambda n'}(\bq)\right|^2}{ 
\lambda\hbar v_F\sqrt{n}-\lambda^{\prime}\hbar v_F\sqrt{n'}+\hbar\omega+i\delta}\\
&&+\,(\omega^+\rightarrow-\omega^-)
\eeqn
where $\omega^+\rightarrow\omega^-$ indicates the replacement $\hbar\omega+i\delta\rightarrow-\hbar\omega-i\delta$ and

\beqn
\Pi_{\lambda n}(\bq,\omega) &=& \sum_{\lambda'}\sum_{n'=0}^{n-1}\Pi_{\lambda n,\lambda'n'}(\bq,\omega)\\
\nn
&&+\sum_{\lambda'}\sum_{n'=n+1}^{N_c}\Pi_{\lambda n,\lambda' n'}(\bq,\omega)
+\Pi_{\lambda n,-\lambda n}(\bq,\omega)
\eeqn
which verify
$\Pi_{\lambda n}(\bq,\omega)=-\Pi_{-\lambda n}(\bq,\omega)$. The
vacuum polarization may then be defined as

\begin{equation}
\Pi^{vac}(\bq,\omega)=-\sum_{n=1}^{N_c}\Pi_{+ n}(\bq,\omega)
\end{equation}
where $N_c$ is a cutoff that delimits the validity of the continuum approximation. Notice that, already
in the absence of a magnetic field, the validity of the continuum
approximation is delimited by a maximal energy $\sim t$. One may then introduce an upper level cutoff with the help of
$\epsilon_{N_c}=\hbar(v_F/l_B)\sqrt{2N_c}\sim t$, that leads
to $N_c\sim 10^4/B[T]$, which is a rather high value even for strong
magnetic fields. However, due to the fact that the separation
between LLs in graphene decreases with $n$, it is always possible to
obtain reliable {\it semi-quantitative} results from smaller values of
$N_c$.

\begin{figure}
\centering
\includegraphics[width=9.0cm,angle=0]{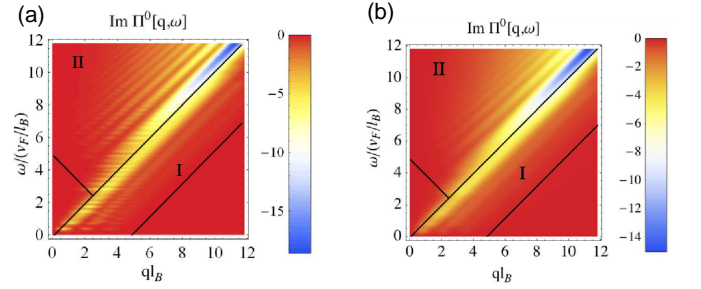}
\caption{\footnotesize{Particle-hole excitation spectrum for graphene in a perpendicular magnetic field. 
We have chosen $N_F=3$ in the CB and a LL broadening of $\delta=0.05 \hbar v_F/l_B$ \textit{(a)} and $\delta=0.2 \hbar v_F/l_B$
\textit{(b)}. The ultaviolet cutoff is chosen such that $N_c=70$.
}}
\label{fig:PHESB}
\end{figure}

The spectral function $S(\bq,\omega)=-{\rm Im}\, \Pi_B^0(\bq,\omega)/\pi$ is shown in Fig. \ref{fig:PHESB} for $N_F=3$ and
for two different values of the 
phenomenologically introduced LL broadening $\delta$. One notices first that the spectral weight
is restricted to the two regions I and II corresponding to the intraband and interband particle-hole continuum, respectively,
in the zero-field limit. This is not astonishing because the electron-hole pair wave vector remains a good quantum number 
also in the presence of a magnetic field and because the overlap between the electron and hole wave functions is largest in
these regions; if one considers the pair with its overall charge neutrality, its motion 
is unaffected by the magnetic field. Indeed, the pair momentum may be viewed as the sum of the pseudomomenta associated
with the guiding-center variable for the electron, $\bR\times \be_z/l_B^2$, and the hole $-\bR'\times \be_z/l_B^2$,
respectively. Each of the pseudomomenta is naturally a constant of motion because so is the guiding center, as we have discussed
in Sec. \ref{Sec:QMtreat}. One therefore obtains the relation 
\beq\label{eq:PseudoMom}
\bq=\Delta\bR\times\be_z/l_B^2 \qquad {\rm or} \qquad \Delta R = |\bq|l_B^2,
\eeq
where $\Delta\bR=\bR-\bR'$ is the distance between the guiding center of the electron and that of the hole. The boundaries of the PHES in
Fig. \ref{fig:PHESB} may then be obtained from the decomposition (\ref{eq:PosOp}), which yields $\eta'-\eta \leq \Delta R\leq
\eta' + \eta$, with the help of the average values $\eta \equiv \langle|\etab|\rangle =l_B\sqrt{2n+1}$ and
$\eta'=l_B\sqrt{2n'+1}$,
\begin{equation}\label{EqBoundaries}
\sqrt{2n'+1}-\sqrt{2n+1}\le ql_B\le \sqrt{2n'+1}+\sqrt{2n+1}.
\end{equation}
Because the energy scales also with $\sqrt{n}$, one obtains the linear boundaries of the particle-hole continua as in the
zero-field case mentioned above.

In contrast to these similarities with the zero-field PHES, one notices a structure in the spectral weight that is due to 
the strong magnetic field. As a consequence of the relativistic LL quantization, the spectral weight corresponds to 
inter-LL transitions at energies $\omega=\sqrt{2}\hbar (v_F/l_B)(\sqrt{n} - \lambda\sqrt{n'})$, where $n>N_F$ and $n'\leq N_F$
(for $\lambda=+$) or $n'>0$ (for $\lambda=-$). For larger values of $N_F$, or quite generally when increasing the energy, 
the level density increases due to the $\sqrt{n}$ scaling of the LLs and the transitions. The LL structure is therefore only
visible in the lower part of PHES, in the clean limit $\delta =0.05\hbar v_F/l_B $ [Fig. \ref{fig:PHESB}(a)], whereas
the inter-LL transitions are blurred at larger energies or even for the lower transitions in the case of less clean
samples [Fig. \ref{fig:PHESB}(b), for $\delta=0.2 \hbar v_F/l_B$].\footnote{The value $\delta=0.2 \hbar v_F/l_B$ is a reasonable
estimate for today's exfoliated graphene samples on an SiO$_2$ substrate \cite{andoDelt}.}

In addition to the (blurred) LL structure in the PHES, one notices another structure of the spectral weight, which is
organised in lines parallel to the central diagonal $\omega=\hbar v_F |\bq|$. This weight is again decreased when 
approaching the right boundary of the intraband continuum (region I) and the left one of the interband continuum (region II),
due to the above-mentioned chirality properties of electrons in graphene. 
The emergence of diagonal lines is a consequence of the graphene form factors (\ref{eq:form2}) the modulus square of
which intervenes in the polarization function. Indeed, these form factors $\Fmath_{\lambda (n+m),\lambda' n}(\bq)$ are
(associated) Laguerre polynomials with $n+1$ zeros \cite{gradsht} due to the overlap between the wave function of the 
hole in the level $\lambda' n$ and that of the electron in the LL $\lambda (n+m)$ \cite{RGF}. These zeros 
in the inter-LL transitions are organised
in lines that disperse parallel to the central diagonal and thus give rise to zones of vanishing spectral weight.
Interestingly, it is this structure of diagonal lines that survives in more disordered samples in which the horizontal
lines associated with inter-LL transitions start to overlap, i.e. once the LL spacing is smaller than the level broadening
$\delta$.\footnote{This behavior is in stark contrast to that of non-relativistic 2D electrons, where the LL spacing is 
constant and given by the cyclotron energy $\hbar eB/m_b$. If this quantity is larger than the level broadening $\delta$,
there is no qualitative difference between low and high energies, and the horizontal lines associated with the inter-LL
excitations (multiples of the cyclotron energy) remain well separated.}

\subsubsection{Electron-electron interactions in the random-phase approximation: upper-hybrid mode and linear magnetoplasmons}

The PHES of non-interacting electrons in graphene gives already insight into the collective modes which one may expect once
electron-electron interactions are taken into account. Indeed the regions of large spectral weight evolve into coherent
collective excitations as a consequence of these interactions. Because the regions of large spectral weight are organised 
in lines parallel to the central diagonal $\omega=\hbar v_F|\bq|$, as mentioned above, one may expect that the dominant collective
excitations are roughly linearly dispersing modes instead of the more conventional weakly dispersing \textit{magneto-excitons}, 
which emerge from the inter-LL transitions \cite{KH}. It has though been argued that such magneto-excitons may play a 
role at low energies in clean samples with low doping \cite{iyengar} and that they may renormalize the 
cyclotron energy at zero wave vector \cite{bychkov}. 

\begin{figure}
\centering
\includegraphics[width=9.0cm,angle=0]{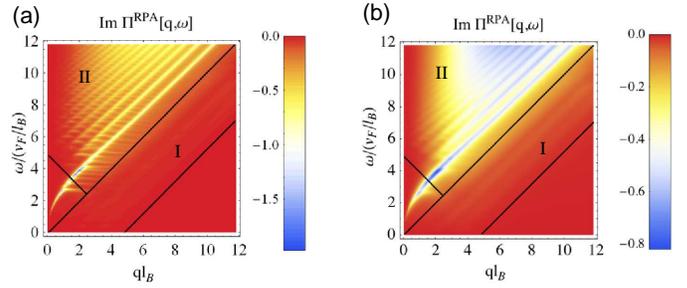}
\caption{\footnotesize{Particle-hole excitation spectrum for graphene in a perpendicular magnetic field. The Coulomb interaction
is taken into account within the RPA.
We have chosen $N_F=3$ in the CB and a LL broadening of $\delta=0.05 \hbar v_F/l_B$ \textit{(a)} and $\delta=0.2 \hbar v_F/l_B$
\textit{(b)}. The ultaviolett cutoff is chosen such that $N_c=70$.
}}
\label{fig:PHESRPA}
\end{figure}

As in the zero-field case, we take into account the Coulomb interaction within the RPA [see Eq. (\ref{eq:PolRPA})]. The 
resulting spectral function is shown in Fig. \ref{fig:PHESRPA} for the same choice of parameters as in the non-interacting
case (Fig. \ref{fig:PHESB}). Furthermore, we have chosen a dimensionless interaction parameter $\alpha_G=1$, here, which corresponds
to a dielectric constant of $\eps\simeq 2$.

One notices the prominent mode that evolves in the originally forbidden region for particle-hole excitations. This mode,
which is called \textit{upper-hybrid mode}, is the magnetic-field descendent of the 2D plasmon mode (\ref{EqPlasmonB02DEG})
and acquires a density-dependent cyclotron gap $\omega_C=eBv_F^2/\epsilon_F=eBv_F/\hbar v_F\sqrt{\pi n_{el}}$. Its 
dispersion relation in the small-$q$ limit is then given by 
\beqn\label{eq:UHdisp}
\omega_{uh}(q)&\simeq&\sqrt{\omega_{pl}^2(q) + \omega_C^2}\\
\nn
&\simeq&\sqrt{\frac{2\hbar e^2 v_F \sqrt{\pi n_{el}}}{\hbar^2\eps}q+ 
\left(\frac{eBv_F^2}{\hbar v_F\sqrt{\pi n_{el}}}\right)^2
},
\eeqn
as may be shown easily within a hydrodynamic approach that has been successfully applied to the upper-hybrid mode in 
non-relativistic 2D electron systems \cite{chiu74}.
It is apparent from Fig. \ref{fig:PHESRPA} that this mode remains 
coherent also within the region II, which corresponds to the $B=0$ interband particle-hole continuum. 

In addition to the upper-hybrid mode, one notices linearly dispersing coherent modes in the regions I and II that emerge
from the lines of large spectral weight in the non-interacting PHES, as expected from the qualitative discussion above. We may
call these modes \textit{linear magneto-plasmons} \cite{RFG,RGF}
in order to distinguish them clearly from the upper-hybrid mode (\ref{eq:UHdisp})
and the weakly dispersing magneto-excitons at low doping \cite{iyengar}. These modes are more prominent in the interband
than in the intraband region although they are also visible there. They are genuine to graphene with its characteristic 
$\sqrt{Bn}$ LLs that inevitably overlap in energy, above a critical LL $n$, if level broadening is taken into account, and
they may not be captured in the usual magneto-exciton approximation where the collective modes are adiabatically connected 
to the inter-LL excitations \cite{iyengar,KH,bychkov}.

\subsubsection{Dielectric function and static screening}
\label{Sec:Diel}

\begin{figure}
\centering
\includegraphics[width=9.0cm,angle=0]{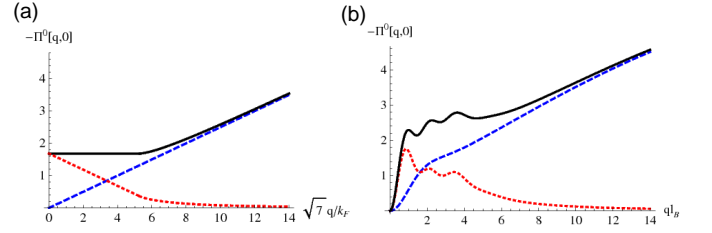}
\caption{\footnotesize{
Static polarization function $\Pi^0(\bq,\omega=0)$ for non-interacting electrons in graphene without \textit{(a)} 
and with \textit{(b)} a magnetic field. To compare both cases, we have chosen a Fermi wave vector
$q_F=\sqrt{2N_F+1}/l_B=\sqrt{7}/l_B$ that corresponds to $N_F=3$. The full black line represents the total
polarizability, whereas the red dotted and the blue dashed lines show the intraband and the interband contributions,
respectively. From \cite{RGF}.
}}
\label{fig:StatPol}
\end{figure}

We terminate this section with a brief discussion of the dielectric function (\ref{eq:dielFon}) in the static limit
\beq\label{eq:dielFonStat}
\eps^{RPA}(\bq)\equiv \eps^{RPA}(\bq,\omega=0)=1-\frac{2\pi e^2}{\eps |\bq|}\Pi^0(\bq,\omega=0),
\eeq
by comparing the $B\neq 0$ to the zero-field case, as shown in Fig. \ref{fig:StatPol}. As mentioned above,
one may distinguish two separate contributions to the static polarizability, the vacuum polarizability 
$\Pi^{vac}(\bq)\equiv\Pi^{vac}(\bq,\omega=0)$ due to interband excitations and the intraband contribution 
$\Pi^{dop}(\bq)\equiv\Pi^{dop}(\bq,\omega=0)$, which is only present in doped graphene,
\beq\label{eq:PolStat}
\Pi^0(\bq,\omega=0) = \Pi^{vac}(\bq) + \Pi^{dop}(\bq).
\eeq

One notices that up to $2q_F$, the zero-field static polarizability [Fig. \ref{fig:StatPol}(a)] remains constant. Indeed, 
the interband contribution (blue dashed line) increases linearly with the wave vector \cite{gonzalez99,ando06}
\beq\label{eq:PolVac}
-\Pi^{vac}(\bq) = \frac{q}{4\hbar v_F}
\eeq
and thus cancels the linear decrease of the intraband contribution (red dotted line), 
\beq\label{eq:PolDop}
-\Pi^{dop}(|\bq| \lesssim 2q_F)\simeq \rho(\epsilon_F)\left(1-\frac{q}{2q_F}\right),
\eeq
where 
\beq\label{eq:DOS}
\rho(\epsilon_F)=\frac{\epsilon_F}{2\pi\hbar^2 v_F^2}
\eeq
is the density of states per unit area at the Fermi energy.
At wave vectors larger than $2q_F$, the intraband contribution dies out, and the polarizability is dominated by interband 
excitations. 

\begin{figure}
\centering
\includegraphics[width=7.5cm,angle=0]{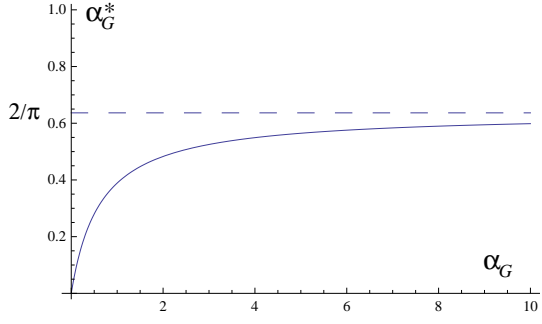}
\caption{\footnotesize{Effective coupling constant $\alpha_G^*$ as a function of the bare coupling $\alpha_G$. The dashed line indicates
the asymptotic value $2/\pi$ obtained for large values of the bare coupling ($r_s\gg 1$).
}}
\label{fig:EffCoupl}
\end{figure}

With the help of these two contributions, we may rewrite the static dielectric function (\ref{eq:dielFonStat}) as
\beq
\eps^{RPA}(\bq)=\eps_{\infty}\left(1 + \frac{\pi \alpha_G}{2\eps_{\infty}}\frac{\Pi^{dop}(\bq)}{\Pi^{vac}(\bq)}\right),
\eeq
where we have defined 
\beq\label{eq:epsInf}
\eps_{\infty}\equiv \eps^{RPA}(|\bq|\rightarrow \infty) = 1 + \frac{\pi}{2}\alpha_G,
\eeq
i.e. the value that the static dielectric function approaches at large wave vectors. Notice that this is precisely the RPA result
for the intrinsic dielectric constant for undoped graphene \cite{gonzalez99}. The above form 
of the static dielectric function may be cast into a Thomas-Fermi form,
\beq\label{eq:dielTF}
\eps^{TF}(\bq)\simeq \eps_{\infty}\left(1+\alpha_G^* \frac{q_{F}}{q}\right),
\eeq
in terms of the \textit{effective coupling constant}
\beq\label{eq:EffCoupl}
\alpha_G^*=\frac{\alpha_G}{\eps_{\infty}}=\frac{\alpha_G}{1+\pi \alpha_G/2},
\eeq
which is plotted in Fig. \ref{fig:EffCoupl}. 

One notices that interband excitations yield a contribution to the dielectric constant that originally takes into account
the dielectric environment in which the graphene sheet is embedded, $\eps\rightarrow \eps^*=\eps \eps_{\infty}$. This is a direct
consequence of the linear behavior of the vacuum polarization (\ref{eq:PolVac}) as a function of the wave vector and thus
specific to graphene. Furthermore, one may also define an effective Thomas-Fermi wave vector $q_{TF}^*=q_{TF}/\eps_{\infty}=\alpha_G^*q_F$,
which describes the screening length in the presence of the vacuum polarization. As a consequence of the saturation of the effective
coupling constant (\ref{eq:EffCoupl}) at large values of $\alpha_G$, the effective Thomas-Fermi vector is thus always on the 
order of the Fermi wave vector unless $\alpha_G\ll 1$, where $\alpha_G^*\sim \alpha_G$. The relevant effective parameters 
are summarized in the table below for freestanding graphene and graphene on mainly used substrates.\footnote{The dielectric constant
$\eps$ is then the average of the dielectric constant of the substrate material and that of the vacuum.}

\begin{center}
\begin{table}[htbp]
{
\begin{tabular}{|c||c|c||c|c|}
\hline
graphene & $\eps$ & $\alpha_G$  & $\eps_{\infty}$ & $\alpha_G^*$\\
\hline \hline
in vacuum & $1$ & $2.2$ & $4.5$ & $0.5$  \\ \hline
on SiO$_2$ & $2.5$ & $0.9$ & $2.4$ & $0.38$ \\ \hline
on h-BN & $2.3$ & $1$ & $2.4$ & $0.39$ \\ \hline
on SiC & $5.5$ & $0.4$ & $1.6$ & $0.25$ \\ \hline
\end{tabular}}
\label{Tab:rs}
\caption{\footnotesize Dielectric constant $\eps$, $\eps_{\infty}$, bare coupling $\alpha_G$, and effective coupling $\alpha_G^*$ for
graphene in vacuum and popular substrates. 
}
\end{table}
\end{center}

Finally, the screened Coulomb interaction potential may be approximated as 
\beq\label{eq:ScrCoul}
v_{scr}(q)\simeq \frac{2\pi e^2}{\eps\eps_{\infty}(1+\alpha_G^* q_F/q) q}.
\eeq
One notices from this expression that processes at wave vectors $q\ll q_F$, where the interband excitations play a negligible
role [see Fig. \ref{fig:StatPol}(a)], are still governed by the bare coupling constant $\alpha_G\sim v_{scr}(q\ll q_F)/\hbar v_F q$.
However, processes at or above the Fermi wave vector, such as those that are relevant in the electronic transport, 
are characterized by the effective coupling constant $\alpha_G^*\sim v_{scr}(q\gtrsim q_F)/\hbar v_F q$, which saturates at a value
of $2/\pi$ as mentioned above. If we consider the value (\ref{rs:graph}), $\alpha_G\simeq 2.2$, for the bare coupling constant of
graphene in vacuum, the effective coupling is roughly four times smaller $\alpha_G^*\simeq 0.5$, such that the electrons in doped
graphene approach the weak-coupling limit. 
The situation is different in undoped graphene, where recent renormalization-group \cite{juricic1,juricic2,juricic3} and
lattice gauge theoretical calculations \cite{DL1,DL2} 
indicate a flow towards strong coupling at moderate values of $\alpha_G$.

\begin{figure}
\centering
\includegraphics[width=7.5cm,angle=0]{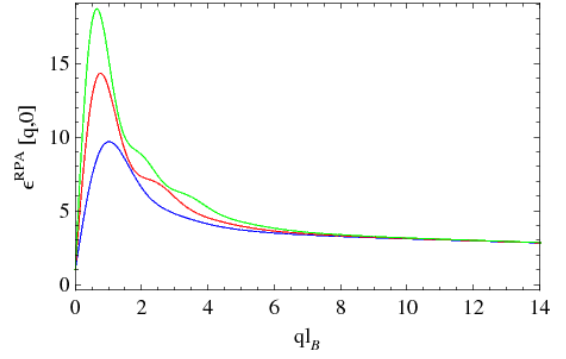}
\caption{\footnotesize{Static dielectric function for graphene in the IQHE regime for $N_F=1,2$ and 3 
(blue, red and green curves, in increasing order).
From \cite{RGF}.
}}
\label{fig:dielB}
\end{figure}

In Fig. \ref{fig:StatPol}(b), we have plotted the static polarizability for graphene in the IQHE regime. Qualitatively, the result
agrees with the zero-field behavior, with a (roughly) linearly increasing vacuum polarizability and a decreasing intraband
contribution, apart from some superimposed oscillations due to the overlap functions that are reflected by the form 
factors (\ref{eq:form2}). An important difference is manifest in the small wave-vector limit of the polarizability. In contrast
to the zero-field case, where the polarizability saturates at a non-zero density of states, the system is gapped in the IQHE regime
with a resulting vanishing density of states at the Fermi energy. This gives rise to a $q^2$ behavior of the polarizability at
small wave vectors. Furthermore, the static dielectric function, which is shown in Fig. \ref{fig:dielB} \cite{shizuya,RGF},
does no longer diverge in this limit, contrary to the zero-field Thomas-Fermi result (\ref{eq:dielTF}). Indeed, the small-$q$ behavior
may be approximated as 
\beq
\varepsilon^{RPA}(q)-1\propto \alpha_GN_F^{3/2}ql_B,
\eeq
which is the same as for non-relativistic 2D electrons \cite{Aleiner95}.\footnote{Notice, however, that the expression becomes exact only
in the large-$N_F$ limit and that in non-relativistic 2D electron systems, the coupling constant $r_s$ also depends on $N_F$,
$r_s\sim N_F^{-1/2}$.} The maximum of the static dielectric function is obtained at $ql_B\sim 1/q_F l_B\sim 1/\sqrt{2N_F+1}$, 
at the value $\varepsilon_{\max}\simeq \varepsilon^{RPA}(q\sim 1/l_B\sqrt{2N_F+1})\sim \alpha_GN_F$. It therefore scales as
$\eps_{max}\propto N_F$ in contrast to a $\sqrt{N_F}$ scaling in non-relativistic 2D systems \cite{Aleiner95}. At large wave
vectors, the static dielectric function saturates at the same value $\eps_{\infty}$ as in zero magnetic field.



\section{Magneto-Phonon Resonance in Graphene}
\label{Chap:MPhon}

In the previous section, we have discussed the role of electron-electron interactions in the IQHE regime, where a 
perturbative (diagrammatic) approach may be applied. Similarly, one may treat the electron-phonon interaction
in a perturbative manner in this regime. This is the subject of the present section, before discussing again
electron-electron interactions in the strong coupling limit of partially filled LLs (Sec. \ref{Chap:FQHE}).

As a consequence of the honeycomb-lattice structure of graphene, with two inequivalent sublattices, there are 
four in-plane phonons, two acoustic and two optical ones. Each phonon type occurs in a longitudinal 
(longitudinal acoustic, LA, and longitudinal optical, LO) and a transverse (transverse acoustic, TA, and transverse 
optical, TO) mode. In addition to these four phonons, one finds two out-of-plane phonons, one acoustic (ZA) and
one optical (ZO) [for a review of phonons in graphene, see Refs. \cite{SDD} and \cite{wirtz}]. Here, we concentrate on 
the in-plane optical phonons LO and TO, which couple to the electronic degrees of freedom. More specifically,
we discuss these phonons at the $\Gamma$ point (E$_{2g}$ modes) in the center of the first BZ, which
are attributed to the G-band at $\hbar\omega_{ph}\simeq 0.2$ eV
in the Raman spectra [see e.g. Refs. \cite{pisana,yan07,ferrari06,gupta,graf}].

One of the most prominent effects of electron-phonon coupling in metals and semiconductors is the so-called
Kohn anomaly \cite{kohnAn}, which consists of a singularity in the phonon dispersion due to a singularity 
in the electronic density-density response function.
The analog of the Kohn anomaly in graphene yields a logarithmic divergence of the phonon frequency when the 
bare phonon frequency coincides with twice the Fermi energy \cite{ando06ElP,Lazz06,ACN07}.
We investigate how this renormalization manifests itself in graphene in
a strong magnetic field \cite{ando07ElP,GFKF}. In Sec. \ref{sec:ElPhon}, 
we consider the specific form of the electron-phonon coupling and
discuss its consequences for the renormalization of the optical phonons at the $\Gamma$ point in Sec. \ref{sec:PhonRen}.
More specifically, we consider both non-resonant (Sec. \ref{sec:NonResCoupl}) and resonant coupling (Sec. \ref{sec:ResCoupl}),
the latter being specific to graphene in a magnetic field when the phonon is in resonance with an allowed inter-LL
transition \textit{(magneto-phonon resonance)} \cite{GFKF}.

\subsection{Electron-Phonon Coupling}
\label{sec:ElPhon}

\begin{figure}
\centering
\includegraphics[width=8.5cm,angle=0]{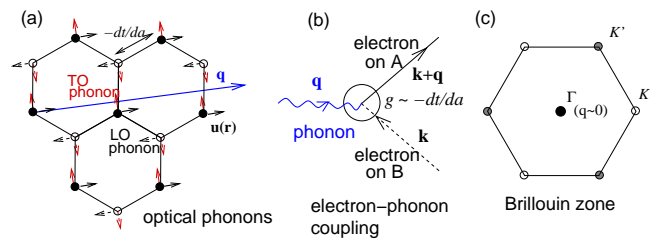}
\caption{\footnotesize{Electron-phonon coupling in graphene.
\textit{(a)} Optical phonons in graphene, with a wave vector $\bq$ in the vicinity of the $\Gamma$ point at the center of the
first BZ [see part \textit{(c)}]. The amplitude of the LO phonon is in the direction of propagation (black arrows), that of 
the TO phonon perpendicular (red arrows). The optical phonons modify the bond lengths of the honeycomb lattice.
\textit{(b)} As a consequence of the modified bond lengths, the electronic hopping is varied, and the electron-phonon coupling is
off-diagonal in the sublattice index.
}}
\label{fig:opticalPhon}
\end{figure}

The LO and TO phonons in graphene are schematically represented in Fig. \ref{fig:opticalPhon}(a). As already mentioned above,
we concentrate on phonons at small wave vectors, in the vicinity of the $\Gamma$ point. The origin of the electron-phonon 
coupling may easily be understood from the variation of the bond length caused by the phonon, which affects the electronic
hopping amplitude between \textit{nn} carbon atoms. As we have discussed in Sec. \ref{DefGraph}, the effect may be 
quantified with the help of Harrison's law \cite{harrison}, which yields $\partial t/\partial a \simeq -4.3$ eV/\AA~[see 
Eq. (\ref{eq:def_t})]. The order of magnitude for the 
bare electron-phonon energy is then obtained by multiplying this variation with the typical length
scale $\sqrt{\hbar/M \omega_{ph}}$, which characterizes the amplitude of a lattice vibration of frequency $\omega_{ph}$ within the 
harmonic approximation. The intervening mass $M$ is that of the carbon atom. Indeed, a tight-binding calculation 
\cite{ando06ElP,andoElPCN} corroborates this argument, apart from a numerical prefactor $3/2$, 
and yields a bare electron-phonon coupling
\beq\label{eq:bareElP}
g=-\frac{3}{2}\frac{\partial t}{\partial a}\sqrt{\frac{\hbar}{M\omega_{ph}}}\simeq 0.26\, {\rm eV}.
\eeq
This value agrees well with ab-initio calculations \cite{piscanec}, although it is though
slightly lower than the value determined experimentally,
which ranges from $g\simeq 0.3$ eV \cite{pisana,faugeras} to $g\simeq 0.36$ eV \cite{yan07}.

Furthermore, one notices that, because the electron-phonon coupling is mediated by a bond variation between sites that belong to 
two different sublattices, the coupling constant is \textit{off-diagonal} in the sublattice basis. This is diagrammatically
depicted in Fig. \ref{fig:opticalPhon}(b).

\subsubsection{Coupling Hamiltonian}

The above considerations help us to understand the different terms in the Hamiltonian, 
$$
H=H_{el} + H_{ph} + H_{coupl}
$$
which serves as the basis for the analysis of the electron-phonon coupling. The Hamiltonian for 2D electrons in a
magnetic field,
\beqn\label{eq:HamElPart}
\nn
H_{el} &=& \sum_{\xi}\int d^2r\, \psi_{\xi}^{\dagger}(\br) \Hmath_B^{{\rm eff},\xi}\psi_{\xi}(\br) \\
&&= \sum_{\lambda n,m;\xi}\epsilon_{\lambda, n}c_{\lambda n,m;\xi}^{\dagger}c_{\lambda n,m;\xi}
\eeqn
may be written, in second quantization, in terms of the one-particle Hamiltonian (\ref{DiracB}) and
the fermionic fields
$$
\psi_{\xi}(\br)=\sum_{\lambda n,m}\psi_{\lambda n,m;\xi}(\br) 
c_{\lambda n,m;\xi},
$$
where $\psi_{\lambda n,m;\xi}(\br)$ is the wave function in position space associated with the spinor 
(\ref{eq:QstateB}).

The lattice vibration is characterized by the relative displacement $\bu(\br)$ between the two sublattices,
which may be decomposed in terms of the bosonic operators $b_{\mu,\bq}$ and
$b_{\mu,\bq}^{\dagger}$,
\beq\label{eq:phononField}
\bu(\br) = \sum_{\mu ,\mathbf{q}}\sqrt{\frac{\hbar}{
2N_{uc}M\omega_{\mu}(\bq) }}\left( b_{\mu ,\mathbf{q}}+b_{\mu ,-\mathbf{q}}^{\dagger
}\right) \mathbf{e}_{\mu ,\mathbf{q}}e^{-i\mathbf{q}\cdot \mathbf{r}},
\eeq
where $\be_{\mu,\bq}$ denotes the two possible linear polarizations ($\mu=\text{LO,TO}$) at the wave vector $\bq$ and 
$N_{uc}=\Amath/(3\sqrt{3}a^2/2)$ is the number of unit cells in the system. The phonon Hamiltonian then 
reads, on the level of the harmonic approximation,
\beq\label{eq:HamPhon}
H_{ph}=\sum_{\mu,\bq}\hbar \omega_{\mu}(\bq)b_{\mu,\bq}^{\dagger}b_{\mu,\bq},
\eeq
in terms of the phonon dispersion $\omega_{\mu}(\bq)$. Notice that, at the $\Gamma$ point, the frequencies of the LO and
TO phonons coincide, and one has $\omega_{ph}\equiv \omega_{\mu}(\bq=0)$. It is then convenient to describe the
phonon modes in terms of circularly polarized modes, $u_{\circlearrowleft}(\br) = [u_x(\br) + iu_y(\br)]/\sqrt{2}$ and
$u_{\circlearrowright}(\br)=u_{\circlearrowleft}^*(\br)$.

Finally, taking into account the above considerations on the electron-phonon coupling,
the coupling Hamiltonian reads \cite{ando06ElP,ACN07,andoElPCN}
\beq\label{eq:HamCoupl}
H_{coupl}=g\sqrt{\frac{2M\omega_{ph}}{\hbar}} \sum_{\xi}\int d^2r\,\psi_{\xi}^{\dagger}(\br) 
\left[\sigmab\wedge \bu(\br)\right] \psi_{\xi}(\br),
\eeq
where $\sigmab\wedge \bu(\br)=[\sigmab\times \bu(\br)]_z=\sigma^x u_y(\br) - \sigma^y u_x(\br) $ is the 2D cross product 
between the Pauli matrices and the displacement field.

\subsubsection{Hamiltonian in terms of magneto-exciton operators}

As a consequence of the off-diagonal character of the electron-phonon coupling (\ref{eq:HamCoupl}), one notices that 
the intervening matrix elements are proportional to $\delta_{n,n\pm1}$, and one thus obtains the selection rules
\beq
\lambda n \rightarrow \lambda' (n\pm 1),
\eeq
in analogy with the magneto-optical selection rules discussed in Sec. \ref{Sec:QMtreat}. Furthermore, if we fix the 
energy of the dipole transition (\ref{eq:dipoleT}) to be\footnote{We consider mainly interband transitions here, which may
have a chance to be in resonance with the phonon of energy $\hbar\omega_{ph}\sim 0.2$ eV.} 
$\Delta_n\equiv \Delta_{n,\lambda=-}=\sqrt{2}\hbar(v_F/l_B)(\sqrt{n+1}+\sqrt{n})$, there are two possible transitions, which may be
distinguished by the circular polarization of the phonon they are coupled to. Indeed, the form of the 
electron-phonon coupling (\ref{eq:HamCoupl}) indicates that the $\circlearrowleft$-polarized phonon is coupled to
the transition $- (n+1)\rightarrow + n$, whereas the $\circlearrowright$-polarized phonon couples to the 
$- n\rightarrow +(n+1)$ interband transition \cite{GFKF}.

It is then convenient to introduce \textit{magneto-exciton} operators, associated with the above-mentioned inter-LL
transitions
\begin{eqnarray}
\phi _{\circlearrowleft }^{\dagger }(n,\xi ) &=&\frac{i\sqrt{1+\delta _{n,0}}%
}{\mathcal{N}_{n}^{\circlearrowleft }}\sum_{m}c_{+n,m;\xi }^{\dagger
}c_{-(n+1),m;\xi },~~~~  \notag  \label{eq:MEx} \\
\phi _{\circlearrowright }^{\dagger }(n,\xi ) &=&\frac{i\sqrt{1+\delta _{n,0}%
}}{\mathcal{N}_{n}^{\circlearrowright }}\sum_{m}c_{+(n+1),m;\xi }^{\dagger
}c_{-n,m;\xi },~~~~
\end{eqnarray}%
where the index $\mathcal{A}=\circlearrowleft ,\circlearrowright $
characterizes the angular momentum of the excitation and where the 
normalization factors 
\beqn
\nn
\mathcal{N}_{n}^{\circlearrowleft } &=& \sqrt{2(1+\delta _{n,0})N_{B}\left[\bar{\nu%
}_{-(n+1)}-\bar{\nu}_{+n}\right]}\\
\nn
{\rm and} \qquad \mathcal{N}_{n}^{%
\circlearrowright } &=& \sqrt{2(1+\delta _{n,0})N_{B}\left[\bar{\nu}_{-n}-\bar{\nu}%
_{+(n+1)}\right]}
\eeqn 
are used to ensure the bosonic commutation relations of
the exciton operators, $[\phi _{\mathcal{A}}(n,\xi ),\phi _{\mathcal{A}%
^{\prime }}^{\dagger }(n^{\prime },\xi ^{\prime })]=\delta _{\mathcal{A},%
\mathcal{A}^{\prime }}\delta _{\xi ,\xi ^{\prime }}\delta _{n,n^{\prime }}$, including
the two-fold spin degeneracy. These commutation relations are obtained
within the mean-field approximation with $\langle c_{\lambda n,m;\xi
}^{\dagger }c_{\lambda^{\prime } n^{\prime },m^{\prime };\xi ^{\prime
}}\rangle =\delta _{\xi ,\xi ^{\prime }}\delta _{\lambda ,\lambda ^{\prime
}}\delta _{n,n^{\prime }}\delta _{m,m^{\prime }}(\delta _{\lambda ,-}+\delta
_{\lambda ,+}\bar{\nu}_{\lambda n})$, where $0\leq\bar{\nu}_{\lambda n}\leq 1
$ is the partial filling factor of the $n$-th LL, normalized to 1.

One notices that the magneto-exciton operators are, apart from a normalization constant, nothing
other than the reduced density operators (\ref{eq:RedDens}), 
$\phi _{\circlearrowleft }^{\dagger }(n,\xi )\propto \rhobar_{+ n,-(n+1)}^{\xi,\xi}(\bq=0)$ 
and $\phi _{\circlearrowright }^{\dagger }(n,\xi )\propto \rhobar_{+ (n+1),- n}^{\xi,\xi}(\bq=0)$,
respectively, at zero wave vector. Notice furthermore that, because of the relative sign $\xi$ between
the kinetic part (\ref{eq:HamElPart}) and the electron-phonon coupling Hamiltonian (\ref{eq:HamCoupl}),
the optical phonons couple to the \textit{valley--anti-symmetric} magneto-exciton combination $\phi _{%
\mathcal{A},as}(n)=[\phi _{\mathcal{A}}(n,\xi=+)-\phi _{\mathcal{A}%
}(n,\xi=-)]/\sqrt{2}$. This needs to be contrasted to the magneto-optical coupling \cite{sadowski,iyengar,abergel},
where the photon couples to the \textit{valley-symmetric} mode $\phi _{%
\mathcal{A},s}(n)=[\phi _{\mathcal{A}}(n,\xi=+)+\phi _{\mathcal{A}%
}(n,\xi=-)]/\sqrt{2}$.

The magneto-exciton operators (\ref{eq:MEx}) allow one to rewrite the electron-phonon Hamiltonian 
at the $\Gamma$ point ($\bq=0$) in a bosonic form as \cite{GFKF}
\beqn\label{eq:HamELPBos}
\nn
H &=&\sum_{\tau =s,as}\sum_{\mathcal{A},n}\Delta _{n}\phi _{\mathcal{A},\tau
}^{\dagger }(n)\phi _{\mathcal{A},\tau }(n)+\sum_{\mathcal{A}}\hbar\omega_{ph} b_{%
\mathcal{A}}^{\dagger }b_{\mathcal{A}}  \\
&&~ +\sum_{\mathcal{A},n}g_{\mathcal{A}}(n)\left[ b_{%
\mathcal{A}}^{\dagger }\phi _{\mathcal{A};as}(n)+b_{\mathcal{A}}\phi _{%
\mathcal{A};as}^{\dagger }(n)\right],
\eeqn
in terms of the \textit{effective coupling constants} 
\beqn\label{eq:EffCouplC}
\nn
g_{\circlearrowleft }(n) &=& g\sqrt{(1+\delta _{n,0})\gamma }\sqrt{%
\bar{\nu}_{-(n+1)}-\bar{\nu}_{+n}},\qquad \\
{\rm and} ~~ g_{\circlearrowright }(n) &=& g\sqrt{(1+\delta _{n,0})\gamma }\sqrt{%
\bar{\nu}_{-n}-\bar{\nu}_{+(n+1)}},\qquad
\eeqn
with the constant $\gamma\equiv 3\sqrt{3}a^{2}/2\pi l _{B}^{2}$. One therefore remarks that, although
the bare coupling constant $g$ is rather large [see Eq. (\ref{eq:bareElP})], the effective coupling is reduced
by a factor of $a/l_B$, 
\beq\label{eq:EffElP}
g_{\Amath}(n)\sim g\frac{a}{l_B}\sim 1 \hdots 2\, {\rm meV}\sqrt{B{\rm [T]}}.
\eeq

\subsection{Phonon Renormalization and Raman Spectroscopy}
\label{sec:PhonRen}

The Hamiltonian (\ref{eq:HamELPBos}) shows that a phonon may be destroyed by exciting
a magneto-exciton, and the associated Dyson equation 
for the dressed phonon propagator $D(\omega)$ reads
\beq\label{eq:Dyson}
D_{\Amath}(\omega) = D_0(\omega) + D_0(\omega) \chi_{\Amath}(\omega) D_{\Amath}(\omega),
\eeq
in terms of the bare bosonic phonon propagator
\beq\label{eq:BarePhonProp}
D_0(\omega) = \frac{1}{\hbar}\frac{2\omega}{\omega^2 - \omega_{ph}^2}
\eeq
and 
\beq\label{eq:MEProp}
\chi_{\Amath}(\omega)=\sum_{n=N_F+1}^{N_c}\frac{2\Delta_n g_{\Amath}^2(n)}{\hbar^2\omega^2 - \Delta_n^2} + 
\frac{2\tilde{\Delta}_{N_F} g_{\Amath}^2(n)}{\hbar^2\omega^2 - \tilde{\Delta}_{N_F}^2}.
\eeq
The form of the last expression is transparent; the magneto-exciton is a boson, and its propagator is therefore of the same 
form as that of the bare phonon. It is equivalent to a particle-hole propagation associated with a polarization 
bubble [see Fig. \ref{fig:PHESB0RPA}(a)], but the expression (\ref{eq:MEProp})
also takes into account the square of the effective coupling constant which is due to the double occurence of the electron-phonon
coupling -- first when the phonon is converted into a magneto-exciton and the second time when the magneto-exciton is destroyed by
creating a phonon. The last term in Eq. (\ref{eq:MEProp}) takes into account the (only) possible intra-band magneto-exciton
from the last filled LL $N_F$ to $N_F+1$ with energy $\tilde{\Delta}_{N_F}=\sqrt{2}(\hbar v_F/l_B)(\sqrt{N_F+1} - \sqrt{N_F})$,
which we have omitted in the Hamiltonian (\ref{eq:HamELPBos}) because it is irrelevant for resonant coupling. The parameter
$N_c$ is the same high-energy cutoff, defined by
$\epsilon_{N_c}=\hbar(v_F/l_B)\sqrt{2N_c}\sim t$, as in Sec. \ref{Sec:PHES} of the preceding chapter. 

The renormalized phonon frequencies $\tilde{\omega}_{\Amath}$ may be obtained from the Dyson equation (\ref{eq:Dyson}), by
searching the poles of the dressed phonon propagator
\beq
D_{\Amath}(\omega)^{-1} = 0 = D_{\Amath}(\otilde_{\Amath})^{-1} - \chi_{\Amath}(\otilde_{\Amath}),
\eeq
and one finds \cite{ando07ElP,GFKF}
\begin{equation}
\tilde{\omega}_{\mathcal{A}}^{2}-\omega_{ph}^{2}=\frac{4\omega_{ph}}{\hbar} \left[
\sum_{n=N_{F}+1}^{N_c}\frac{\Delta_{n}g_{\mathcal{A}}^{2}(n)}{\hbar^2\tilde{\omega}_{%
\mathcal{A}}^{2}-\Delta_{n}^{2}}+ \frac{\tilde{\Delta}_{N_F}g_{\mathcal{A}}^2(N_F)}{
\hbar^2\tilde{\omega}_{\mathcal{A}}^2-\tilde{\Delta}_{N_F}^2} \right].  \label{eq:PhononFreq}
\end{equation}%

\subsubsection{Non-resonant coupling and Kohn anomaly}
\label{sec:NonResCoupl}

Before discussing resonant coupling, i.e. when the phonon frequency is in resonance with a possible inter-LL excitation,
in a strong magnetic field, we comment on the relation between Eq. (\ref{eq:PhononFreq}) and the 
(non-resonant) renormalization of the phonon frequency in zero magnetic field. The zero-field limit may indeed be 
obtained from Eq. (\ref{eq:PhononFreq}) if one replaces the sum $\sum_n$ by an integral $\int dn$, i.e. if the spacing
between the LLs vanishes, $\tilde{\Delta}_{N_F}\rightarrow 0$. Linearizing  Eq. (\ref{eq:PhononFreq}), if one replaces
$\otilde_{\Amath}\rightarrow \omega_{ph}$ in the denominators, and using the approximation $\sqrt{n}+\sqrt{n+1}\approx 2\sqrt{n}$ 
yields 
\beq\label{eq:PhononFreq0B}
\tilde{\omega} \simeq \tilde{\omega}_{0}+\ltilde \left[ \sqrt{2N_{F}}\frac{
v_F}{l_{B}}-\frac{\omega_{ph} }{4}\ln \left( \frac{\omega_{ph} +2\sqrt{2N_{F}}%
v_F/l_{B}}{\omega_{ph} -2\sqrt{2N_{F}}v_F/l_{B}}\right) \right], 
\eeq
where $\ltilde =(2/\sqrt{3}\pi )(g/t)^{2}\simeq 3.3\times 10^{-3}$ is
the dimensionless electron-phonon coupling constant introduced in Refs. \cite{ando06ElP,ando07ElP}, and
\beq
\tilde{\omega}_{0} \simeq \omega_{ph} +\frac{2}{\hbar}\int_{0}^{N_c}dn\frac{\Delta_{n}g_{%
\mathcal{A}}^{2}(n)}{\hbar^2\omega_{ph} ^{2}-\Delta_{n}^{2}}
\eeq
is the physical phonon frequency at zero doping. Indeed, the frequency $\omega_{ph}$ is not relevant in a physical 
measurement in graphene even if it occurs in the Hamiltonian, but one measures $\tilde{\omega}_0$ at zero doping and 
$B=0$. Equation (\ref{eq:PhononFreq0B}) coincides precisely with the zero-field result \cite{ando06ElP,Lazz06,ACN07}
if one identifies the chemical potential with the energy of the last filled LL,
$\mu=\sqrt{2 N_F}\hbar v_F/l_B$ \cite{GFKF}. 

\subsubsection{Resonant coupling}
\label{sec:ResCoupl}

\begin{figure}
\centering
\includegraphics[width=5.5cm,angle=0]{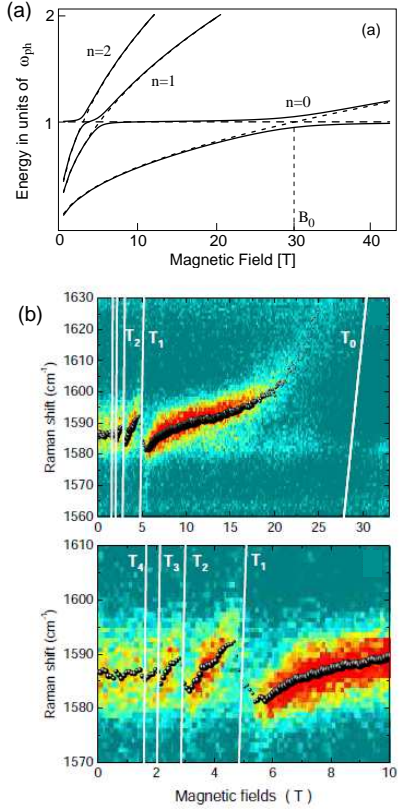}
\caption{\footnotesize{\textit{(a)} From Ref. \cite{GFKF}. 
Anticrossing of coupled phonon-magneto-exciton modes as a function of the magnetic field.
\textit{(b)} From Ref. \cite{faugeras}. Colour plot of the Raman spectra as a function of the magnetic field. The continuous white
lines indicate the magnetic field for which the phonon is in resonance with an inter-LL excitation of energy $\Delta_n$. \textit{Top:} 
data of the full $B$-field range. \textit{Bottom:} zoom on the range from $0$ to $10$ T.
}}
\label{fig:AntiCross}
\end{figure}

Apart from the non-resonant coupling discussed in the preceding section, the high-field electron-phonon coupling reveals
a linear effect when the phonon is in resonance with a particular magneto-exciton, $\hbar\omega_{ph}\simeq\Delta_n$. 
In this case, the sum 
on the right-hand side in Eq. (\ref{eq:PhononFreq}) is dominated by a single term and may be approximated by
$2(\omega_{ph}/\hbar)g_{\Amath}^2(n)/(\hbar\otilde_{\Amath}-\Delta_n)$. 
This results in a fine structure of mixed phonon--magneto-exciton modes, $%
\phi _{\mathcal{A},as}(n)\cos \theta +b_{\mathcal{A}}\sin \theta $ with
frequency $\tilde{\omega}_{\mathcal{A}}^{+}$ and $\phi _{\mathcal{A}%
,as}(n)\sin \theta -b_{\mathcal{A}}\cos \theta $ with frequency $\tilde{%
\omega}_{\mathcal{A}}^{-}$ [where $\cot 2\theta =(\Delta _{n}-\hbar\tilde{\omega}%
_{0})/2g_{\mathcal{A}}$]. The frequencies of these mixed boson modes read \cite{ando07ElP,GFKF}
\begin{equation}
\tilde{\omega}_{\mathcal{A}}^{\pm }(n)=\frac{1}{2}\left( \frac{\Delta _{n}}{\hbar}+\tilde{%
\omega}_0\right) \mp \sqrt{\frac{1}{4}\left(\frac{\Delta _{n}}{\hbar}-\tilde{\omega}_0\right)^{2}+g_{%
\mathcal{A}}^{2}(n)},  \label{eq:ResFreq}
\end{equation}
and the resulting phonon--magneto-exciton anticrossing is depicted in Fig. \ref{fig:AntiCross}(a).

The above-mentioned anticrossing of the coupled phonon--magneto-exciton modes has been observed in 
recent Raman experiments on epitaxial graphene. Remember that Raman spectroscopy is sensitive 
to the phonon component of the mixed modes \cite{faugeras}.
The results are shown in Fig. \ref{fig:AntiCross}(b) and corroborate the theoretically expected behavior
\cite{ando07ElP,GFKF}. Indeed, one may obtain the oscillating behavior from a numerical solution of Eq. (\ref{eq:PhononFreq})
if one expresses the equation in terms of $\tilde{\omega}_0$ instead of $\omega_{ph}$ and if one takes into account a
finite broadening of the levels.
If the phonon is out of resonance with an inter-LL transition, its frequency is
essentially field-independent and coincides with the energy of the 
E$_{2g}$ line at $1586$ cm$^{-1}\,\simeq\, 0.2$ eV. When it approaches the resonance (by increasing the magnetic
field), its energy is shifted upwards as a consequence of the anticrossing but rapidly dies out in intensity once the
magneto-exciton component becomes dominant in the $\otilde_{\Amath}^+$ mode. Upon further increase of the magnetic
field, the $\otilde_{\Amath}^-$ mode becomes more phonon-like and therefore visible in the Raman spectra.

The fine structure of the high-field resonant electron-phonon coupling may furthermore be investigated by sweeping
the chemical potential when the magnetic field is held fixed at resonance. The effect is most pronounced for the 
resonance $\hbar\omega_{ph}\simeq \Delta_{n=0}$, which is expected at $B\simeq 30$ T [see Fig. \ref{fig:AntiCross}(a)]. 
In this case, the mode 
consists of an equal-weight superposition of the phonon and the magneto-exciton ($\cos\theta = \sin\theta=1/\sqrt{2}$),
and the E$_{2g}$ band would appear as two lines, 
at the energies $\hbar\otilde^{\pm}=\hbar \otilde_0 \pm g_{\Amath}$, for the case of undoped graphene.\footnote{Notice, however,
that only an oscillation of the phonon mode, and not a splitting, has been observed in the experiment by Faugeras \textit{et al.}
\cite{faugeras}.} 
With the above estimation
(\ref{eq:bareElP}) for the bare electron-phonon coupling constant, one obtains for the line splitting 
$2g_{\mathcal{A}}\sim 16$ meV ($\sim 130$ cm$^{-1}$), which
largely exceeds the G-band width observed in Refs. \cite{pisana,yan07,ferrari06,gupta,graf}.

It is apparent from the expressions (\ref{eq:EffElP}) for the effective coupling constants $g_{\circlearrowleft}$ and
$g_{\circlearrowright}$ that the splitting may be controled by the LL filling factor. 
Exactly at zero doping, both coupling constants coincide, $g_{\circlearrowleft}=g_{\circlearrowright}$, but upon electron 
doping the transition $-1\rightarrow 0$ associated with the $\circlearrowleft$-polarization 
becomes weaker due to the reduced number of final states in $n=0$, whereas the $0\rightarrow +1$ transition (with
polarization $\circlearrowright$) is strengthened. As a consequence, the associated coupling constants are increased and
decreased, respectively, until the coupling constant $g_{\circlearrowleft}$ vanishes at
$\nu=2$. 

The above-mentioned filling-factor dependence has a direct impact on the Raman lines \cite{GFKF}. Whereas at $\nu=0$, one 
expects two lines separated by the energy $2g_{\circlearrowleft}=2g_{\circlearrowright}$, the degeneracy in the circular
polarization is lifted between $0<\nu<2$.\footnote{We present the argument for a Fermi energy in the CB, i.e. $\nu>0$, but
the situation is generic, and the argument also applies in the VB if one interchanges the polarizations.}
One therefore expects to observe four lines instead of two, where the inner ones are associated with the polarization
$\circlearrowleft$, whereas the outer ones with increased splitting correspond to the opposite polarization
$\circlearrowright$. The separation between the inner lines vanishes then at $\nu=2$,
where the splitting of the outer lines is maximal and where one expects to observe three lines.



\section{Electronic Correlations in Partially Filled Landau Levels}
\label{Chap:FQHE}

This last section is devoted to the physics of interacting electrons in the strong-correlation limit of a partially filled LL.
The motivation stems from non-relativistic quantum Hall systems in GaAs heterostructures, where these correlations lead to
the formation of incompressible quantum-liquid phases, which display the fractional quantum Hall effect (FQHE) 
\cite{TSG}, as well as of exotic electron-solid phases, such as the high-field Wigner crystal \cite{andrei88,williams91} or the 
theoretically predicted bubble and stripe phases \cite{KFS96,FKS96,MC96}. The latter are likely to be at the origin of highly anisotropic
transport properties at half-filled higher LLs \cite{lilly99,du99}, particular electron transport under microwave irradiation 
\cite{lewis02,lewis04,lewis05}, and an intriguing reentrance of the IQHE in $n=1$ and $n=2$ \cite{cooper99,eisenstein02}.

It is therefore natural to ask whether such strongly-correlated phases exist also in graphene and if so what
the differences are with respect to non-relativistic 2D electrons. Moreover, the fact that the electrons reside at the surface
opens up the possibility to probe these phases by spectroscopic means, such as scanning tunneling spectroscopy, which has already been
applied successfully in the analysis of the electron density distribution of exfoliated \cite{martin08} and epitaxial 
graphene \cite{mallet07}, as well as graphene on graphite substrates \cite{li09}.

After a short discussion of the Coulomb interaction in graphene as compared to non-relativistic 2D electrons, we
introduce the basic model of interacting electrons in a partially
filled relativistic LL (Sec. \ref{sec:Model}). 
This model yields a qualitative understanding of the above-mentioned correlated electronic
phases in the context of graphene, as compared to non-relativistic electrons. In Sec. \ref{sec:QHFM}, we apply this model 
to the quantum Hall ferromagnetism with an internal SU(4) symmetry that is the relevant symmetry in graphene LLs and discuss
its relation with the experimentally observed degeneracy lifting of the zero-energy LL $n=0$ \cite{zhang}. We terminate this
section with a review of the specific FQHE in graphene (Sec. \ref{sec:FQHE}), which has recently been observed in the two-terminal
\cite{grapheneFQHE1,grapheneFQHE2} as well as in the four-terminal geometry \cite{deanFQHE10,ghahariFQHE10}.

\subsection{Electrons in a Single Relativistic Landau Level}
\label{sec:Model}

Quite generally, the origin of strongly-correlated electron phases 
is a quenched kinetic energy, where the partially filled LL is separated by the cyclotron
gap from the neighboring ones such that inter-LL excitations constitute high-energy degrees of freedom. The Coulomb interaction, which
may though be small with respect to the cyclotron gap, remains then as the only relevant energy scale which dominates the low-energy
degrees of freedom if we can neglect disorder effects. This leads to the seemingly counter-intuitive finding of strongly-correlated
phases in weakly-correlated matter.

In order to quantify the degree of separation between the energy scales, one may use a similar argument as the one that led us
to the definition of the dimensionless interaction parameter (\ref{rs}), introduced at the beginning of Sec. \ref{Chap:Ints}. 
Indeed, one needs to compare the Coulomb interaction energy $E_{int}=e^2/\eps R_C$ at the characteristic length 
scale 
$R_C=l_B\sqrt{2n+1}$ to the LL spacing $\hbar \omega_C=\hbar eB/m_b$, where we concentrate on non-relativistic electrons first,
\beq\label{rsBnr}
r_s^B=\frac{e^2}{\hbar\eps v_F(n,B)}, \qquad {\rm with} \qquad v_F(n,B)\equiv R_C\omega_C.
\eeq
If one identifies the Fermi wave vector $k_F\simeq \sqrt{2n}/l_B$, one obtains the same expression as for the zero-field
coupling constant (\ref{rs:nonrel}),
\beq
r_s^B=r_s=\frac{m_be^2}{\hbar^2\eps}k_F^{-1}\sim \frac{1}{a_0^*k_F}.
\eeq
This means that the degree of LL mixing is still governed by $r_s$, and the inter-LL excitations are well separated from the low-energy
intra-LL degrees of freedom unless $r_s$ becomes very large. Notice, however, that $r_s\sim 1$ in most 2D electron systems.

\begin{figure}
\centering
\includegraphics[width=8.5cm,angle=0]{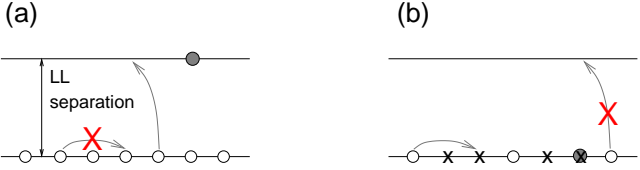}
\caption{\footnotesize{\textit{(a)} Completely filled topmost LL. Due to the Pauli principle,
the only possible excitations are inter-LL transitions. \textit{(b)} Partially filled LL. For sufficiently 
small values of $r_s$ (or $\alpha_G$), 
the inter-LL excitations constitute high-energy degrees of freedom that may be omitted at low energies, 
where the relevant degrees couple states within the same LL.
}}
\label{fig:LLcorrel}
\end{figure}

In the case of partially filled relativistic LLs in graphene, one is tempted to apply the same argument 
-- if the Coulomb interaction $e^2/\eps R_C$ is sufficiently small
as compared to the LL spacing $\tilde{\Delta}_n$, the relevant degrees of freedom are those which couple quantum states 
in the same LL, whereas inter-LL excitations may be considered as frozen out (see Fig. \ref{fig:LLcorrel}). Although this
seems a reasonable assumption for the lowest LLs, one is confronted with the apparent problem that the LL spacing 
rapidly decreases once the Fermi level resides in higher LLs, 
\beq\label{eq:cyclGap}
\tilde{\Delta}_n=\sqrt{2}\frac{\hbar v_F}{l_B}(\sqrt{n+1} - \sqrt{n})\simeq \frac{\hbar v_F}{l_B\sqrt{2n}}.
\eeq 
Notice, however, that this decrease is balanced by the $1/\sqrt{2n}$ scaling of the characteristic Coulomb interaction, such
that even in higher LLs the separation between low- and high-energy degrees of freedom is governed by the dimensionless 
coupling constant
\beq\label{rsBrel}
\alpha_G^B=\frac{e^2/\eps l_B\sqrt{2n}}{\hbar v_F/l_B\sqrt{2n}} = \frac{e^2}{\hbar \eps v_F}=\alpha_G,
\eeq
which coincides with the scale-invariant zero-field coupling constant (\ref{rs:graph}). From the interaction point of view,
the restriction of the electron dynamics to a single partially filled LL in the large-$n$ limit is therefore as justified
as for the lowest relativistic LLs. Naturally, this statement only holds true in the absence of disorder that induces stronger
LL mixing for $n\gg 1$ than in $n=0$ or $\pm 1$. 

\subsubsection{SU(4)-symmetric model}

Formally, the above-mentioned separation into high- and low-energy degrees of freedom may be realized with the help of the 
reduced density operators (\ref{eq:RedDens}). For the moment, we only consider the case where $\xi=\xi'$, i.e. we concentrate
on the valley-symmetric model, in which case the reduced (intra-valley) density operators (\ref{eq:RedDens2}) fall into
two distinct classes: for $n\neq n'$ or $\lambda\neq \lambda'$, the operators $\rhobar_{\lambda n,\lambda'n'}(\bq)$ 
describe density fluctuations corresponding to inter-LL transitions of an energy equal to or larger than the LL separation
$\tilde{\Delta}_n$, whereas the \textit{projected density operators}
\beqn\label{eq:ProjDens}
\rhobar(\bq) &\equiv& \rhobar_{\lambda n,\lambda n}(\bq) \\
\nn
&=& \sum_{\xi=\pm}\sum_{\sigma=\ua,\da}\sum_{m,m'}\left\langle m\left| e^{-i\bq\cdot\bR} \right| 
m'\right\rangle c_{\lambda n,m;\xi,\sigma}^{\dagger}c_{\lambda n,m';\xi,\sigma}
\eeqn
describe the density fluctuations inside the LL $\lambda n$ that interest us here. Notice that we have dropped the 
index $\lambda n$ in the definition of the projected density operators; they satisfy the quantum-mechanical 
commutation relations \cite{GMP}
\beq\label{eq:GMP}
\left[\rhobar(\bq),\rhobar(\bq')\right] = 2i\sin\left(\frac{\bq'\wedge\bq\, l_B^2}{2}\right)\rhobar(\bq+\bq'),
\eeq
where $\bq'\wedge\bq=(\bq'\times\bq)_z=q_x'q_y - q_y'q_x$ is the 2D vector product between $\bq'$ and $\bq$, and
these commutation relations are \textit{independent} of the LL index $\lambda n$.
The information about the LL is indeed waved to the effective interaction potential
\beq\label{eq:LLpot}
v_n(\bq)=\frac{2\pi e^2}{\eps q}\left[\Fmath_n(\bq)\right]^2,
\eeq
in terms of the LL form factors [see Eq. (\ref{eq:form2}) and their explicit form (\ref{eqnA02}), discussed in Appendix \ref{app:ME}]
\beqn\label{eq:LLform}
\nn
\Fmath_n(\bq) &=& \frac{1}{2}\left[(1-\delta_{n,0}) L_{n-1}\left(\frac{q^2l_B^2}{2}\right) \right.\\
&&\left.+ (1+\delta_{n,0})L_{n}\left(\frac{q^2l_B^2}{2}\right)\right]
e^{-q^2l_B^2/4},
\eeqn
independent of the band index $\lambda$ \cite{GMD,nomura}. The Hamiltonian resulting from Eq. (\ref{eq:IntHamSU4})
\beq\label{eq:LLHamSU4}
H_n=\frac{1}{2}\sum_{\bq} v_n(\bq) \rhobar(-\bq)\rhobar(\bq)
\eeq
therefore defines, together with the commutation relation (\ref{eq:GMP}) the model of strongly-correlated electrons restricted to a 
single relativistic LL. The model respects the SU(4) spin-valley symmetry, and naturally, there is no kinetic energy scale because
all processes involve states within the same LL.

\paragraph{Algebraic properties.}

The SU(4) spin-valley symmetry is formally described with the help of the spin and valley-pseudospin operators 
\beq\label{eq:SIdens}
\Sbar^{\mu}(\bq) = \left(S^{\mu}\otimes\bone\right)\otimes \rhobar(\bq) ~~ {\rm and} ~~
\Ibar^{\mu}(\bq) = \left(\bone\otimes I^{\mu}\right)\otimes \rhobar(\bq),
\eeq
respectively, that are tensor products between the projected density operators (\ref{eq:ProjDens}) and the operators
$S^{\mu}$ and $I^{\mu}$, which are (up to a factor $1/2$) Pauli matrices and that describe the spin and valley degrees of
freedom, respectively.  
The operators $(S^{\mu}\otimes\bone)$ and $(\bone\otimes I^{\mu})$
are the generators of the SU(2)$\times$SU(2) subgroup of SU(4). 
However, once combined in a tensor product with the projected density operators 
$\rhobar(\bq)$, the SU(2)$\times$SU(2)-extended magnetic translation 
group is no longer closed due to the non-commutativity of the 
Fourier components of the projected density operators. The commutators $[\Sbar^{\mu}(\bq),\Ibar^{\nu}(\bq')]$
yield the remaining generators of the SU(4)-extended magnetic translation group \cite{hasebe02,tsitsish03,DGLM}.

Physically, the operators introduced in Eq. (\ref{eq:SIdens}) play the role of projected spin and valley-pseudospin 
densities, where the LL projection is induced by the projected charge-density operator $\rhobar(\bq)$. Their 
non-commutativity with the projected charge density, $[\Sbar^{\mu}(\bq),\rhobar(\bq')]\neq 0$ and
$[\Ibar^{\mu}(\bq),\rhobar(\bq')]\neq 0$, which are due to the commutation relation (\ref{eq:GMP}),
is at the origin of the (pseudo-)spin-charge entanglement in quantum
Hall systems: as we discuss in more detail in Sec. \ref{Sec:Skyrm}, this entanglement yields (pseudo-)spin-texture
states that carry an electric in addition to their topological charge.

\paragraph{Validity of the model.}

With the help of the Hamiltonian (\ref{eq:LLHamSU4}), we may render more transparent the model assumption of electrons
restricted to a single relativistic LL. 
We need to show that the energy scale that governs the model (\ref{eq:LLHamSU4}) and its resulting
phases is indeed given by $e^2/\eps R_C$ and not $e^2/\eps l_B$. As an upper bound for the energy scale, one may use the energy of 
a completely filled LL described by $\langle c_{\lambda n,m;\xi,\sigma}^{\dagger}c_{\lambda n,m';\xi,\sigma}\rangle = \delta_{m,m'}$,
the mean-field energy $\langle H_n\rangle/N$ of which is simply the exchange energy,\footnote{The direct energy is compensated by
the positively charged background (``jellium model'') \cite{mahan}}
\beq\label{eq:ExchEn}
E_X^n=-\frac{1}{2}\sum_{\bq} v_n(\bq)= - \frac{e^2}{2\eps}\int_0^{\infty}dq \left[\Fmath_n(q)\right]^2.
\eeq
In order to estimate the integral in the large-$n$ limit, one may use the scaling form \cite{gradsht,abramowitz}
of the Laguerre polynomials 
\beq\label{eq:scalingForm}
L_n\left(\frac{q^2 l_B^2}{2}\right)e^{-q^2l_B^2/4}\simeq J_0(ql_B\sqrt{2n+1}),
\eeq 
in terms of the Bessel function $J_0(x)$, such that 
one obtains by a simple change of the integration variable 
$\int_0^{\infty}dq[\Fmath_n(q)]^2\simeq (l_B\sqrt{2n})^{-1}\int_0^{\infty}dx[J_0(x)]^2 = c/l_B\sqrt{2n}$, 
where $c$ is a numerical factor of order one. 
The exchange energy of a completely filled LL $n$ therefore scales with $n\gg 1$ as
\beq
E_X^n \simeq -c\frac{e^2}{\eps l_B \sqrt{2n}} \simeq -c\frac{e^2}{\eps R_C},
\eeq
in agreement with the model assumption of a separation between high- and low-energy degrees of freedom and
the definition (\ref{rsBrel}) of the coupling constant $\alpha_G^B$.

\subsubsection{Symmetry-breaking long-range terms}
\label{sec:sbTerms}

When decomposing the Coulomb interaction in the two-spinor basis (Sec. \ref{Sec:CoulB}), we have seen that the 
SU(4)-symmetric model yields the leading energy scale, whereas the only relevant symmetry-breaking term is associated with
backscattering processes at an energy scale roughly $a/l_B$ times smaller than the leading one. When restricted to a 
single relativistic LL $\lambda n$, these backscattering terms yield a contribution 
\beq\label{eq:Ham_sb}
H_{n}^{sb} = \frac{1}{2}\sum_{\xi=\pm}\sum_{\bq} v_{n}^{sb}(\bq)\rhobar^{\xi,-\xi}(-\bq)\rhobar^{-\xi,\xi}(\bq),
\eeq
in terms of the effective \textit{backscattering} potential 
\beqn\label{eq:Int_sb}
\nn
v_n^{sb}(\bq) &=& \frac{2\pi e^2}{\eps q} \left|\Fmath_{\lambda n,\lambda n}^{+,-}(\bq)\right|^2\\
&=& \frac{2\pi e^2}{\eps q}\frac{(1-\delta_{0,n})}{2n}(q_y - K_y)^2l_B^2 \\
\nn
&&\times \left[L_{n-1}^1\left(\frac{|\bq-\bK|^2l_B^2}{2}\right)\right]^2
e^{-|\bq -\bK|^2l_B^2/2},
\eeqn
where we have made use of Eq. (\ref{eq:form}) and of the explicit expressions for the intervening matrix elements (\ref{eqnA04}).

The effect of this symmetry-breaking term will be discussed in more detail in Sec. \ref{sec:QHFM} in the context of the 
SU(4) quantum Hall ferromagnetism. The term (\ref{eq:Ham_sb}) is only relevant in relativistic LLs $n\neq 0$ 
as a consequence of the factor $(1-\delta_{n,0})$ in the expression (\ref{eq:Int_sb}) for the backscattering potential \cite{GMD}.
This is a consequence of the chiral symmetry of the zero-energy LL \cite{hatsugai} where the sublattice index is confunded with
the valley pseudospin, as may be seen from the expression (\ref{spinN0}) for the associated wave functions.
Notice, however, that there may occur other symmetry-breaking terms in $n=0$ as a consequence of short-range interactions 
on the lattice scale \cite{alicea,herbut,doretto}.

\subsubsection{Qualitative expectations for correlated electron phases}
\label{Sec:qual}

\begin{figure}
\centering
\includegraphics[width=5.5cm,angle=0]{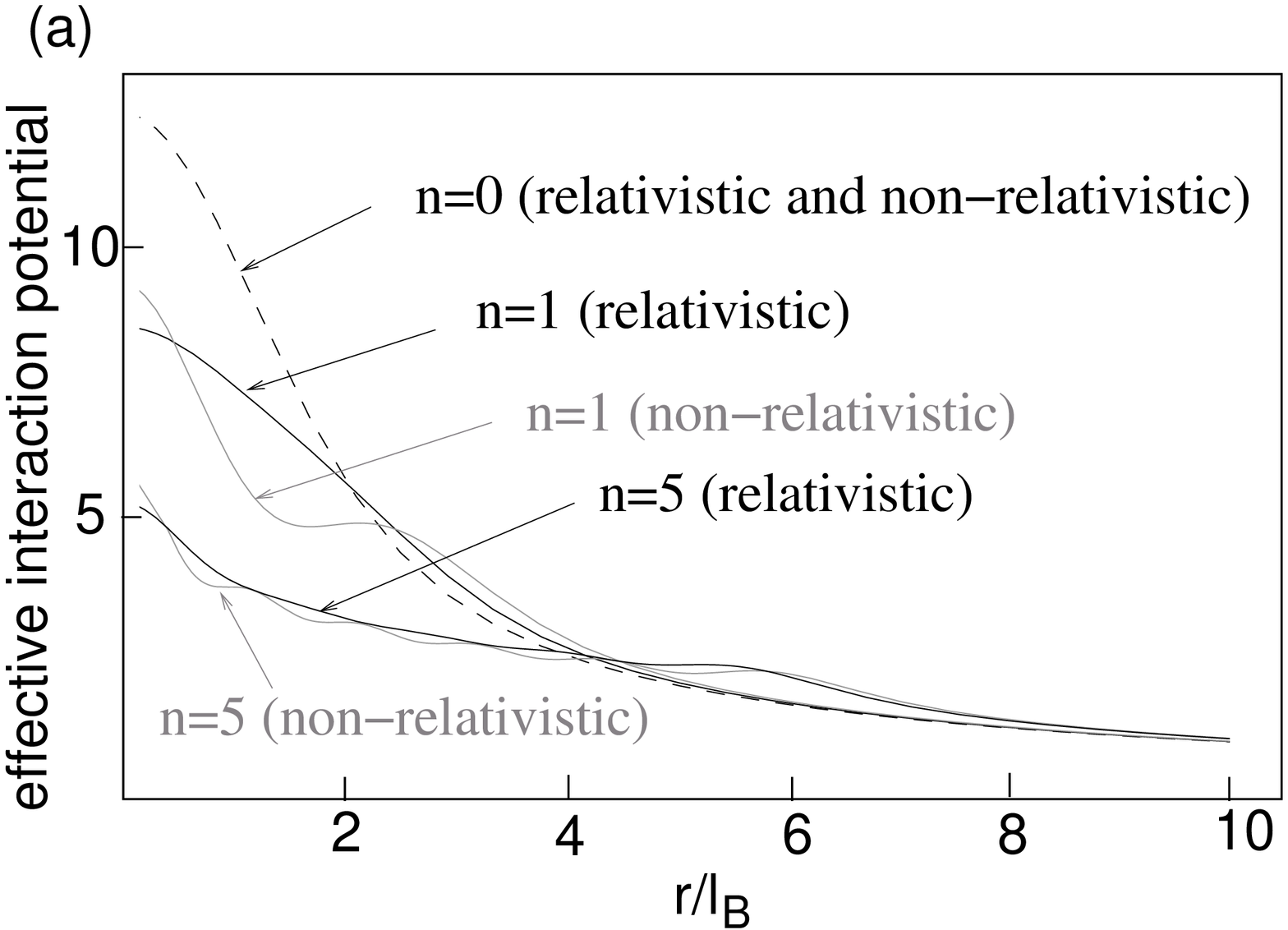}
\includegraphics[width=5.5cm,angle=0]{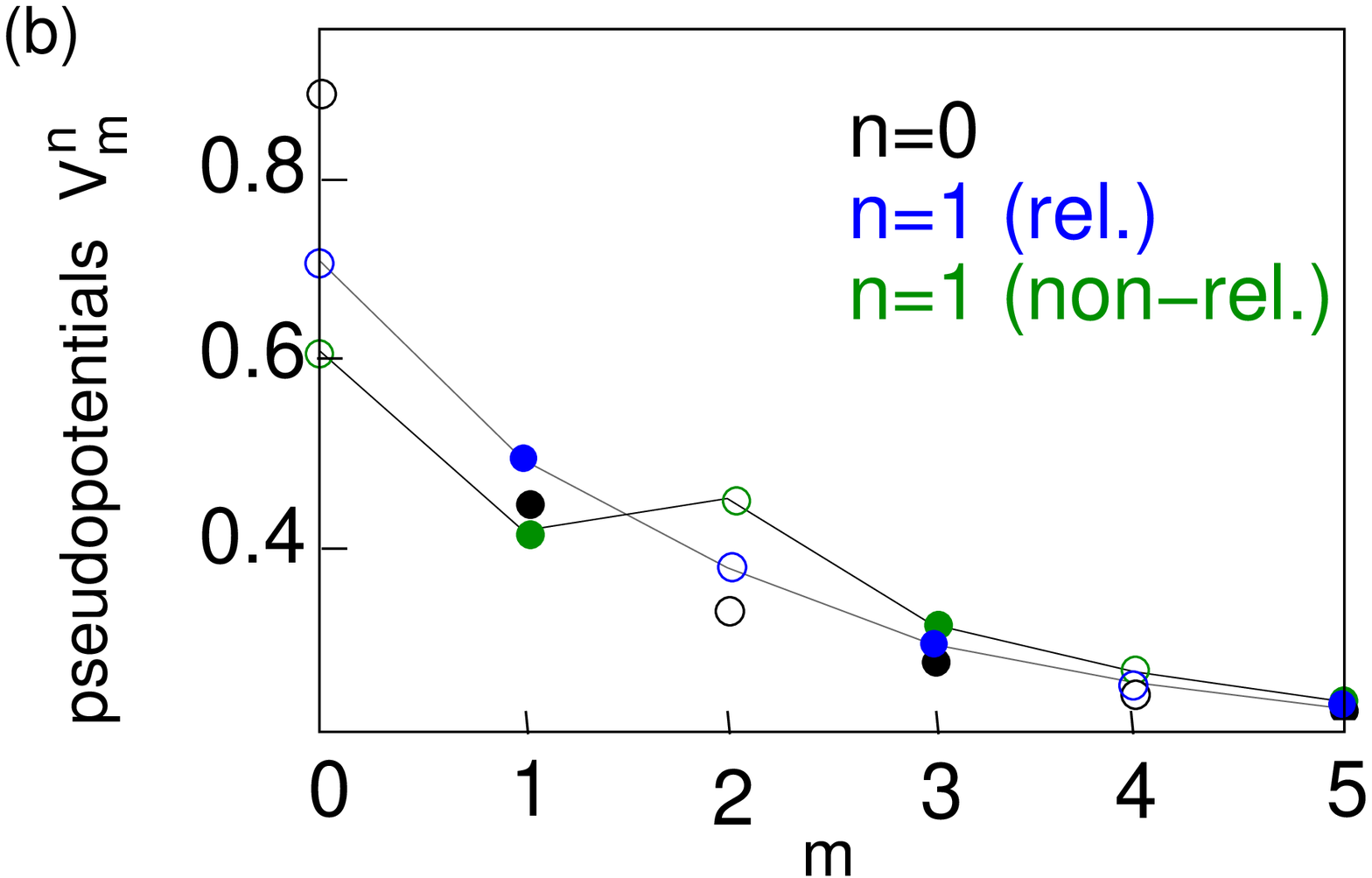}
\caption{\footnotesize{\textit{(a)} From Ref. \cite{GMD}; comparaison between the relativistic (black curves) and non-relativistic 
(grey curves) potentials for the LLs $n=0,1$, and $5$ in real space. 
The dashed line shows the potential in $n=0$, which is the same in both the relativistic and the non-relativistic
case. 
\textit{(b)} Pseudopotentials for
$n=0$ (black circles), $n=1$ relativistic (blue) and $n=1$ non-relativistic
(green). The lines are a guide to the eye.
The open circles represent pseudopotentials with even relative pair angular momentum that are irrelevant
in the case of completely spin-valley polarized electronic states. The energies are given in units of $e^2/\epsilon l_B$.
}}
\label{fig:IntPotGraph}
\end{figure}

The model of interacting electrons in a single relativistic LL has the same structure as the one for non-relativistic LLs -- 
in both cases, one has an interaction Hamiltonian that is quadratic in the projected density operators (\ref{eq:ProjDens}) which
satisfy the commutation relations (\ref{eq:GMP}). This is a noticeable result in the sense that, whereas non-relativistic 
2D electron systems are governed by Galileian invariance, the electrons in graphene are embedded in a Lorentz-invariant ``space-time''.
However, once restricted to a single LL, the electrons forget about their original spatial symmetry properties and are governed by the 
magnetic translation algebra, which is at the origin of the commutation relations (\ref{eq:GMP}). As a consequence
and in contrast to the IQHE, the differences
between strongly-correlated electrons in graphene and non-relativistic 2D electrons do not stem from their respective space-time
properties, as one would expect from a mean-field Chern-Simons approach \cite{peres06,khvesh}.

The differences between graphene and non-relativistic 2D electrons are rather to be seeked in the 
larger internal symmetry -- instead of an SU(2) spin symmetry, one has an SU(4) spin-valley symmetry if one neglects the 
small symmetry-breaking term (\ref{eq:Ham_sb}) in the interactions. Another difference arises 
from the different effective interaction potential
(\ref{eq:LLpot}), instead of 
\beq\label{eq:LLpotNR}
v_n^{\rm non-rel}(\bq)= 
\frac{2\pi e^2}{\eps q}
\left[L_n\left(\frac{q^2l_B^2}{2}\right)\right]^2e^{-q^2l_B^2/2}
\eeq
for the usual 2D electron gas. As one may see from the graphene 
form factors (\ref{eq:LLform}), the effective interaction potential in graphene for $n\neq 0$ is the average 
of the non-relativistic ones in the adjacent LLs $n$ and $n-1$, whereas for $n=0$ there is no difference between the relativistic 
and the non-relativistic case (see Fig. \ref{fig:IntPotGraph}), as a consequence of the above-mentioned chiral properties. 

One notices, furthermore, that the difference between the relativistic and non-relativistic effective interaction potentials 
become less prominent in the large-$n$ limit [see $n=5$ in Fig. \ref{fig:IntPotGraph}(a)]. This may be understood from
the approximate expression (\ref{eq:scalingForm}) of the form factors, which yields 
$\Fmath_n(\bq)\simeq [J_0(ql_B\sqrt{2n+1}) + J_0(ql_B\sqrt{2n-1})]/2\simeq J(ql_B\sqrt{2n}) + \Omath(1/n)$. 
This result agrees indeed to leading order in $1/n$ with the scaling expression of the form factors (\ref{eq:form2DEG}) for 
usual non-relativistic 2D electrons. 

The strongest difference in the interaction potentials is thus found for $n=1$, which in graphene is quite reminiscent to that 
in $n=0$, apart from a reduced repulsion at very short distances, whereas for non-relativistic 2D electrons it has an additional
structure [see Fig. \ref{fig:IntPotGraph}(a)]. The behavior of the effective interaction potential may also be analyzed 
with the help of Haldane's pseudopotentials \cite{haldane}
\beq\label{eq:HPP}
V_{\ell}^n = \frac{1}{2\pi}\sum_{\bq}v_n(\bq)L_{\ell}(q^2l_B^2)e^{-q^2l_B^2/2},
\eeq
which represent the interaction between pairs of electrons in a magnetic field, in a relative angular momentum state with
quantum number $\ell$. This quantum number is related to the average distance $l_B\sqrt{2\ell+1}$ between the two particles 
constituting the pair and is a good quantum number for any two-particle interaction potential $v(\br_i-\br_j)$. The 
pseudopotentials for graphene are shown in Fig. \ref{fig:IntPotGraph}(b) 
for $n=0$ and $1$.

Haldane's pseudopotentials are an extremely helpful quantity in the understanding of the possible 
FQHE states which one may expect in 2D electron systems. One notices first that as a consequence of the anti-symmetry of a two-particle
wave function under fermion exchange, the relative angular-momentum quantum number $\ell$ must be an \textit{odd} integer, i.e. 
only the pseudopotentials with odd values of $\ell$ play a physical role in the description of two interacting electrons of the 
same type (spin or valley). Even-$\ell$ pseudopotentials become relevant if the SU(4) spin-valley pseudospin is not completely polarized,
in the treatment of two electrons with different internal quantum number $\sigma$ or $\xi$. One then notices that the $n=1$ 
pseudopotentials, apart from a the difference in $V_{\ell=0}$, are much more reminiscent of those in $n=0$ than of those
for non-relativistic 2D electrons in the same LL $n=1$ [see Fig. \ref{fig:IntPotGraph}(b)]. If one considers polarized electrons, 
one therefore expects essentially the same strongly-correlated electronic phases in graphene for $n=1$ as for $n=0$ \cite{GMD},
as also corroborated by numerical studies for FQHE states \cite{AC,toke1,toke2,GR,papic09} and electron-solid phases 
\cite{ZJ1,ZJ2,Popl}. Because the pseudopotentials (\ref{eq:HPP}) are systematically larger in $n=1$ than in $n=0$ (apart
from the short-range component for $\ell=0$), the gaps of the FQHE states in $n=1$ are larger than the corresponding ones
in $n=0$, as one may also see from numerical calculations \cite{AC,toke1}.

As much as we have emphasized the similarity between the $n=0$ and $n=1$ LLs in graphene, we need to stress the difference 
between the $n=1$ LL in graphene as compared to $n=1$ in non-relativistic 2D electron systems. Remember that the physical
phase diagram in the non-relativistic $n=1$ LL is extremely rich; an intriguing even-denominator FQHE has been observed at $\nu=5/2$ 
\cite{willett} and probably possesses non-Abelian quasiparticle excitations \cite{MR,GWW}. Furthermore, a particular 
competition between FQHE states and electron-solid phases, which is characteristic of the non-relativistic $n=1$ LL 
\cite{GLMS1,GLMS2}, is at the origin of the reentrance phenomena observed in transport measurements \cite{lewis05,eisenstein02}.
These phenomena are absent in the $n=0$ LL, and the similarity between the $n=0$ and the relativistic $n=1$ LL thus
leads to the expectation that FQHE states corresponding to the 5/2 state in non-relativistic quantum Hall systems and the
above-mentioned reentrance phenomena are absent in the $n=1$ LL in graphene. This expectation
has recently been corroborated in exact-diagonalization studies on the non-Abelian $5/2$ state \cite{WMC10}.

\subsubsection{External spin-valley symmetry breaking terms}
\label{Sec:Zeeman}

Before we consider the different phases due to electron-electron interactions, we start with an analysis of the 
different \textit{external} effects,\footnote{By external effects we mean
those that are not caused by the mutual Coulomb repulsion between the electrons.} 
which are capable of lifting the four-fold spin-valley degeneracy. 

The probably most familiar external symmetry-breaking term is the Zeeman effect, which lifts the two-fold spin degeneracy
while maintaining the SU(2) symmetry associated with the valley pseudospin. The size of the Zeeman splitting is given by
the energy $\Delta_Z\sim 1.2 B{\rm [T]}$ K, for a $g$-factor that has been experimentally determined as $g\sim 2$ \cite{zhang}.
If we adopt a compact eight-spinor notation to take into account the four different spin-valley components, in addition to 
the two sublattice components, the Zeeman term has the form
\beqn\label{eq:MCZeemanS}
\Delta_Z^{\rm spin} &\sim& \Psi^{\dagger} \left(\bone_{\rm valley} \otimes \bone_{AB}\otimes \tau_{\rm spin}^z\right) \Psi\\
\nn
&\sim& \psi_{K,\ua}^{A\dagger}\psi_{K,\ua}^{A} - \psi_{K,\da}^{A\dagger}\psi_{K,\da}^{A} + 
\psi_{K',\ua}^{A\dagger}\psi_{K',\ua}^{A}- \psi_{K',\da}^{A\dagger}\psi_{K',\da}^{A}  \\
\nn
&&+\, (A\leftrightarrow B),
\eeqn
where the tensor product consists of the valley pseudospin (represented by the Pauli matrices $\tau_{\rm valley}^{\mu}$ and 
$\bone_{\rm valley}$), the sublattice pseudospin ($\tau_{AB}^{\mu}$ and $\bone_{AB}$), and the true spin ($\tau_{\rm spin}^{\mu}$ 
and $\bone_{\rm spin}$). For a better understanding, we have given the explicit expression in terms of spinor components 
in the second line of Eq. (\ref{eq:MCZeemanS}).

A possible valley-degeneracy lifting, that one could describe with the help of a ``valley Zeeman effect'' similarly to Eq.
(\ref{eq:MCZeemanS}),
\beq\label{eq:MCZeemanV}
\Delta_Z^{\rm valley} \sim \Psi^{\dagger} \left(\tau_{\rm valley}^z \otimes \bone_{AB}\otimes \bone_{\rm spin}\right) \Psi,
\eeq
is more involved because there is no physical field that couples directly to the valley pseudospin, as suggested by the 
otherwise intuitive form (\ref{eq:MCZeemanV}). There have however been proposals that such an effect may be achieved with
the help of strain-induced disordered gauge fields that mimic large-scale ripples \cite{meyer} and 
that yield an easy-plane anisotropy in $n=0$ \cite{abanin07}, similarly to the backscattering term (\ref{eq:Int_sb}) in higher LLs.
Quite generally, a valley-degeneracy lifting may be achieved indirectly in the zero-energy LL $n=0$ via fields that couple 
to the sublattice index. This is due to the fact that the 
components $\psi_{K,\sigma}^A$ and $\psi_{K',\sigma}^B$ vanish as a consequence of the chiral properties, which identify
the sublattice and the valley pseudospins in $n=0$, as we have discussed in Sec. \ref{Sec:RelLL}.

In order to illustrate this indirect lifting of the valley degeneracy, we consider the term \cite{haldane88}
\beq\label{eq:MCmassT}
\mathcal{M}_H = M \Psi^{\dagger} \left(\bone_{\rm valley} \otimes \tau_{AB}^z\otimes \bone_{\rm spin}\right) \Psi,
\eeq
which breaks the lattice inversion symmetry and opens a mass gap at the Dirac point in the absence of a magnetic field. 
In the presence of a $B$-field, the LL spectrum (\ref{RelLLs}) is modified by the term (\ref{eq:MCmassT}) and reads
\beq\label{RelLLsM}
\epsilon_{\lambda,n;\xi} = \lambda \sqrt{M^2 + 2\frac{\hbar^2 v_F^2}{l_B^2}n},
\eeq
for $n\neq 0$, independent of the valley index $\xi$, whereas the fate of the $n=0$ LL depends explicitly on $\xi$,
\beq\label{RelLL0M}
\epsilon_{n=0;\xi}=\xi M,
\eeq
such that the valley degeneracy is effectively lifted. Notice, however, that due to the vanishing components 
$\psi_{K,\sigma}^A$ and $\psi_{K',\sigma}^B$, the mass term (\ref{eq:MCmassT}) is now indistinguishable (in $n=0$) from the 
above-mentioned valley Zeeman term (\ref{eq:MCZeemanV}),
\beqn\nn
\mathcal{M}_H^{n=0} &\sim& \Psi^{\dagger} \left(\bone_{\rm valley} \otimes \tau_{AB}^z\otimes \bone_{\rm spin}\right) \Psi\\
&\sim& \Psi^{\dagger} \left(\tau_{\rm valley}^z \otimes \bone_{AB}\otimes \bone_{\rm spin}\right) \Psi,
\eeqn
whereas this is not the case for the LLs $n\neq 0$, where the valley degeneracy is only lifted by an explicit valley
Zeeman effect. A mass term of the form (\ref{eq:MCmassT}) typically arises 
in the presence of a frozen out-of-plane phonon that yields a crumbling of the graphene sheet \cite{fuchs}. 

More recent studies have concentrated on a spontaneous deformation of the graphene sheet due to frozen in-plane 
phonons that yield a Kekul\'e-type distortion \cite{nomura09,hou10}. 
This distortion, which is associated with a characteristic wave vector $2\bK$ and which therefore couples the two
valleys, directly breaks the valley degeneracy, via a term $\mathcal{M}_K=\mathcal{M}_x + \mathcal{M}_y$, with
\beq\label{eq:KekMass}
\mathcal{M}_{x,y} = \frac{\Delta_{x,y}}{2} 
\Psi^{\dagger} \left(\tau_{\rm valley}^{x,y} \otimes \bone_{AB}\otimes \bone_{\rm spin}\right) \Psi.
\eeq
Such a term yields the same energy spectrum (\ref{RelLLsM}) and (\ref{RelLL0M}) as the mass term (\ref{eq:MCmassT}) if 
one replaces $M$ by $\Delta_{kek}/2$, with the characteristic energy scale 
$\Delta_{kek}=\sqrt{\Delta_x^2 + \Delta_y^2}\simeq 2 B{\rm [T]}$ K, \cite{hou10,ajiki}. Notice that this energy
scale is slightly larger than, but roughly on the same order as, the Zeeman energy scale. 

Finally, we mention another class of terms that break the spin-valley degeneracy and that have received recent 
interest in the framework of research on topological insulators [for recent reviews see \cite{HasanRev10,ZhangRev}]. In an original
work, Haldane argued that a time-reversal-symmetry breaking term with an inhomogeneous flux distribution inside each hexagon 
opens a gap in a honeycomb lattice
with zero magnetic field \cite{haldane88}. Most saliently, he showed
that one may thus achieve a quantum Hall effect without an external magnetic field, a system that is now often referred to 
as the ``quantum anomalous Hall insulator'' \cite{HasanRev10,ZhangRev}. A similar situation arises when spin-orbit
interactions are taken into account, which are of the form
\beq\label{eq:MassSO}
\Hmath_{SO}=\frac{\Delta_{SO}}{2} \Psi^{\dagger} \left(\tau_{\rm valley}^z \otimes \tau_{AB}^z\otimes \tau_{\rm spin}^z\right) \Psi
\eeq
and which provide again the same LL spectrum (\ref{RelLLsM}) and (\ref{RelLL0M}) if one replaces $M$ by $\Delta_{SO}/2$
\cite{KM05}. In spite of the conceptually appealing prospect of the quantum spin Hall effect, which is revealed by this model
because the spin orientation is locked to a particular valley index via the term (\ref{eq:MassSO}), the associated energy
scale $\Delta_{SO}\sim 10$ mK turns out to be vanishingly small in graphene, whereas an extrinsic Rashba-type
spin-orbit coupling in graphene can be on the order of 1 K \cite{min06}.

\subsubsection{Hierarchy of relevant energy scales}

These energy scales associated with external fields need to be compared to the characteristic (bare) interaction energy
$e^2/\epsilon l_B\simeq 625(\sqrt{B{\rm [T]}}/\eps)$ K, which is, for experimentally accessible magnetic fields, much larger than 
$\Delta_Z$ or $\Delta_{kek}$. As we have discussed in Sec. \ref{Sec:Diel}, inter-band LL excitations screen the bare Coulomb 
interaction and yield a contribution to the dielectric constant. In the absence of a quantizing magnetic field, we have seen
that this dielectric constant is given by [see Eq. (\ref{eq:epsInf})]
\beq\label{eq:epsInfb}
\eps_{\infty}=1+\frac{\pi}{2}\alpha_G=1+\frac{\pi}{2}\frac{e^2}{\hbar \eps v_F},
\eeq
where we remember that $\eps$ is the \textit{extrinsic} dielectric constant of the surrounding medium. As one may notice 
from Fig. \ref{fig:StatPol}, the vacuum contribution $\Pi^{vac}(\bq)$ (thick dashed lines) is only marginally modified 
by the magnetic field, such that one may use $e^2/\eps\eps_{\infty} l_B$ as an approximation for the interaction-energy 
scale for graphene taking into account inter-band screening. 

The relevant energy scales are summarized in the table below for different values of the magnetic field, in 
comparison with the interaction energy scales, taking into account the effective dielectric constants for several
widely used substrates from Tab. I.

\begin{widetext}
\begin{center}
\begin{table}[htbp]
{
\begin{tabular}{|c||c|c|c|}
\hline
energy & value for arbitrary $B$ & for $B=6$ T & for $B=25$ T\\
\hline \hline
$\Delta_{Z}$ & $1.2 B{\rm [T]}$ K & $7$ K & $30$ K \\ \hline
$\Delta_{kek}$ & $2 B{\rm [T]}$ K & $12$ K & $50$ K \\ \hline\hline
$e^2/\eps l_B$ (bare) & $625(\sqrt{B{\rm [T]}}/\eps)$ K & $(1550/\eps)$ K & $(3125/\eps)$ K \\ \hline\hline
$e^2/\eps\eps_{\infty} l_B$ (vacuum) & $139\sqrt{B{\rm [T]}}$ K & $344$ K & $694$ K
\\ \hline
$e^2/\eps\eps_{\infty} l_B$ (on SiO$_2$) & $104\sqrt{B{\rm [T]}}$ K & $258$ K & $521$ K
\\ \hline
$e^2/\eps\eps_{\infty} l_B$ (on h-BN) & $109\sqrt{B{\rm [T]}}$ K & $270$ K & $543$ K
\\ \hline
$e^2/\eps\eps_{\infty} l_B$ (on SiC) & $71\sqrt{B{\rm [T]}}$ K & $176$ K & $355$ K
\\ \hline\hline
$\Delta_{sb}<(e^2/\eps l_B)(a/l_B)$ & $<1 B[{\rm T}]$ K & $<6$ K & $<25$ K\\
\hline
\end{tabular}}
\label{Tab:Zeeman}
\caption{\footnotesize Energy scales at different magnetic fields. The first two lines show the energy scales associated with the
major external symmetry-breaking fields (Zeeman and Kekul\'e-type lattice distortion, $\Delta_Z$ and $\Delta_{kek}$, respectively), 
which scale linearly in $B$. 
Below are shown the interaction-energy scales ($\propto \sqrt{B}$), the bare one with an unspecified dielectric constant and
the ones for different substrates taking into account inter-band screening via the term $\eps_{\infty}$ [Eq. (\ref{eq:epsInfb})]. 
The last line 
yields the interaction-energy scale associated with the intrinsic symmetry breaking due to inter-valley coupling, discussed in
Sec. \ref{sec:sbTerms}.
}
\end{table}
\end{center}
\end{widetext}

In view of the above discussion, one may conclude that the SU(4)-symmetric part of the Coulomb interaction yields the leading
energy scale in the problem of electrons in partially filled lower LLs, whereas external terms, such as the Zeeman effect or
spontaneous lattice distortions, play a subordinate role. The remainder of this section is therefore concerned with a detailed
discussion of strongly-correlated electron phases that are formed to minimize the Coulomb interaction.

\subsection{SU(4) Quantum Hall Ferromagnetism in Graphene}
\label{sec:QHFM}

A prominent example of the above-mentioned strongly-correlated phases is the 
generalized quantum Hall ferromagnet. It arises in systems with a discrete internal degree of freedom
described by an SU($\Nmath$) symmetry,
such that each single-particle quantum state $\psi_{n,m}$ occurs in $\Nmath$ copies. Prominent examples are the non-relativistic 
quantum Hall systems when taking into account the electronic spin $\sigma=\ua,\da$ ($\Nmath=2$) or bilayer quantum Hall systems that 
consist of two parallel 2D electron gases, where the layer index may be viewed as a ``spin'' 1/2 [$\Nmath=2$, or $\Nmath=4$ if one also
takes into account the physical 
spin \cite{hasebe02,tsitsish03}].\footnote{For a review on non-relativistic multi-component systems see Ref. \cite{moon,ezawa}.}
In this sense, graphene may be viewed as an SU(4) quantum Hall system as a consequence of its four-fold spin-valley 
degeneracy.

\subsubsection{Ferromagnetic ground state and Goldstone modes}
\label{Sec:FMGM}

Quite generally, quantum Hall ferromagnetism arises when the filling factor, defined from the bottom of the 
LL,\footnote{Remember that the filling factor in graphene is defined with respect to the center of the $n=0$ LL. There is thus
a shift of 2 in the filling factor as compared to the non-relativistic case.}
is an integer that is not a multiple of $\Nmath$ \cite{arovas}. From the point of view of the kinetic Hamiltonian, 
one is thus  confronted with a 
macroscopic ground-state degeneracy. Even if one has an integer filling factor, the situation is thus much more reminiscent of the
FQHE, i.e. the relevant energy scale is the Coulomb interaction, and
the system may be described in the framework of the model (\ref{eq:LLHamSU4}) of interacting electrons in a single 
(relativistic) LL.
For the moment, we consider that there are no symmetry-breaking terms, such as the backscattering term (\ref{eq:Ham_sb}) 
or Zeeman-type terms that are discussed below in Sec. \ref{Sec:Zeeman}.

Qualitatively, one may understand the formation of a ferromagnetic ground state as a consequence of the repulsive Coulomb interaction.
In order to minimize this interaction, the electrons prefer to form a state described by a maximally anti-symmetric orbital wave
function that must then be accompanied by a fully symmetric SU($\Nmath$) spin wave function to satisfy an overall fermionic 
(anti-symmetric) wave function. In a usual metal with a finite band dispersion, this ferromagnetic ordering (e.g. all electrons in 
the spin-$\ua$ states) is accompanied by a cost in kinetic energy -- indeed, the Fermi energy for spin-$\ua$ electrons is increased
whereas that of spin-$\da$ electrons is lowered. The competition between the gain in interaction and the cost in kinetic energy 
defines the degree of polarization, i.e. how ferromagnetic the electrons effectively are.
In the quantum Hall effect, however, we are confronted with a highly degenerate LL that may be viewed as an infinitely flat band, such 
that the kinetic-energy cost for complete spin polarization is zero. 

As an example of an SU($\Nmath$) quantum Hall ferromagnet, one may consider the state 
\beq\label{eq:FMGS}
|{\rm FM}\rangle = \prod_{i=1}^k\prod_{m=0}^{N_B-1} c_{m,i}^{\dagger}|{\rm vac}\rangle,
\eeq
which consists of $k<\Nmath$ arbitrarily chosen completely filled subbranches [$i\in \{(K,\ua),(K',\ua),(K,\da),(K',\da)\}$, 
for the SU($\Nmath=4$) symmetry in graphene LLs], where we have omitted the LL index $\lambda n$ at the fermion operators to simplify the
notation. The arbirariness in the choice of the SU($\Nmath$) spin subbranches may be viewed as a spontaneous symmetry breaking
that accompanies the ferromagnetism. Indeed, the state (\ref{eq:FMGS}) is no longer invariant under an SU($\Nmath$) rotation, but only 
under a rotation described by the subgroup SU($k$)$\times$SU($\Nmath-k$), where the first factor indicates a symmetry transformation
in the fully occupied subbranches $i=1,...,k$ and the second factor one in the empty subbranches $i=k+1,...,\Nmath$.
Therefore the quantum Hall ferromagnet (\ref{eq:FMGS}) is associated with an order parameter with a spontaneous symmetry breaking
described by the coset space ${\rm SU}(\Nmath)/\SU(k)\times \SU(\Nmath-k)\times \U(1)\sim \U(\Nmath)/\U(k)\times \U(\Nmath-k)$, where the
additional $\U(1)$ is due to the phase difference between the occupied and the unoccupied subbranches \cite{arovas,yang}.

The coset space, with its $\Nmath^2-k^2-(\Nmath-k)^2=2k(\Nmath-k)$ complex generators, defines also the Goldstone modes, which are
nothing other than the $k(\Nmath-k)$ spin-wave excitations of the ferromagnetic ground state (\ref{eq:FMGS}).\footnote{The complex
generators come in by pairs of conjugate operators, and each pair corresponds to one mode.} The number of the 
spin-wave modes may also have been obtained from a simple inspection into the LL-subbranch spectrum. Indeed, a spin wave can be described
with the help of the components of the projected density operators (\ref{eq:ProjDens}),
\beq
\rhobar_{ij}(\bq) = \sum_{m,m'}\left\langle m\left|e^{-i\bq\cdot\bR}\right|m'\right\rangle c_{m,i}^{\dagger}c_{m',j},
\eeq
which represent coherent superpositions at wave vector $\bq$ of excitations from the occupied subbranch $j$ to the 
empty subbranch $i$. One 
has then $k$ possibilities for the choice of the initial subbranch $j$ and $\Nmath-k$ for the final one, and one obtains therefore
$k(\Nmath-k)$ different spin-wave excitations, in agreement with the above group-theoretical analysis.

Notice that all spin-wave excitations have the same dispersion, which may be calculated within a mean-field 
approximation \cite{alicea,doretto,KH,yang}, 
\beqn\label{eq:SWdisp}
\nn
E_{\bq} &=& \langle {\rm FM}| \rhobar_{ij}(-\bq) H_n \rhobar_{ij}(\bq) - H_n|{\rm FM}\rangle\\
&=& 2
\sum_{\bk}v_n(\bk)\sin^2\left(\frac{\bq\wedge\bk\,l_B^2}{2}\right),
\eeqn
which saturates at large values of $q=|\bq|$,
\beq\label{eq:SWactgap}
E_{q\rightarrow\infty}=2E_X^n=\sum_{\bk}v_n(\bk),
\eeq
i.e. at twice the value of the exchange energy (\ref{eq:ExchEn}). This result is not astonishing insofar as the large-$q$
limit corresponds, as we have discussed in Sec. \ref{Sec:PolBnon0} [see Eq. (\ref{eq:PseudoMom})], to an electron-hole pair where
the electron is situated far away from the hole. The energy (\ref{eq:SWactgap}) is therefore nothing other than the cost in
exchange energy to create a \textit{spin-flip} excitation, i.e. 
an electron with reversed spin and a hole in the ferromagnetic ground state. Because of the large distance between the electron
and the hole in such an excitation and the resulting decoupled dynamics, one may be tempted to 
view this energy as the activation gap of the quantum-Hall state at $\nu=k$, but we see below in Sec. \ref{Sec:Skyrm}
that there exist elementary charged excitations (skyrmions) that have, in some LLs, a lower energy than these spin-flip excitations.

In the opposite limit of small wave vectors ($ql_B\ll 1$), one may not understand the excitation in terms of decoupled holes and 
electrons, and the excitation can therefore not contribute to the charge transport.
A Taylor expansion of the sine in the spin-wave dispersion (\ref{eq:SWdisp})
yields the usual $q^2$ dispersion of the spin-wave Goldstone modes,
\beq\label{eq:SWsmallQ}
E_{q\rightarrow 0} = \frac{\rho_s^n}{2} q^2l_B^2,
\eeq
in terms of the \textit{spin stiffness} 
\beq\label{eq:spinstiff}
\rho_s^n=\frac{1}{4\pi}\sum_{\bk}v_n(\bk)|\bk|^2l_B^2 .
\eeq
One notices that the above results for the excitation energies do not depend on the size of the internal symmetry group, but they
can be derived within the SU(2) model of the quantum Hall ferromagnetism \cite{moon,sondhi} -- the enhanced internal symmetry of graphene
(or of a general $\Nmath$-component system) affects only the degeneracies of the different modes.

\subsubsection{Skyrmions and entanglement}
\label{Sec:Skyrm}

\begin{figure}
\centering
\includegraphics[width=8.5cm,angle=0]{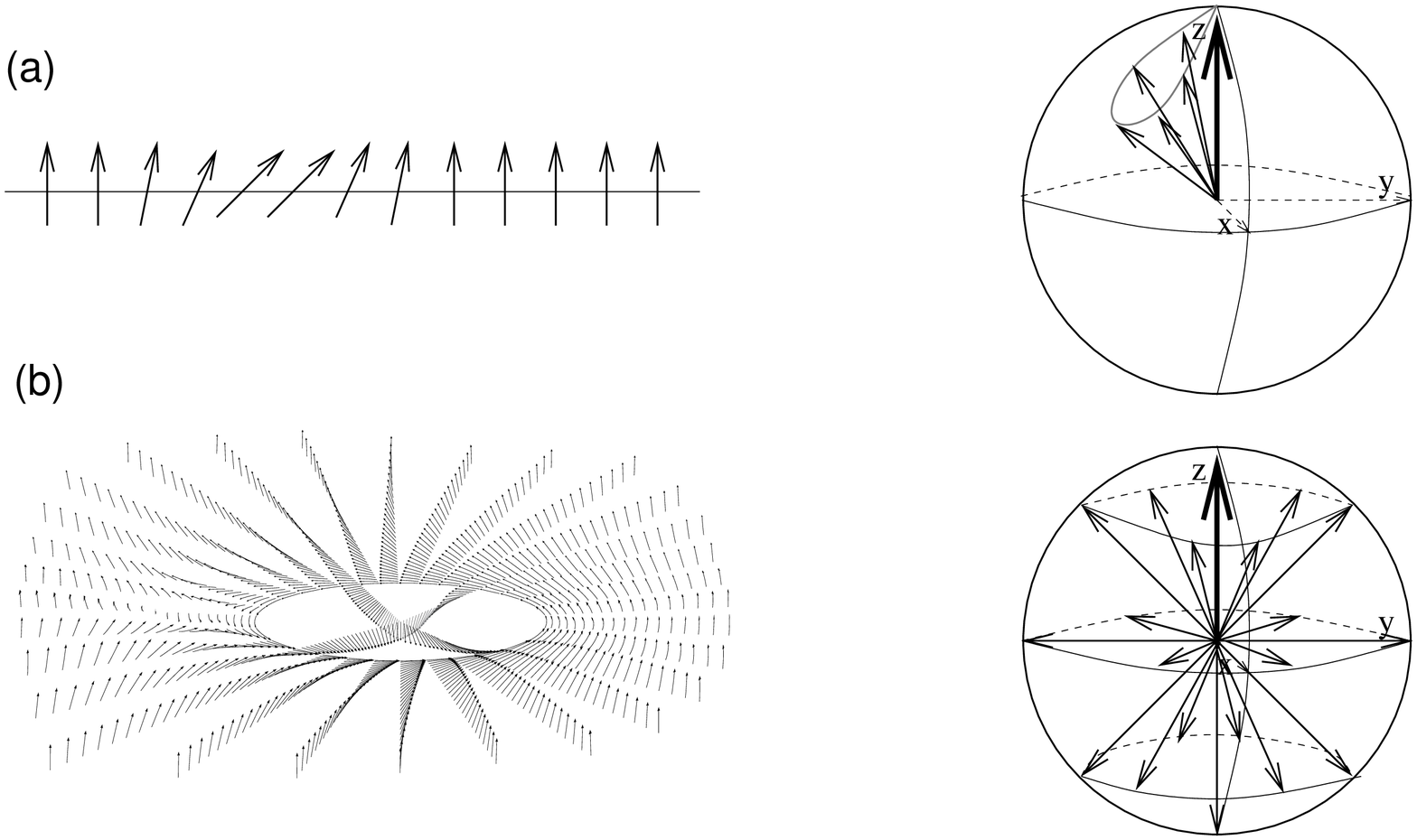}
\caption{\footnotesize{Excitations of the SU(2) ferromagnetic state. 
\textit{(a)} Spin waves. Such an excitation can be continuously deformed into the ferromagnetic
ground state [spin represented on the Bloch sphere (right) by fat arrow] -- the grey curve
can be shrinked into a single point
\textit{(b)} Skyrmion with non-zero topological charge [from Ref. \cite{GirvinLH}]. 
The excitation consists of a a reversed spin at the origin $z=0$, and the ferromagnetic state is
recovered at large distances $|z/\zeta|\rightarrow \infty$. Contrary to spin-wave excitations, the
spin explores the whole surface of the Bloch sphere and cannot be transformed by a continuous
deformation into the majority spin (fat arrow).
}}
\label{fig:Skyrm}
\end{figure}

In addition to the above-mentioned spin-wave modes, the SU($\Nmath$) ferromagnetic ground state (\ref{eq:FMGS}) is characterized 
by a particular elementary excitation that consists of a topological spin texture, the so-called \textit{skyrmion} \cite{sondhi}.
Similarly to a spin-wave in the limit $ql_B\ll 1$,
the variation of the spin texture in a skyrmion excitation is small on the scale of the magnetic length
$l_B$, such that its energy is determined by the spin stiffness (\ref{eq:spinstiff}) in the small-$q$ limit. Indeed, one may
show that its energy is given by \cite{moon,ezawa,sondhi}
\beq\label{eq:SkyrmEn}
E_{sk}=4\pi \rho_s^n|Q_{top}|,
\eeq
in terms of the topological charge $Q_{top}$, which may be viewed as the number of times the Bloch sphere is covered by
the spin texture [see Fig. \ref{fig:Skyrm}(b)] and which we discuss in more detail below.
Skyrmions are the relevant elementary excitations of the quantum Hall ferromagnetism
if the energy (\ref{eq:SkyrmEn}) is lower than that of an added electron (or hole) with reversed spin, which is nothing other than
the exchange energy (\ref{eq:ExchEn}), i.e. if $E_{sk}<E_X^n$. Whereas this condition is fulfilled in non-relativistic LLs only
in the lowest one $n=0$, skyrmions are the lowest-energy elementary excitations in the graphene LLs $n=0,1$ and 2 \cite{toke1,yang},
as a consequence of the difference in the form factors.

As in the case of the spin waves discussed above, the skyrmion energy is independent of the
size of the internal symmetry group, and we will first illustrate the skyrmion texture in an effective SU(2) model, where the
texture is formed only from states within the last occupied ($k$) and the first unoccupied ($k+1$) LL subbranch. The skyrmion may then
be described with the help of the wave function, in terms of the complex coordinate $z=(x-iy)/l_B$,
\beq\label{eq:SkyrmWF}
|\Smath_{k,k+1}\rangle =\frac{1}{\sqrt{|\zeta|^2 + |z|^2}}\left(z\left|\ua^k(z)\right\rangle + \zeta \left|\da^{k+1}(z)\right\rangle
\right)
\eeq
where $|\ua^k(z)\rangle$ corresponds to states in the subbranch $k$ and $|\da^{k+1}(z)\rangle$ to those in $k+1$,
at the position $z$. One notices 
that at the origin $z=0$ the ``spin'' associated with these two components is $\da$ because the first component of Eq. (\ref{eq:SkyrmWF})
vanishes, whereas the spins are $\ua$ at $|z/\zeta|\rightarrow \infty$ (see Fig. \ref{fig:Skyrm}), where the ferromagnetic ground
state is then recovered. The parameter $\zeta$ plays the role of the skyrmion size, measured in units of $l_B$ -- 
indeed, for $|z|=|\zeta|$, both components are of the same weight and the spin is therefore oriented in the $xy$-plane. 

The skyrmion excitation (\ref{eq:SkyrmWF}) can also be illustrated on the so-called Bloch sphere on the surface of which the 
(normalized) spin moves (see Fig. \ref{fig:Skyrm}). The angles ($\theta$ for the azimuthal and $\phi$ for the polar angle) 
of the spin orientation on the Bloch sphere correspond to the SU(2) parametrization $|\psi\rangle =\cos(\theta/2)|\ua\rangle
+\sin(\theta/2)\exp(i\phi)|\da\rangle$, and the spin orientation at the circle $|z|=|\zeta|$ in the complex plane
describes the equator of the Bloch sphere. The topology of the skyrmion excitation becomes apparent by the number of full circles
the spin draws when going around the origin of the $xy$-plane on the circle $|z|=|\zeta|$. More precisely, the topological
charge $Q_{top}$ is not defined in terms of such closed paths, but it
is the number of full coverings of the Bloch sphere in a skyrmion excitation [$Q_{top}=1$ in the example
(\ref{eq:SkyrmWF})]. Notice that a spin-wave excitation has a topological charge $Q_{top}=0$ and corresponds to an excursion of the
spin on the Bloch sphere that is not fully covered and that can then be reduced continuously to a single point describing
the ferromagnetic ground state [Fig. \ref{fig:Skyrm}(a)].

The above considerations may be generalized to systems with larger internal symmetries, i.e. to SU($\Nmath$) quantum Hall 
ferromagnets. The state (\ref{eq:SkyrmWF}) is invariant under the SU($\Nmath$) subgroup $\SU(k-1)\times\SU(\Nmath-k-1)$, where 
the first factor describes a rotation of the occupied subbranches that do not take part in the skyrmion excitation and the second
factor is associated with a symmetry transformation of the corresponding unoccupied subbranches $k+2, ..., \Nmath$. A similar
group-theoretical analysis as the one presented in Sec. \ref{Sec:FMGM} yields the number of residual symmetry transformations
\cite{yang} $2k(\Nmath-k) + 2(\Nmath -1)$, where the first term describes the Goldstone modes of the ferromagnetic ground state, 
and the second one corresponds to the $\Nmath-1$ internal modes of the skyrmion excitation.

In addition to the topological charge, skyrmions in quantum Hall systems
carry an electric charge that coincides, for $\nu=k$ with the topological charge. 
Indeed, the skyrmion state (\ref{eq:SkyrmWF}) describes an electron that is expelled from the origin $z=0$ in the $\ua$-component,
and its net electric charge is therefore that of a hole. This means that skyrmions are excited when sweeping the filling factor away
from $\nu=k$, and the net topological charge is given by $Q_{tot}=|\nu-k|N_B$. The number of internal modes is then
$Q_{tot}(\Nmath -1)$, in addition to one mode per charge that corresponds to a simple translation $z\rightarrow z + a$ of 
the excitation \cite{DGLM}. As a consequence of the Coulomb repulsion, it
is energetically favorable to form a state in which $Q_{tot}$ skyrmions of charge $1$ are homogeneously distributed over the 2D
plane than a single defect with charge $Q_{tot}$ \cite{moon}. A natural (semi-classical) 
candidate for the ground state of $Q_{tot}$ skyrmions is then a skyrmion crystal \cite{brey95} that has recently been revisited
in the framework of the SU(4) symmetry in graphene \cite{cote07,cote08}. In this case, the $Q_{tot}(\Nmath -1)$ internal modes, 
which are dispersionless zero-energy modes in the absence of electronic interactions or Zeeman-type symmetry-breaking terms, are
expected to yield $\Nmath-1$ Bloch bands of Goldstone-type, in addition to the $Q_{tot}$ translation modes that form a 
magnetic-field phonon mode of the skyrmion crystal with a characteristic $\omega\propto q^{3/2}$ dispersion \cite{fuku75}.

\paragraph{Skyrmions and activation gaps in graphene.}

Quite generally, the activation gap in quantum Hall states is the energy required to create a quasi-particle--quasi-hole pair,
in which the two partners are sufficiently well separated to contribute independently to the charge transport. 
In the framework of the quantum Hall ferromagnet, the activation gap may be viewed as the energy to create a skyrmion of
topological charge $Q=1$ and an anti-skyrmion of charge $Q=-1$ that are well-separated from each other such that one may neglect their 
residual interaction. The energy of such a skyrmion--anti-skyrmion pair is then given, in the absence of symmetry-breaking
terms, by twice the energy in Eq. (\ref{eq:SkyrmEn}),
\beq\label{eq:ActGapSkyrmSym}
\Delta_a^{\rm sym} = 8\pi \rho_s^n.
\eeq
For graphene, the energies of the theoretical activation gaps for $n=0$ and $n=1$ are shown in the table below.
\begin{center}
\begin{table}[htbp]
{
\begin{tabular}{|c||c|c|}
\hline
 ~ & activation gap & arbitrary value of $B$ \\
\hline \hline
$n=0$  & $\frac{1}{2}\sqrt{\frac{\pi}{2}}\frac{e^2}{\eps\eps_{\infty} l_B}$ & $400(\sqrt{B{\rm [T]}}/\eps\eps_{\infty} )$ K  \\ \hline
$n=1$ & $\frac{7}{32}\sqrt{\frac{\pi}{2}}\frac{e^2}{\eps\eps_{\infty} l_B}$ & $175(\sqrt{B{\rm [T]}}/\eps\eps_{\infty} )$ K  \\ \hline
\end{tabular}}
\label{Tab:Skyrm}
\caption{\footnotesize Theoretical estimates for the activation gaps in the $n=0$ and 1 graphene LLs
due to well-separated skyrmion--anti-skyrmion pairs.}
\end{table}
\end{center}

For further illustration, we consider the scenario in which the Zeeman effect is the only SU(4)-symmetry breaking 
term.\footnote{The energetic argument remains
valid in the case where the dominant term is a valley-Zeeman effect if one interchanges the role of the spin and the valley pseudospin.} 
Due to the Zeeman effect, spin-$\da$ electrons are energetically favored. 
If only one spin-valley branch of a particular LL is filled ($k=1$), 
the spin magnetization of the spin-valley ferromagnet is preferentially 
oriented in this direction whereas the valley polarization may point in any direction. The activation gap would then be 
dominated by valley (anti-)skyrmions with no reversed
physical spin such that one would not expect any dependence of the gap on the in-plane component of the magnetic field,
in agreement with the experimental findings \cite{jiang07}.

The situation is different when both valley branches of the spin-$\da$
branch are occupied; an excitation of the SU(4) ferromagnet with a full spin polarization would then 
necessarily comprise
reversed spins, and the corresponding Zeeman energy must be taken into account in the energy of the (spin) skyrmion--anti-skyrmion
pair (\ref{eq:ActGapSkyrmSym}),
\beq\label{eq:ActGapSkyrmZ}
\Delta_a^{\rm Z} = 8\pi \rho_s^n + 2N_{rs}\Delta_Z,
\eeq
where $N_{rs}\sim |\zeta|^2$ is the number of reversed spins in a single (anti-)skyrmion. 
Notice that this number depends on the competition between the Zeeman effect itself, which tries to reduce the skyrmion size 
$\zeta$, and the cost in exchange energy due to the strong variation in small textures \cite{sondhi,moon}.\footnote{This energy
cost may be evaluated from a gradient expansion of the energy in the magnetization fields. At leading order, one obtains, however,
a non-linear sigma model that is scale-invariant, such that the energy cost must be calculated at higher orders \cite{moon}.}
The energy of a skyrmion--anti-skyrmion pair in the spin channel (with two completely filled valley sublevels) 
is therefore larger than that (\ref{eq:ActGapSkyrmSym}) of a pair in the valley channel when only one valley subbranch of the LL is
completely filled. Notice that this energy increase may even be significant for large skyrmions because of the larger number 
of reversed spins.
As a thumbrule, the stability of a quantum Hall state is proportional to the activation gap, which has in the present
case been  identified with the skyrmion--anti-skyrmion energy and which is dominated by the Coulomb interaction energy. 
Additional external symmetry-breaking
terms, such as those discussed in Sec. \ref{Sec:Zeeman}, may enhance this stability although they provide only a small correction
to the activation energy.

\paragraph{Spin-valley entanglement in graphene.}

\begin{figure}
\centering
\includegraphics[width=6.5cm,angle=0]{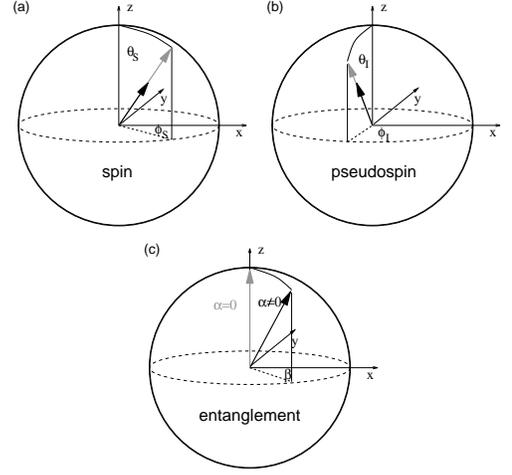}
\caption{\footnotesize{From Ref. \cite{DGLM}; Bloch spheres for entangled spin-pseudospin systems. 
Bloch sphere for the spin \textit{(a)}, pseudospin \textit{(b)}, and a third type
of spin representing the entanglement \textit{(c)}. In the case of spin-pseudospin
entanglement ($|\cos\alpha|\neq 1$), the (pseudo)spin-magnetizations explore
the interior of their spheres, respectively (black arrows). 
}} 
\label{fig:BlochSp}
\end{figure}

In an experimental measurement, one typically has not direct access to the full SU(4) spin that describes the internal degrees
of freedom in graphene LLs, but only to the SU(2) part associated with the physical spin, e.g. in a magnetization measurement. 
It is therefore useful to parametrise the SU(4) spin in a manner such as to keep track of the two SU(2) copies associated with 
the physical spin and with the valley pseudospin, respectively. This may be achieved with the so-called Schmidt decomposition
of the four-spinor
\beq\label{eq:schmidt}
|\Psi(z)\rangle = \cos\frac{\alpha}{2}|\psi_S\rangle|\psi_I\rangle
+\sin\frac{\alpha}{2} e^{i\beta}|\chi_S\rangle|\chi_I\rangle\ ,
\eeq
where $\alpha$ and $\beta$ are functions of the complex position $z$, and the local two-component
spinors $|\psi_{S}\rangle$, $|\chi_{S}\rangle$, $|\psi_{I}\rangle$, and
$|\chi_{I}\rangle$ are constructed according to 
\beq
|\psi\rangle=
\left( 
{\cos\frac{\theta}{2}}\atop{\sin\frac{\theta}{2}e^{i\phi}}
\right) \qquad {\rm and} \qquad
|\chi\rangle = \left(
{-\sin\frac{\theta}{2}e^{-i\phi}}\atop{\cos\frac{\theta}{2}}
\right)\ .
\eeq
The angles $\theta$ and $\phi$ define the usual unit vector 
\beq
\bn(\theta,\phi)=\left(\sin\theta \cos\phi,\sin\theta
\sin\phi,\cos\theta\right),
\eeq
which explores the surface of the Bloch sphere depicted in Fig. \ref{fig:Skyrm}. Notice that one has two Bloch spheres, one
for the unit vector $\bn(\theta_S,\phi_S)$ associated with the spin angles $\theta_S$ and $\phi_S$ and a second one 
for $\bn(\theta_I,\phi_I)$ for the valley-pseudospin angles $\theta_I$ and $\phi_I$ (see Fig. \ref{fig:BlochSp}). 
In addition, one may associate a third
Bloch sphere with the angles $\alpha$ and $\beta$ that describe the degree of factorizability of the wave functions and 
thus the degree of entanglement between the spin and the valley pseudospin \cite{DGLM}.

With the help of the Schmidt decomposition (\ref{eq:schmidt}), one obtains immediately the reduced density matrices for 
the spin and the valley-pseudospin sectors
\beqn
\nn
\rho_S &=& {\rm Tr}_I\left(|\Psi\rangle\langle \Psi|\right)=
\cos^2\frac{\alpha}{2}|\psi_S\rangle\langle \psi_S|+\sin^2\frac{\alpha}{2}
|\chi_S\rangle\langle \chi_S|\ ,\\
\nn
\rho_I &=& {\rm Tr}_S\left(|\Psi\rangle\langle \Psi|\right)=
\cos^2\frac{\alpha}{2}|\psi_I\rangle\langle \psi_I|+\sin^2\frac{\alpha}{2}
|\chi_I\rangle\langle \chi_I|\ ,\\
\eeqn
respectively, and the local spin and valley-pseudospin densities are simply 
\beq\label{LSdens}
m_{S}^{a}={\rm Tr}(\rho_S S^{a})=\cos\alpha\langle\psi_S|S^{a}|\psi_S
\rangle=\cos\alpha \, n^a(\theta_S,\phi_S)
\eeq
and
\beq\label{LPdens}
m_{I}^{\mu}={\rm Tr}(\rho_I I^{\mu})=\cos\alpha\langle\psi_I|I^{\mu}|\psi_I
\rangle=\cos\alpha \, n^{\mu}(\theta_I,\phi_I),
\eeq
where $S^a$ and $I^{\mu}$ represent the components of the spin and valley-pseudospin operators,
respectively [see Eq. (\ref{eq:SIdens})]. One notices from these expressions that, for the case $\alpha\neq 0$ or $\pi$
(i.e. $\cos^2\alpha<1$), the local (pseudo)spin densities are no longer normalized,
but they are of length $|{\bf m}_{S/I}|^2=\cos^2\alpha$. Thus, in a semiclassical
picture, the (pseudo)spin dynamics is no longer restricted to the surface of
the Bloch  sphere, but explores the entire volume enclosed by the sphere
(Fig. \ref{fig:BlochSp}) \cite{DGLM}. This result indicates that one may be confronted, in the case
of full entanglement (e.g. $\alpha=\pi/2$), with an SU(4) quantum Hall ferromagnet the 
(spin) magnetization of which completely vanishes, as one would naively expect for an \textit{unpolarized}
state.

\subsubsection{Comparison with magnetic catalysis}

An alternative scenario proposed for the degeneracy lifting in $n=0$ is that of the 
magnetic catalysis \cite{herbut,gusynin,herbut08,gorbar08,mezawa}, which was discussed even before the discovery
of graphene \cite{khvesh01,gorbar02}. According to this scheme, the Coulomb interaction spontaneously
generates a mass term for the (originally massless) 2D electrons once the magnetic field increases the density of states
at zero energy by the formation of the highly degenerate $n=0$ LL. As a consequence of this mass generation, the particles condense
in a state of coherent particle-hole pairs (excitonic condensation). The effect is at first sight reminiscent of 
the excitonic condensation at $\nu=1$ in non-relativistic bilayer quantum Hall systems \cite{fertig,WZ,EI}. Its superfluid behavior
gives rise to a zero-bias anomaly in the tunneling conductance between the two layers \cite{spielman} as well as to a simultaneous
suppression of the longitudinal and the Hall resistance in a counterflow experiment \cite{kellogg,tutuc}. The 
bilayer excitonic condensate may be described as an easy-plane quantum Hall ferromagnet \cite{moon}, where the spin mimics the 
layer index. The origin of this easy-plane anisotropy stems from the difference in the interactions between electrons in the same
layer as compared to the weaker one for electron pairs in different layers. 

This comparison with non-relativistic 2D electrons in bilayer systems indicates that there may exist a close relation between 
the quantum Hall ferromagnetism and the scenario of the magnetic catalysis also in graphene in a strong magnetic field. Notice,
however, that the excitonic state in graphene is not in the same universality class as that of the quantum-Hall bilayer -- in 
the latter case the symmetry of the (interaction) Hamiltonian is U(1) as a consequence of the easy-plane anisotropy, and the symmetry
breaking is associated with a superfluid mode that disperses linearly with the wave vector, $\omega\propto q$. In contrast to
this system, the interaction Hamiltonian (\ref{eq:LLHamSU4}) has the full SU(4) symmetry, and even for a sufficiently strong
Zeeman effect, the symmetry is quite large with $\SU(2)_{\ua}\times\SU(2)_{\da}$, i.e. each spin projection $\ua$ and $\da$ is governed 
by the residual SU(2) valley symmetry and has the characteristic $\omega\propto q^2$ pseudospin-wave modes.

The connection between the two scenarios becomes transparent within a mean-field treatment of the Coulomb interaction Hamiltonian.
The quantum Hall ferromagnetic states discussed in the previous subsections may be described equivalently with the help of the 
mean-field order parameters
\beq\label{eq:FM_meanfieldOP}
\left\langle\Psi^{\dagger} \left(\tau_{\rm valley}^{\nu} \otimes \bone_{AB}\otimes \tau_{\rm spin}^{\mu}\right) \Psi\right\rangle,
\eeq
where $\Psi$ denotes the same eight-spinor as in Sec. \ref{Sec:Zeeman}.
Remember that a pure spin quantum Hall ferromagnet is obtained for $\tau_{\rm valley}^{\nu}=\bone_{\rm valley}$, whereas
a pure valley-pseudospin ferromagnet is described by an order parameter (\ref{eq:FM_meanfieldOP}) with 
$\tau_{\rm spin}^{\mu}=\bone_{\rm spin}$. The remaining order parameters describe states with a certain degree of spin-valley
entanglement, as discussed above. 

Notice, however, that the choice of order parameters is not restricted to those in Eq. (\ref{eq:FM_meanfieldOP}). Indeed, one may also
opt for a mean-field calculation of the interaction Hamiltonian with the order parameters \cite{gusynin,gorbar08}
\beq\label{eq:MCmassS}
\mathcal{M}_s = \left\langle\Psi^{\dagger} \left(\tau_{\rm valley}^z \otimes \tau_{AB}^z\otimes \bone_{\rm spin}\right) \Psi\right\rangle
\eeq
and 
\beq\label{eq:MCmassTb}
\mathcal{M}_t = \left\langle\Psi^{\dagger} \left(\bone_{\rm valley} \otimes \tau_{AB}^z\otimes \bone_{\rm spin}\right) \Psi\right\rangle,
\eeq
which describe mass gaps. Indeed, we already encountered a term of the form (\ref{eq:MCmassTb}) in Sec. \ref{Sec:Zeeman} and
showed that it lifts the valley degeneracy of the $n=0$ LL. Whereas such a term arises naturally in the context of an out-of-plane 
distortion of the graphene lattice, here, it is generated dynamically via the 
repulsive electron-electron interaction. The difference between the two mass terms $\mathcal{M}_s$ and 
$\mathcal{M}_t$ stems from the residual symmetry
of the SU(2)$_{\sigma}$ groups. The term (\ref{eq:MCmassS}), which may be viewed as a \textit{singlet mass term} 
explicitly breaks this symmetry, whereas the term (\ref{eq:MCmassTb}) has been coined \textit{triplet mass} \cite{gusynin,gorbar08}.

In Sec. \ref{Sec:Zeeman}, we have argued that mass terms of the above form only lift the valley degeneracy in the zero-energy
LL $n=0$, whereas they simply renormalize the LL energy for $n\neq 0$. Furthermore we have seen that as a consequence of
the vanishing spinor components $\psi_{K,\sigma}^A$ and $\psi_{K',\sigma}^B$, the mass term $\mathcal{M}_t$ is indistinguishable, 
in $n=0$, from a valley-pseudospin ferromagnetic state,
\beqn
\nn
\mathcal{M}_t^{n=0} &=& 
\left\langle\Psi^{\dagger} \left(\bone_{\rm valley} \otimes \tau_{AB}^z\otimes \bone_{\rm spin}\right) \Psi\right\rangle\\
&&\sim \left\langle\Psi^{\dagger} \left(\tau_{\rm valley}^z \otimes \bone_{AB}\otimes \bone_{\rm spin}\right) \Psi\right\rangle,
\eeqn
whereas the singlet mass term simply renormalizes the overall chemical potential, 
\beqn
\nn
\mathcal{M}_s^{n=0} &=& \left\langle 
\Psi^{\dagger} \left(\tau_{\rm valley}^z \otimes \tau_{AB}^z\otimes \bone_{\rm spin}\right) \Psi \right\rangle\\
&&\sim \left\langle \Psi^{\dagger} \left(\bone_{\rm valley} \otimes \bone_{AB}\otimes \bone_{\rm spin}\right) \Psi\right\rangle.
\eeqn
These arguments lead to the conclusion that the magnetic catalysis in $n=0$, i.e. the spontaneous generation of a mass gap due to 
electron-electron interactions, may be fully described in the framework of the SU(4) quantum Hall ferromagnetism. 
Furthermore, the recent observation of a fully lifted spin-valley degeneracy in the $n=1$ graphene LL \cite{deanFQHE10} is naturally understood in the framework of quantum Hall 
ferromagnetism, whereas the mass terms (\ref{eq:MCmassS}) and (\ref{eq:MCmassTb}), obtained from magnetic catalysis, would not provide a fully lifted spin-valley 
degeneracy.

\subsubsection{The quantum Hall effect at $\nu=\pm 1$ and $\nu=0$}
\label{Sec:QHFMexp}

Before discussing the experimental results on the quantum Hall effect, 
a clarification on the filling factor is required. In the preceding parts of this section,
which were concerned with general aspects of the quantum Hall ferromagnet in LLs with internal degrees of freedom, 
the filling factor $\nu=k$ has been defined
with respect to the \textit{bottom} of the partially filled LL. However, in graphene, 
this is at odds with the natural definition of the filling factor (\ref{filling}) in terms of the electronic 
density measured from the charge neutrality point in undoped graphene -- a zero filling factor therefore corresponds to
\textit{two} completely filled spin-valley subbranches ($k=2$) of the $n=0$ LL. In the remainder of this section, we therefore 
make a clear distinction between the two filling factors, and $\nu$ denotes the filling of the $n=0$ LL measured from the
bottom of the level, whereas the \textit{natural} filling factor (\ref{filling}) is from now on denoted by $\nu_G$. Explicitly, the 
relation between the two filling factors reads
\beq\label{eq:FFDico}
\nu=\nu_G+2.
\eeq

Early transport measurements in exfoliated graphene on a SiO$_2$ have revealed broken spin-valley-symmetry states at 
at $\nu_G=0, \pm 1$ and $\pm 4$ \cite{zhang,jiang07}, where the latter corresponds to the LLs $\pm 1$. More recent
experiments on exfoliated graphene on a h-BN substrate have furthermore revealed quantum Hall states at $\nu_G=\pm 3$
\cite{deanFQHE10}, thus completing the full resolution of the spin-valley quartet, not only in $n=0$, but also in $\pm 1$.

The observed states may generally be understood in the framework of the quantum Hall ferromagnetism, but the understanding
of the situation at $\nu_G=0$ requires an additional consideration of the subleading external symmetry-breaking terms discussed in
Sec. \ref{Sec:Zeeman}. The two-stage picture, which we adopt here based on the above discussions, 
may be summarized as follows. (a) The quantum Hall ferromagnetic states are formed to minimize the leading energy given by the
Coulomb interaction. However, because of the (approximate) SU(4) symmetry of the interaction, the orientation
of the quantum Hall ferromagnets is not fixed -- a polarization in the spin channel is as probable as one in the valley channel,
and this yields the high degeneracy of the Goldstone modes described in Sec. \ref{Sec:FMGM}. (b) Therefore, in 
spite of the small energy scale of the external fields, the latter are relevant for the orientation of the ferromagnets and 
for the degeneracy lifting of the Goldstone modes. 

\paragraph{The quantum Hall effect at $\nu_G=\pm 1$.}

For $\nu_G=-1$, only one spin-valley branch is completely filled by electrons.\footnote{For $\nu_G=+1$, the same arguments
apply in terms of holes due to particle-hole symmetry.} The Zeeman effect would give a small energetic advantage to spin-$\da$ 
electrons, such that the two spin Goldstone modes associated with collective excitations to the spin-$\ua$ acquire a $q=0$
gap, given by $\Delta_Z$. In contrast to the spin excitations, the Goldstone mode, which couples the two valleys in the 
spin-$\da$ branch of $n=0$, remains gapless, and the ground state may thus be viewed as a valley-pseudospin ferromagnet
in the spin-$\da$ branch. The activation gap would be given by Eq. (\ref{eq:ActGapSkyrmSym}) for pseudospin skyrmion--anti-skyrmion
pairs, and its associated scaling $e^2/\eps l_B\propto \sqrt{B}$ has indeed been observed experimentally \cite{jiang07}. The
residual valley SU(2) symmetry may be broken by the lattice distortions, which we have discussed in Sec. \ref{Sec:Zeeman}. 
Whereas an out-of-plane lattice distortion would yield a gapped 
valley-pseudospin-wave mode, a Kekul\'e-type in-plane distortion orients the pseudospin ferromagnet in the $XY$-plane,
associated with a gapless U(1) superfluid mode \cite{nomura09}. Notice that the lattice distortion, characterized by the 
energy scale $\Delta_{kek}$ is not in competition, at $\nu_G=\pm 1$, with the Zeeman effect, such that the resulting
ferromagnetic state is the same for $\Delta_Z>\Delta_{kek}$ as for $\Delta_Z<\Delta_{kek}$. In the remainder of this 
section we restrict the discussion of the valley-pseudospin degeneracy lifting to in-plane distortions that seem to be energetically
more relevant than out-of-plane distortions, but the overall picture remains unchanged if the latter are more relevant.

\paragraph{The quantum Hall effect at $\nu_G=0$.}

\begin{figure}
\centering
\includegraphics[width=8.5cm,angle=0]{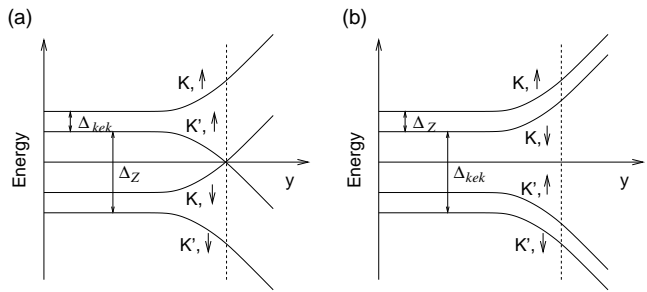}
\caption{\footnotesize{Possible scenarios for the lifted spin-valley degeneracy at $\nu_G=0$.
\textit{(a)} $\Delta_Z>\Delta_{kek}$ in the bulk. When approaching the edge, the energy difference between the two valleys 
increases drastically, and two levels ($K',\ua$) and ($K,\da$) cross the Fermi energy at the edge depicted by the dashed line
(Quantum Hall state).
\textit{(b)} $\Delta_{kek}>\Delta_Z$ in the bulk. 
The $K$ subbranches are already located above the Fermi energy, and those 
of $K'$ below, such that the energy difference is simply increased when approaching the edge with no states crossing the
Fermi energy (Insulator).
}} 
\label{fig:Abanin}
\end{figure}

The situation is more subtle at $\nu_G=0$, where it is not possible to fully polarize both the spin and the valley pseudospin and
where the Zeeman effect is in competition with a lattice distortion that orients
the valley pseudospin. For $\Delta_Z>\Delta_{kek}$, it is favorable to fill both valley sublevels of the spin-$\da$ branch
and the resulting state is a spin ferromagnet with gapped spin-wave excitations. For $\Delta_Z<\Delta_{kek}$, a 
pseudospin-ferromagnetic ground state is favored with both spin sublevels completely filled.  The two different
situations are depicted in Fig. \ref{fig:Abanin}. Most saliently, the two phases reveal drastically different transport properties as
one may see from their behavior at the sample edges.

The electronic behavior at the edges may be described within a model of electron
confinement, in which the sample edge is described via a \textit{mass} confinement term
$M(y)\tau_{AB}^z$ in the Hamiltonian, which has the symmetry of the term
(\ref{eq:MCmassT}) or else, in $n=0$, that of a valley-Zeeman term (\ref{eq:MCZeemanV}), as argued in Sec. \ref{Sec:Zeeman}.
The parameter $M(y)$ is zero in the bulk and increases drastically at the edge at a certain value of the coordinate $y$.\footnote{
For the present argument, we consider translation invariance in the $x$-direction.} Although the model is a simplification to
treat the graphene edges in the continuum description of the Dirac equation, a more sophisticated treatment that takes into 
account the geometry of the edges yields, apart from a fine structure of the levels at the edge, qualitatively similar
results \cite{BreyFertig}. The mass term $M(y)$ modifies the valley coupling due to the lattice distortion and yields a $y$-dependent
term $\sqrt{\Delta_{kek}^2 + M(y)^2}$, which therefore equally diverges at the sample edge.\footnote{In the case of an out-of-plane
distortion, the term $M(y)$ simply adds up to the energy scale $\Delta_Z^{\rm valley}$ [see Eq. (\ref{eq:MCZeemanV})],
but the physical picture remains unaltered.}

These preliminary considerations on the gap behavior at the edges allow us to appreciate the difference 
in the expected electronic transport between a spin
ferromagnet and a valley-pseudospin ferromagnet at $\nu_G=0$.
Indeed, for $\Delta_Z>\Delta_{kek}$, one obtains a quantum Hall state at $\nu_G=0$ that is characterized by a bulk
gap associated with two counter-propagating edge states [Fig. \ref{fig:Abanin}(a)]. In the bulk, where $M(y)=0$,
both valley sublevels of the spin-$\da$ branch are filled (spin ferromagnet).
When approaching the edge, however, the energy term $\sqrt{\Delta_{kek}^2 + M(y)^2}$
is enhanced by the rapidly increasing contibution from $M(y)$, and the ($K,\ua$) level eventually crosses the ($K',\da$)
one at the edge at the Fermi energy \cite{abanin07nu0}. This situation corresponds to a quantum Hall state
with a bulk insulator and (two counter-propagating) conducting channels. 
In contrast to usual quantum Hall states, the edge states are not chiral, but the chiralitiy, i.e. 
the transport direction, of each channel is linked
to its spin orientation.\footnote{These helical edges are the signature of a quantum spin Hall effect \cite{HasanRev10,ZhangRev}.}
The quantum Hall state therefore remains stable unless magnetic impurities couple the two chiralities \cite{shimshoni}.
One notices furthermore a change in the spin polarization at the edge;
whereas the spin polarization in the bulk is complete, the system becomes spin unpolarized at the edge. If one takes into 
account the exchange interaction, the change in the polarization takes place over a certain distance, and the conducting
properties may be described in terms of spin-carrying one-dimensional edge excitations \cite{shimshoni}. 

In the opposite limit with $\Delta_Z<\Delta_{kek}$ in the bulk [Fig. \ref{fig:Abanin}(b)], the system 
at $\nu_G=0$ is already valley-polarized, and an increase of $\sqrt{\Delta_{kek}^2+M(y)^2}$ when approaching the edge does not induce 
a level crossing at the Fermi energy. Thus, there are no zero-energy states at the edge, and the system would be insulating 
both in the bulk and at the edge.

From an experimental point of view, it is not fully settled which of the two phases describes the state at $\nu_G=0$. Whereas 
early experiments in exfoliated graphene on SiO$_2$ samples were discussed in the framework of a dominant Zeeman effect
\cite{zhang,jiang07,abanin07nu0}, more recent experiments at very large magnetic fields \cite{ong1,ong2} and on suspended graphene
samples with increasing mobility
\cite{grapheneFQHE1} favor the insulating scenario of Fig. \ref{fig:Abanin}(b) with a dominant valley 
degeneracy lifting. Especially the high-field measurements hint at an easy-plane or $XY$ (valley-pseudospin) ferromagnetic
ground state because the transition between the metallic and the insulating state is 
reminiscent of a Kosterlitz-Thouless phase transition \cite{KT} if one replaces the temperature by the magnetic field as the
parameter driving the transition \cite{ong1,ong2,nomura09,hou10}. However, it has also been argued
that this effect may be understood within the above scenario of a Zeeman-dominated
quantum Hall ferromagnet in the bulk, in the framework of a Luttinger-liquid description of the domain wall seperating the
polarized from the unpolarized region at the edge \cite{shimshoni}.

One notices that both the Zeeman effect and the Kekul\'e-type
lattice distortion are very close in energy (see Tab. II) such that one may speculate that other effects, as e.g. impurities, strongly 
influence the formation and the orientation of the quantum Hall ferromagnet. Further experimental and theoretical studies 
therefore seem to be necessary to clearly identify the leading symmetry-breaking mechanisms, which 
need not be universal, in the zero-energy LL at $\nu_G=0$ and $\pm1$.

We finally mention scanning-tunneling spectroscopic results for the level splitting at $\nu_G=0$ that were performed on
graphene on a graphite substrate \cite{liSTM09}. Although a gap has been observed as one may expect in the framework of the
above scenario, it saturates as a function of the magnetic field. This is in disagreement with both the $\sqrt{B}$-behavior of
an interaction-dominated gap as well as with the linear dependence of the Zeeman or lattice-distortion effects. 
A probable origin of this gap is the
commensurability of the graphene lattice with the graphite substrate that may break the inversion symmetry between the two 
sublattices by a term of the
type (\ref{eq:MCmassT}). The coupling to the substrate being essentially electrostatic, one would then expect no or only a weak
magnetic-field dependence of the splitting, as observed in the experiment \cite{liSTM09}.

\subsection{Fractional Quantum Hall Effect in Graphene}
\label{sec:FQHE}

The most salient aspect of strongly-correlated 2D electrons in partially filled LLs is certainly the FQHE, which is due
to the formation of incompressible liquid phases at certain \textit{magical} values of the filling factor. As we have already argued
in Sec. \ref{Sec:qual} on the basis of the pseudopotentials, the FQHE is expected to be present in the graphene LLs $n=0$ and $n=1$,
and the main difference with respect to non-relativistic 2D electron systems should arise from the internal SU(4) symmetry
[for a recent theoretical review see Ref. \cite{papic09}]. 

On the experimental level, recent progress in the fabrication of high-mobility samples, e.g. via current annealing
\cite{bolotin08,du08}, has allowed for the observation of several FQHE states in graphene. The first observations 
of a state at $\nu_G=\pm 1/3$ were reported in 2009 on current-annealed suspended graphene samples in the two-terminal configuration 
\cite{grapheneFQHE1,grapheneFQHE2}.\footnote{There are also some weak indications for FQHE states at other filling 
factors than $\nu_G=\pm 1/3$ in these samples.} 
More recently, in 2010, the FQHE has also been observed in the four-terminal geometry, which allows for the simultaneous measurement
of the longitudinal and the Hall resistance, in suspended graphene \cite{ghahariFQHE10} and on graphene on a h-BN
substrate \cite{deanFQHE10}. 

Before commenting in more detail on these first experimental results (indeed, this part of
graphene research has just started), we introduce the theoretical four-component or SU(4) picture of the FQHE in graphene, in terms
of generalized Halperin wave functions. These wave functions, which may be viewed as multi-component generalizations of 
Laughlin's wave function, provide the natural framework for the description of the phenomenon
in view of the model of electrons restricted to a single relativistic LL (Sec. \ref{sec:Model})

\subsubsection{Generalized Halperin wave functions}

The theoretical study of the FQHE is intimitely related to trial $N$-particle wave functions. In 1983, Laughlin proposed 
a one-component wave function \cite{laughlin},
\beq\label{eq:laughlin}
\phi_m^{L}\left(\left\{z_k\right\}\right) = \prod_{k<l}^N\left(z_k-z_l\right)^m e^{ - \sum_k^N|z_k|^2/2},
\eeq
which allows for an understanding of incompressible FQHE states at the filling factors $\nu=1/m$ that are determined
by the exponent $m$ for the particle pairs $k,l$ in Eq. (\ref{eq:laughlin}). The variable $z_k=(x_k - iy_k)/l_B$ is
the complex position of the $k$-th particle, and the form of the Laughlin wave function (\ref{eq:laughlin}) is dictated
by the analyticity condition for wave functions in the lowest LL.\footnote{The lowest-LL condition of analytic wave functions
may seem a very strong restriction when discussing FQHE states in higher LLs. However, the model (\ref{eq:LLHamSU4}) indicates
that all LLs can be treated as the lowest one, $n=0$, if the interaction potential is accordingly modified. We adopt this point
of view here.} Furthermore, the exponent $m$ must be an odd integer as a consequence of the fermionic statistics imposed on
the electronic wave function.
Even if Eq. (\ref{eq:laughlin}) describes only a trial wave function, one can show that it is the \textit{exact}
ground state for a class of model interactions that yield, with the help of Eq. (\ref{eq:HPP}), the pseudopotentials \cite{haldane}
\beq\label{eq:ModelPP}
V_{\ell}>0 ~~ {\rm for} ~~ \ell<m \qquad {\rm and }\qquad V_{\ell}=0 ~~ {\rm for} ~~ \ell\geq m.
\eeq
Although the Coulomb interaction does not fulfill such strong conditions, the pseudopotentials decrease as $1/\sqrt{m}$ for large values 
of $m$. Because the incompressible ground state is protected by a gap that is on the order of $V_1$, one may view the pseudopotentials
$V_{\ell\geq m}$ as an irrelevant perturbation that does not change the nature of the ground state. Indeed, exact-diagonalization
calculations have shown that, for the most prominent FQHE at $\nu=1/3$, the overlap between the true ground state and the Laughlin
state (\ref{eq:laughlin}) is extremely large ($>99\%$) \cite{HaldRez,FOC}.

Soon after Laughlin's original proposal, Halperin generalized the wave function (\ref{eq:laughlin})
to the SU(2) case of electrons with spin, in the absence of a Zeeman effect \cite{halperin83} 
-- one has then two classes of particles, $N_{\ua}$ spin-$\ua$ and $N_{\da}$ spin-$\da$ particles,
which are described by the complex positions $z_{k_\ua}^{(\ua)}$ and $z_{k_\da}^{(\da)}$, respectively. In the (theoretical) absence
of interactions between electrons with different spin orientation, the most natural ground-state candidate would then be a simple
product of two Laughlin wave functions (\ref{eq:laughlin}),
\beq
\phi_{m_{\ua}}^{L}\left(\left\{z_{k_\ua}^{(\ua)}\right\}\right)\phi_{m_{\da}}^{L}\left(\left\{z_{k_\da}^{(\da)}\right\}\right),
\eeq
one for each spin component with the exponents $m_{\ua}$ and $m_{\da}$, respectively, that need not necessarily be identical.
Inter-component correlations may be taken into account by an additional factor 
\beq
\prod_{k_{\ua}}^{N_{\ua}}\prod_{k_{\da}}^{N_{\da}}\left(z_{k_{\ua}}^{(\ua)} - z_{k_{\da}}^{(\da)}\right)^n,
\eeq
where the exponent $n$ can now also be an even integer because the fermionic anti-symmetry condition is concerned only with 
electrons in the same spin state. 

Halperin's idea is easily generalized to the case of more than two components, and the corresponding trial wave function 
for an $\SU(\Nmath)$ quantum Hall system with $\Nmath$ components reads \cite{GR}
\beq\label{eq:GenHalp}
\psi_{m_1,...,m_\Nmath;n_{ij}}^{\SU(\Nmath)}
=\phi_{m_1,...,m_\Nmath}^L \phi_{n_{ij}}^{inter},
\eeq
in terms of the product 
\beq
\phi_{m_1,...,m_\Nmath}^L=\prod_{j=1}^\Nmath\prod_{k_j<l_j}^{N_j}\left(z_{k_j}^{(j)}-z_{l_j}^{(j)}\right)^{m_j}e^{-\sum_{j=1}^{\Nmath}
\sum_{k_j=1}^{N_j}|z_{k_j}^{(j)}|^2/4}
\eeq
of $\Nmath$ Laughlin wave functions and the term
\beq
\phi_{n_{ij}}^{inter}=\prod_{i<j}^{\Nmath}\prod_{k_i}^{N_i}\prod_{k_j}^{N_j}
\left(z_{k_i}^{(i)}-z_{k_j}^{(j)}\right)^{n_{ij}},
\eeq
which describes inter-component correlations. As in the case of Halperin's two-component wave function \cite{halperin83}, the
exponents $m_j$ must be odd integers for fermionic particles whereas the exponents $n_{ij}$ may also be even integers. 
These exponents define a symmetric $\Nmath\times\Nmath$ matrix $M = n_{ij}$, where the diagonal elements are $n_{ii}\equiv m_i$.
This exponent matrix encodes the statistical properties of the quasi-particle excitations, such as their (fractional) 
charge and their statistical angle \cite{WZ,WenZee}.

Moreover, the exponent matrix $M$ determines the component densities $\rho_j$ -- or, equivalently, the component filling factors
$\nu_j=\rho_j/n_B$, 
\beq\label{eq:CompFill}
\left(\begin{array}{c} \nu_1 \\ \vdots \\ \nu_\Nmath \end{array} \right) = M^{-1}
\left(\begin{array}{c} 1 \\ \vdots \\ 1 \end{array} \right),
\eeq
where $\nu=\nu_1 + \hdots +\nu_{\Nmath}$ is the total filling factor measured from the bottom of the lowest LL.
Notice that Eq. (\ref{eq:CompFill}) is only well-defined if the exponent matrix $M$ is invertible. In this case, all
component filling factors $\nu_j$ are completely determined, whereas otherwise some of the component fillings remain unfixed,
e.g. $\nu_1$ and $\nu_2$ for the sake of illustration, although the sum of them ($\nu_1+\nu_2$) is fixed. This is nothing other 
than a consequence of the underlying ferromagnetic properties of the FQHE state that, similarly to the states at $\nu=k$ discussed
in Sec. \ref{sec:QHFM}, are described by subgroups of $\SU(\Nmath)$. 

Finally, we notice that not all SU($\Nmath$) wave functions (\ref{eq:GenHalp}) describe
incompressible quantum liquids with a homogeneous charge density for all components. A generalization of Laughlin's plasma
picture, according to which the modulus square of the trial wave function corresponds to the Boltzmann weight of a classical 2D 
plasma \cite{laughlin}, shows that all eigenvalues of the exponent matrix $M$ must be positive (or zero for states with 
ferromagnetic order). Otherwise, some of the different components phase-separate in the 2D plane because the inter-component
repulsion between them exceeds the intra-component repulsion \cite{dGRG}.

\subsubsection{The use of generalized Halperin wave functions in graphene}
\label{Sec:Halperin}

These general considerations allow us to define the framework for a basic description of the FQHE in graphene where
the SU(4) spin-valley symmetry imposes $\Nmath=4$. Four-component Halperin wave functions are therefore expected to 
play an equally central role in the description of the graphene FQHE as Laughlin's in a one-component or 
Halperin's in two-component systems. In the remainder of this section,
we attribute the four spin-valley components as $1=(\ua,K)$, $2=(\ua,K')$, $3=(\da,K)$, and $4=(\da,K')$.

\paragraph{Fractional SU(4) quantum Hall ferromagnet.}

In a first step, we consider a four-component Halperin wave function in which all components are equal (odd) integers,
$m_j=n_{ij}=m$, regardless of whether they describe intra- or inter-component correlations. One obtains 
then a completely anti-symmetric orbital wave function that is accompanied by a fully symmetric SU(4) spin-valley wave function.

As we have argued in Sec. \ref{Sec:FMGM},
this situation represents precisely a perfect SU(4) quantum Hall ferromagnet -- indeed, for $m=1$, the generalized
Halperin wave function (\ref{eq:GenHalp}) is nothing other than the orbital wave function of the state at $\nu=1$,
i.e. when one of the subbranches is completely filled. The SU(4) symmetry is then spontaneously broken, and the 
group-theoretical analysis presented in Sec. \ref{Sec:FMGM} yields 3 degenerate Goldstone modes that are generalized
spin waves. 

The situation is exactly the same for any other odd exponent $m$, but the orbital wave function (\ref{eq:GenHalp}) 
is then a Laughlin  wave function (\ref{eq:laughlin}) in terms of the particle positions $z_k$ regardless of 
their internal index $j=1,...,4$. The ferromagnetic properties of these wave functions may be described by the
same equations as the spin-wave and skyrmion modes derived in Sec. \ref{sec:QHFM} if one takes into account a 
renormalization of the spin stiffness, as it has been discussed extensively in the literature for SU(2) quantum
Hall ferromagnets \cite{moon,ezawa,sondhi}.
States described by such a wave function are ground-state candidates for the
filling factors $\nu=1/m$, which correspond to the graphene filling factors [see Eq. (\ref{eq:FFDico})] $\nu_G= - 2+1/m$
or hole states at $\nu_G=2- 1/m$. 

There are now two different manners to break the internal SU(4) symmetry explicitly. The simplest one is the same as for
the quantum Hall ferromagnetism at $\nu_G=0$ or $\pm 1$, in terms of external symmetry-breaking fields such as those
discussed in Sec. \ref{Sec:Zeeman}. However, one may also change some of the exponents in the generalized Halperin
wave function (\ref{eq:GenHalp}), in which case one also changes the filling factor. One may for instance consider
the $[m;m-1,m]$ wave function with $m_j=m$ for all $j$, $n_{13}=n_{24}=m-1$, and $n_{12}=n_{14}=n_{23}=n_{34}=m$, which correspond to
a filling factor\footnote{We only discuss electronic states here, but the arguments are equally valid for the
particle-hole symmetric states at $\nu_G = 2 - 2/(2m-1)$.} 
\beq
\nu=\frac{2}{2m-1} \qquad {\rm or} \qquad \nu_G = -2 + \frac{2}{2m-1}.
\eeq
Indeed, the difference in the inter-component exponents explicitly breaks the spin-valley symmetry -- electrons
in different valleys are more weakly correlated (with an exponent $m-1$) than electrons in the same valley (exponent $m$),
regardless of their spin orientation. As a consequence, the filling factors in each of the two valleys, 
$\nu_K=\nu_1+\nu_3$ and $\nu_{K'}=\nu_2+\nu_4$, respectively, are fixed, $\nu_K=\nu_{K'}=1/(2m-1)$, and one
may view the wave function as a state with ferromagnetic spin ordering, but that is valley-pseudospin unpolarized.
Alternatively, the $[m;m-1,m]$ wave function may be interpreted as a tensor product of an SU(2) Halperin $(m,m,m-1)$ 
pseudospin-singlet wavefunction \cite{halperin83} and a completely symmetric (ferromagnetic) two-spinor that describes the 
physical spin. The relevance of the $[m;m-1,m]$ wave function, with $m=3$ ($\nu=2/5$) has been corroborated in recent
exact-diagonalization studes, both in the graphene LLs $n=0$ and $n=1$ \cite{toke2,papic09}.

The SU(4) spin-valley symmetry is fully broken, e.g., in the case of the $[m;m-1,m-1]$ wave function with all $m_j=m$
and off-diagonal $n_{ij}=m-1$. This wave function, which describes a state at 
\beq
\nu=\frac{4}{4m-3} \qquad {\rm or} \qquad \nu_G = -2 + \frac{4}{4m-3},
\eeq
may be viewed as an SU(4) singlet where the filling factors of all spin-valley components are $1/(4m-3)$.
Exact-diagonalization calculations for $N=4$ and 8 particles 
have shown that the $[m;m-1,m-1]$ wave function with $m=3$ (at $\nu=4/9$) describes to great accuracy the ground state 
for a Coulomb interaction (\ref{eq:LLpot}), with overlaps $\Omath_{N=8}=0.992$ in $n=0$ and 
$\Omath_{N=8}=0.944$ and in the $n=1$ graphene LL \cite{papic09}.

\paragraph{A route to understanding the graphene FQHE at $\nu_G=\pm 1/3$.}

The discussion of the above-mentioned states was based on the understanding acquired from quantum Hall systems in
semiconductor heterostructures, where the filling factor is defined with respect to the bottom of the $n=0$ LL. First
experimental observations, however, indicated a prominent FQHE at $\nu_G=\pm 1/3$, which corresponds to 
two completely filled spin-valley sublevels of the graphene $n=0$ LL, and a third one that is 1/3 filled, $\nu=2+1/3$. Such a
state would naturally arise in a system where the SU(4) symmetry is strongly broken, e.g. by a strong Zeeman effect.
However, as argued in Sec. \ref{Sec:Zeeman}, these external fields are weak as compared to the leading
interaction energy scale, and it is therefore natural to ask how such a state may arise from the interaction point
of view in the framework of four-component Halperin wave functions.  

A Halperin wave function that describes the above-mentioned situation is \cite{papic10}
\beqn\label{wave1}
\nn
\psi_{2+1/3}&=&\prod_{\xi=K,K'}\prod_{i<j}\left(z_i^{\ua,\xi}-z_j^{\ua,\xi}\right)^3
\prod_{i,j}\left(z_i^{\ua,K}-z_j^{\ua,K'}\right)^3\\
&&\times\prod_{\xi=K,K'}\prod_{i<j}\left(z_i^{\da,\xi}-z_j^{\da,\xi}\right),
\eeqn
or any permutation of the spin-valley components. One notices that this wave function implicitly breaks the 
SU(4) spin-valley symmetry and, moreover, is not an eigenstate of the full SU(4) pseudospin, such that it cannot
describe the ground state in the total absence of an external symmetry-breaking field. However, exact-diagonalization
calculations have shown that even a tiny external Zeeman field is capable to stabelize the state (\ref{wave1}), which
becomes the ground state for $\Delta_Z^1\simeq 0.01 e^2/\eps l_B$ \cite{papic10}. Furthermore, the state (\ref{wave1})
possesses, in addition to the valley-pseudospin-wave Goldstone mode in the spin-$\ua$ branch, low-lying spin-flip 
excitations for moderately small Zeeman fields, even if the charge (activation) gap is the same as for the usual 
1/3 Laughlin state. This particular interplay between the leading Coulomb energy and subordinate external spin-valley
symmetry breaking terms, illustrated at the $\nu_G=1/3$ example, shows the complexity of the graphene FQHE, and
further surprises may be expected in future experiments.

\subsubsection{Experiments on the graphene FQHE}

We terminate this section on the graphene FQHE with a short discussion of experimental observations in the light of 
the above-mentioned theoretical four-component picture. 

\paragraph{Two-terminal measurements.}

In the first observations of the FQHE, the two-terminal configuration was used, where the voltage (and thus the resistance) 
is measured between the same two contacts used to drive the electric current through the sample \cite{grapheneFQHE1,grapheneFQHE2}.
In this two-terminal configuration, it is not possible to measure simultaneously the Hall and the longitudinal resistance. 
It is nevertheless possible to extract the Hall and the longitudinal conductivities from the two-terminal resistance with the help
of a conformal mapping, as a consequence of the 2D nature of the quantum transport in these systems \cite{AL08,Williams09}. 
This technique has been applied to obtain insight into the longitudinal conductivity the expected activated behavior 
of which yields a rough estimate of the activation gap at $\nu_G=1/3$ ($\Delta_{1/3}\sim 4.4$ K at $B=12$ T) \cite{Abanin10}, which
is an order of magnitude smaller than the theoretically expected value \cite{AC,toke1}.\footnote{Notice that
the theoretical estimates have been obtained within a simplified two-component model, with a completely frozen spin degree of
freedom. In spite of this simplification, the above-mentioned exact-diagonalization calculations with an implemented SU(4) symmetry
have shown that the charge gap, which is responsible for the activated behavior, coincides indeed with that obtained in the
two-component model \cite{papic10}.}

\paragraph{Four-terminal measurements.}

The activation gap of the 1/3 FQHE state has also been measured in suspended graphene in the four-terminal configuration,
in which the longitudinal resistance can be measured directly and independently from the Hall resistance \cite{ghahariFQHE10}. 
In this case, the activation gap has been estimated to be $\Delta_{1/3}\sim 26 ... 50$ K at $B=14$ T, a value that agrees 
reasonably well with the theoretically expected value \cite{AC,toke1} if one considers the energy scale $e^2/\eps\eps_{\infty} l_B$,
which takes into account the RPA dielectric constant $\eps_{\infty}$ for graphene in vacuum (see Sec. \ref{Sec:Diel}).

Finally, we would mention very recent high-field transport measurements in the four-terminal configuration on graphene on
a h-BN substrate \cite{deanFQHE10}. These experiments allowed for a clear identification of several states of the 1/3 family, at
$\nu_G=\pm 1/3,\pm 2/3$, and $\pm 4/3$ corresponding to the zero-energy LL $n=0$, as well as at $\nu_G=\pm 7/3,\pm 8/3,\pm 10/3$, and
$\pm 11/3$ which reside in the $n=1$ LL. The estimation of the activation gap at $\nu_G=4/3$ agrees again reasonably well with 
the theoretical expectation for the 1/3 state. The perhaps most salient (and unexpected) feature of the transport measurement
is the absence (or extreme weakness)
of the $\nu_G=\pm 5/3$ representative of the 1/3 family, which would correspond to the Laughlin state 
($\nu=1/3\leftrightarrow \nu_G=-5/3$ and the corresponding hole state) with a full SU(4) spin-valley ferromagnetic order, 
as argued in Sec. \ref{Sec:Halperin}. 

Whereas the absence of this state remains to be understood, these findings corroborate
the theoretical four-component picture of the graphene FQHE. Indeed, it clearly shows that the 
SU(4) symmetry of the $n=0$ LL is essential because the only correspondence between the FQHE states is particle-hole 
symmetry that maps $\nu_G\leftrightarrow -\nu_G$. If the SU(4) symmetry were broken, e.g. by a sufficiently strong Zeeman effect, the
only symmetry would be the valley-SU(2) symmetry in each spin branch of the $n=0$ LL, in which case there exist the further mappings
$-2+\nu\leftrightarrow -\nu$ in the spin-$\da$ branch and $\nu\leftrightarrow 2-\nu$ in the spin-$\ua$ branch. However, the 
(observed) $\pm 1/3$ state would than be mapped on the (unobserved or extremely weak) $\pm 5/3$ state, and the strong difference in 
the visibility between these two states is therefore difficult to understand. This is also the case if the SU(4) symmetry is fully
broken by strong external spin and valley Zeeman fields, such that all spin-valley sublevels are completely resolved, and $\pm 5/3$
would be mapped on $\pm 4/3$, in the same manner as $\pm 1/3$ on $\pm 2/3$.

\section{Conclusions and Outlook}

We have reviewed the quantum-mechanical properties of relativistic 2D electrons in monolayer graphene exposed to a strong 
magnetic field. The main parts of this review are concerned with the role of electronic interactions in graphene LLs. 
Whereas we have argued that
these interactions may be treated perturbatively in the regime of the relativistic (integer) quantum Hall effect (RQHE), they 
constitute the relevant energy scale in partially filled graphene LLs due to the quenching of the kinetic energy. This is reminiscent
of partially filled LLs in non-relativistic 2D electron systems, and the most prominent consequence of this quenched kinetic 
energy and the macroscopic LL degeneracy is certainly the FQHE. The graphene FQHE is expected to be reminiscent of that of
non-relativistic 2D electrons but it is governed by a larger internal degeneracy described to great accuracy by the SU(4) group. 
The experimental study of the FQHE in graphene is still in its infancy, and novel surprises may be expected. Only recently have been 
reported measurements in the four-terminal geometry which allow for an analysis of prominent characteristics of FQHE states, such as
the activation gaps. In view of the generally accepted universality of the quantum Hall effect,
it will certainly be interesting to make a systematic comparison with the activation gaps of related FQHE states in 
conventional 2D electron gases with a parabolic band.

In the perturbative regime of the RQHE, the theoretical study of electron-electron interactions indicates the presence of fascinating
novel collective modes, such as linear magneto-plasmons, that are particular to graphene and do not have a counterpart in 
non-relativistic 2D electron systems in a perpendicular magnetic field. Also the upper-hybrid mode, which is the magnetic-field
counterpart of the usual 2D plasmon, is expected to behave in a particular manner in graphene as a consequence of the linear
disperison relation and the vanishing band mass. Whereas these studies are at present only theoretical, these collective
modes may find an experimental verification in inelastic light-scattering measurements. 

Similarly to the role of electron-electron interactions in the RQHE regime, the electron-phonon coupling yields exciting 
resonance phenomena in graphene in a strong magnetic field. The electron-phonon interaction in graphene LLs has been discussed in 
the framework of a perturbative approach. Indications for the magneto-phonon resonance, e.g., have recently 
been found in Raman spectroscopy of epitaxial graphene.

The present review has been limited to monolayer graphene, and it is definitely a reasonable research program to ask how the 
effects described here manifest themselves in bilayer graphene. For example, the particular collective excitations described in Sec. 
\ref{Chap:Ints} have been attributed to the lack of equidistant LL spacing and the presence of two bands. Whereas bilayer graphene
also consists of two (particle-hole-symmetric) bands in the low-energy regime, the approximate parabolicity there yields almost 
equidistant LLs. The presence of additional high-energy bands (in the 300 meV range) certainly also affects the plasmonic modes.

\begin{acknowledgments}
I would like to express my deep gratitude to numerous collaborators without whom the realisation of this review would not have been
possible. Above all, I must acknowledge the very fruitful long-term collaborations with Jean-No\"el Fuchs, 
on electron-electron interactions in the IQHE regime and electron-phonon coupling, and with Nicolas Regnault on the FQHE
in graphene. I would furthermore thank my collaborators Claire Berger, Rapha\"el de Gail, Beno\^it Dou\c cot, 
Volodya Fal'ko, Cl\'ement Faugeras, Kostya Kechedzhi, Pascal Lederer, Roderich Moessner, 
Gilles Montambaux, Cristiane Morais Smith, Zlatko Papi\'c, Fr\'ed\'eric Pi\'echon, Paulina Plochocka, Marek Potemski,
Rafael Rold\'an, and Guangquan Wang. Many thanks also to my colleagues H\'el\`ene Bouchiat, Antonio Castro Neto, Jean-No\"el
Fuchs, Christian Glattli, Paco Guinea, Anuradha Jagannathan, Philip Kim, and Bernhard Wunsch 
for fruitful discussions and a careful reading of this review.

\end{acknowledgments}

\appendix

\section{Matrix Elements of the Density Operators}
\label{app:ME}

The matrix elements that intervene in the expression for the density operators (\ref{eq:DensB}) are of the form
$\langle n,m| \exp(-i\bq\cdot\br)| n',m'\rangle$ and may be calculated with the help of the decomposition
of the cyclotron variable $\etab$ and the guiding centre $\bR$ into the ladder operators $\ahat$ and $\bhat$,
respectively [see Eqs. (\ref{ladder}), (\ref{CyclVar}) and (\ref{laddGC})].
We furthermore define the complex wave vectors $q\equiv (q_x+iq_y)l_B$ and $\qbar=(q_x-iq_y)l_B$,\footnote{We use this
notation solely in the present appendix. Throughout the main text, $q$ denotes the modulus
of the wave vector $\bq$, $q=|\bq|$.}
One finds
\begin{eqnarray}
\label{eqnA01}
\nonumber
\langle n,m| e^{-i\bq\cdot\br}| n',m'\rangle&=&
\langle m| e^{-i\bq\cdot\bR}| m'\rangle\otimes 
\langle n| e^{-i\bq\cdot\etab}| n'\rangle \\
&=&\langle m| e^{-\frac{i}{\sqrt{2}}(q\bhat^{\dagger}+\qbar \bhat)}| m'\rangle\\
\nn
&&\otimes 
\langle n| e^{-\frac{i}{\sqrt{2}}(\qbar \ahat^{\dagger}+q\ahat)}| n'\rangle.
\end{eqnarray}
The two matrix elements may be simplified with the
help of the Baker-Hausdorff formula $\exp(A)\exp(B)=\exp(A+B)\exp([A,B]/2)$, for the case
 $[A,[A,B]]=[B,[A,B]]=0$ \cite{CT}. The second matrix element thus becomes, for $n\geq n'$
\begin{eqnarray}
\label{eqnA02}
\langle n| e^{-i\bq\cdot\etab}| n'\rangle&=&\langle n| e^{-\frac{i}{\sqrt{2}}(\qbar \ahat^{\dagger}+q\ahat)}|
n'\rangle\\
\nonumber
&=& e^{-|q|^2/4}\langle n| e^{-\frac{i}{\sqrt{2}}\qbar \ahat^{\dagger}}
e^{-\frac{i}{\sqrt{2}}q \ahat}| n'\rangle\\
\nonumber
&=&e^{-|q|^2/4}\sum_j \langle n| e^{-\frac{i}{\sqrt{2}}\qbar \ahat^{\dagger}}
| j\rangle \langle j| e^{-\frac{i}{\sqrt{2}}q \ahat}| n'\rangle\\
\nonumber
&=&e^{-|q|^2/4}\sqrt{\frac{n'!}{n!}}\left(\frac{-i\qbar}{\sqrt{2}}\right)^{n-n'}\\
\nn
&&\times \sum_{j=0}^{n'}\frac{n!}{(n-j)!(n'-j)!j!}\left(-\frac{|q|^2}{2}\right)^{n'-j}\\
\nn
&=&e^{-|q|^2/4}\sqrt{\frac{n'!}{n!}}\left(\frac{-i\qbar}{\sqrt{2}}\right)^{n-n'}
L_{n'}^{n-n'}\left(\frac{|q|^2}{2}\right),
\end{eqnarray}
where we have used
$$\langle n| e^{-\frac{i}{\sqrt{2}}\qbar \ahat^{\dagger}}| j\rangle=\left\{\begin{array}{c}0\qquad\qquad\qquad\qquad\qquad{\rm for}~j>n\\
\sqrt{\frac{n!}{j!}}\frac{1}{(n-j)!}\left(-\frac{i}{\sqrt{2}}\qbar\right)^{n-j} ~ {\rm for}~j\leq n
\end{array}\right.$$
in the third line and the definition of the associated Laguerre polynomials \cite{gradsht},
$$L_{n'}^{n-n'}(x)=\sum_{m=0}^{n'}\frac{n!}{(n'-m)!(n-n'+m)!}\frac{(-x)^m}{m!}.$$
In the same manner, one obtains for $m\geq m'$
\begin{eqnarray}
\label{eqnA03}
\nonumber
\langle m| e^{-i\bq\cdot\bR}| m'\rangle&=&\langle m| e^{-\frac{i}{\sqrt{2}}(q \bhat^{\dagger}+\qbar \bhat)}| m'\rangle\\
\nn
&=& e^{-|q|^2/4}\sqrt{\frac{m'!}{m!}}\left(\frac{-iq}{\sqrt{2}}\right)^{m-m'}\\
&&\times L_{m'}^{m-m'}\left(\frac{|q|^2}{2}\right).
\end{eqnarray}
With the help of the definition
$$\mathcal{G}_{n,n'}(q)\equiv\sqrt{\frac{n'!}{n!}}\left(\frac{-iq}{\sqrt{2}}\right)^{n-n'}
L_{n'}^{n-n'}\left(\frac{|q|^2}{2}\right),$$
one may rewrite the expressions without the conditions $n\geq n'$ and $m\geq m'$,
\beqn
\label{eqnA04}
\langle n| e^{-i\bq\cdot\etab}| n'\rangle &=&\left[\Theta(n-n')\mathcal{G}_{n,n'}(\qbar)\right.\\
\nn
&&\left.+\Theta(n'-n-1)\mathcal{G}_{n',n}(-q)\right] e^{-|q|^2/4}
\eeqn
and
\beqn
\label{eqnA05}
\langle m| e^{-i\bq\cdot\bR}| m'\rangle &=& \left[\Theta(m-m')\mathcal{G}_{m,m'}(q)\right.\\
\nn
&&\left.+\Theta(m'-m-1)\mathcal{G}_{m',m}(-\qbar)\right] e^{-|q|^2/4}.
\eeqn

\bibliography{refsHabil}

\end{document}